\documentclass[final,5p,times,twocolumn,super]{elsarticle}
\journal{xxx}

\usepackage{amssymb}
\usepackage{siunitx}
\usepackage{amsmath}
\usepackage[utf8]{inputenc} 

\newcommand*{\mytitle}{When and How Ultrasound Enhances Nanoparticle Diffusion in Hydrogels: \\ A Stick-and-Release Mechanism}

\newcommand*{\hedda}{Hedda H. Rønneberg}
\newcommand*{\pablo}{Pablo M. Blanco}
\newcommand*{\rita}{Rita S. Dias}

\affiliation[NTNU]{organization={Department of Physics, NTNU - Norwegian University of Science and Technology},
            city={Trondheim},
            postcode={NO-7491}, 
            country={Norway.}}

\newcommand*{\eg}{\emph{e.g.}}
\newcommand*{\etal}{\emph{et al.}}
\newcommand*{\ie}{\emph{i.e.}}
\newcommand*{\cf}{\emph{cf.}}
\newcommand*{\via}{\emph{via}}
\newcommand*{\reffig}[1]{Fig.~\ref{#1}}
\newcommand*{\refsec}[1]{Section~\ref{#1}}

\newcommand*{\refeq}[1]{Eq.~\ref{#1}}

\newcommand*{\refsecSIswellingeq}{Section~S1.3}

\newcommand*{\refsecfixednetwork}{Section~S2.1}

\newcommand*{\refsecmass}{Section~S4}

\newcommand*{\refeqD}{Eq. 3}

\newcommand*{\kb}{k_\mathrm{B}}

\newcommand*{\kT}{k_\mathrm{B}T}

\newcommand*{\Nbead}{N_\mathrm{bead}}
\newcommand*{\rbead}{d_\mathrm{bead}}
\newcommand*{\rNP}{d_\mathrm{NP}}
\newcommand*{\rmax}{r_{\mathrm{max}}}
\newcommand*{\kfene}{k^{\mathrm{FENE}}}
\newcommand*{\kbend}{k^{\mathrm{bend}}}
\newcommand*{\rcut}{r^{\mathrm{cut}}}
\newcommand*{\roff}{r^{\mathrm{off}}}
\newcommand*{\Psys}{P^{\mathrm{sys}}}
\newcommand*{\Pres}{P^{\mathrm{res}}}
\newcommand*{\Fext}{\mathbf{F}^{\mathrm{ext}}}
\newcommand*{\MSD}{\langle R(\tau)^{2}\rangle}
\newcommand*{\bfxi}{\boldsymbol{\xi}}
\newcommand*{\dt}{\textrm{d}t}
\newcommand*{\tf}{{t_\mathrm{f}}}
\newcommand*{\Dshort}{D^\mathrm{short}}
\newcommand*{\Dlong}{D^\mathrm{long}}
\newcommand*{\dvdt}{\frac{{\rm d} \mathbf{v}}{{\rm d} t}}
\newcommand*{\vmax}{v_{\mathrm{max}}}
\newcommand*{\tc}{\tau_\mathrm{c}}
\newcommand*{\rc}{r_\mathrm{c}}
\newcommand*{\Dt}{D(\tau)}
\newcommand*{\rpore}{r_\mathrm{pore}}
\newcommand*{\mbead}{m_\mathrm{bead}}
\newcommand*{\mNP}{m_\mathrm{NP}}
\newcommand*{\mus}{\,\mu\mathrm{s}}

\newcommand*{\Da}{D_\mathrm{A}}
\newcommand*{\Dm}{D_\mathrm{M}}
\newcommand*{\tus}{t_{\mathrm{US}}}
\newcommand*{\tpulse}{t_{\mathrm{pulse}}}
\newcommand*{\Pmax}{P_{\textrm{max}}}
\newcommand*{\Cp}{C_{\textrm{pol}}}
\newcommand*{\np}{\rho_{\textrm{pol}}}

\begin{document}

\begin{frontmatter}

\title{\mytitle}
\author[NTNU]{\pablo\corref{cor1}}
\cortext[cor1]{pablb@ntnu.no; pablomiguel.blanco@udl.cat}
\author[NTNU]{\hedda}
\author[NTNU]{\rita}

\begin{abstract}
Nanoparticles (NPs) are widely used as drug carriers in cancer therapy due to their ability to accumulate in tumor tissue via the enhanced permeability and retention  effect.
However, their transport within tumors is often hindered by the dense extracellular matrix (ECM), where diffusion dominates. 
Several studies suggest that ultrasound (US) irradiation can enhance NP diffusion in ECM-mimicking hydrogels, yet the underlying molecular mechanisms remain unclear, and experimental findings are often contradictory.

Here, we use coarse-grained Langevin Dynamics simulations to investigate the conditions under which US can enhance NP diffusion in hydrogels. 
After validating our simulation framework against an exact analytical solution for NP motion under US in dilute buffer, we systematically explore NP diffusion in hydrogels with varying degrees of NP-network attraction.

Our results reveal that acoustic enhancement arises from reduced contact time between NPs and the hydrogel matrix. 
This effect becomes significant only when NP-hydrogel interactions are sufficiently strong and US pulses are long enough to disrupt these interactions, following a “stick-and-release” mechanism, in which NPs  stick transiently to the network and are periodically released by the US oscillations.

These findings reconcile previously conflicting experimental observations and explain why acoustic enhancement is observed in some studies but not others.
Overall, our study provides a molecular-level explanation for US-enhanced NP diffusion in hydrogels and establishes design principles for optimizing therapeutic US protocols in drug delivery applications.
\end{abstract}







\end{frontmatter}




\section{\label{sec:introduction}Introduction}
Effective strategies for enhancing targeted drug delivery into tumor tissue remain a critical unmet need in cancer therapy.
One promising approach involves the use of nanoparticles (NPs) as drug carriers to exploit the enhanced permeability and retention effect, which allows NPs to preferentially cross the leaky blood vessel walls of tumors while being largely excluded from healthy tissue.\cite{peer20a,patra18a}

\begin{figure*}[t]
    \centering
    \includegraphics[width=0.95\linewidth]{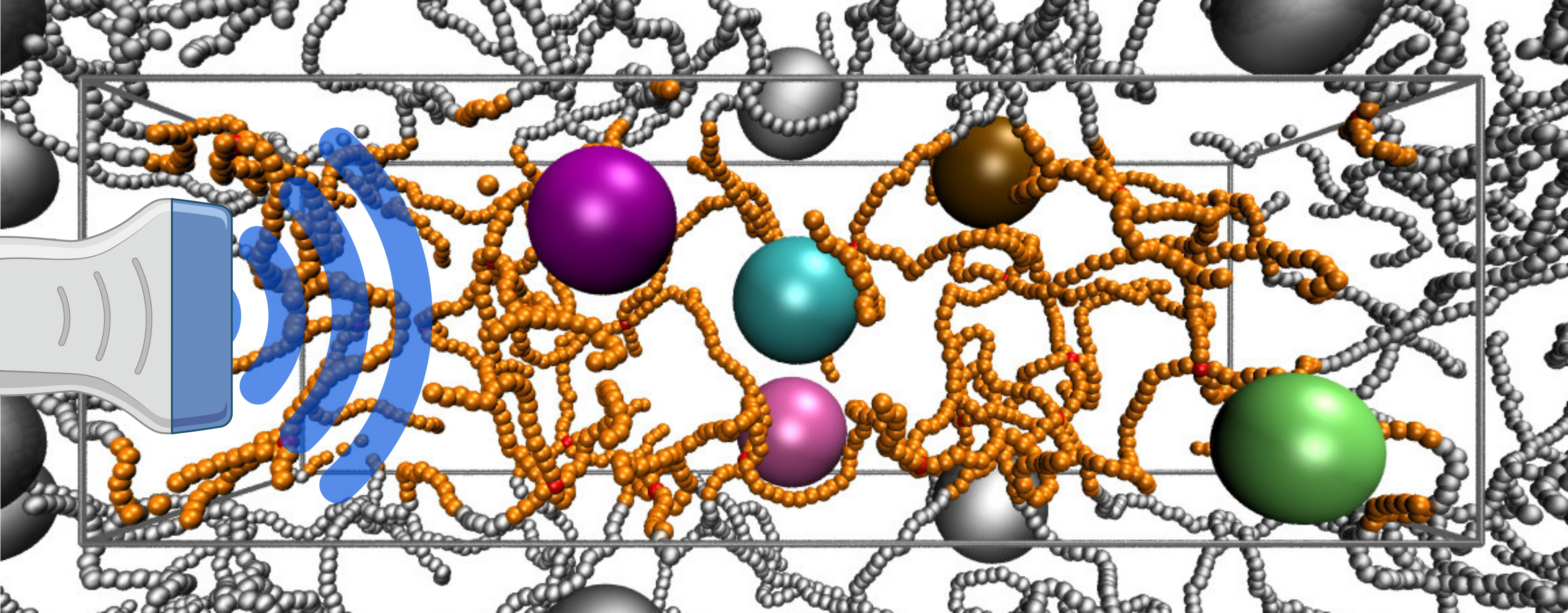}
    \caption{Simulation snapshot of nanoparticles (NPs) diffusing in a hydrogel with a polymer volume fraction of $0.86\%$.
    The system is subjected to an ultrasound (US) wave propagating along the long axis $L_\parallel$ of the simulation box (a square prism with shorter edges $L_\perp = 1/3L_\parallel$), 
    schematically illustrated by a speaker in the left corner of the box (created in  https://BioRender.com).
    This snapshot corresponds to the “steric network” scenario, where NP-hydrogel interactions are purely steric.
    Color code: hydrogel fibril beads (orange), hydrogel node beads (red), NPs (purple, pink, cyan, brown, green), periodic images (gray).}
    \label{fig:snapshot}
\end{figure*}

Following extravasation, however, NPs must traverse the dense and complex extracellular matrix (ECM), a hydrogel-like network composed primarily of collagen and elastin fibers, embedded in a matrix of glycosaminoglycans and proteoglycans.
The composition of the ECM strongly determines its pore architecture, which directly governs the diffusion of nutrients, drugs, and NPs.
Pore sizes vary widely across tissues: from a few nanometers in glomerular and tubular basement membranes to tens of nanometers in the cerebral extracellular space, and up to several hundred nanometers in softer, hydrogel-like matrices such as the vitreous or mucus.\cite{tomasetti18a}
These dimensions are comparable to the typical size of drug-loaded NPs (tens to hundreds of nanometers), meaning that NP diffusion is strongly hindered by both steric obstruction, due to the limited pore size and fiber crowding, and specific adhesive interactions with ECM components such as collagen or hyaluronic acid, which can transiently trap particles within the network.\cite{motezakker24a}
Due to elevated interstitial fluid pressure in tumors, which suppresses convective transport, NP movement within the ECM is dominated by diffusion,\cite{heldin04a}
which, as mentioned, is severely hindered, leading to NP accumulation near capillary walls and poor drug distribution throughout the tumor tissue.\cite{davies04a,yemane19a}

Therapeutic ultrasound (US)  has emerged as a promising modality to enhance drug delivery in tumors.\cite{moradikashkooli23a} 
High-intensity US can disrupt liposomal carriers, leading to the release of their contents, and facilitate transmembrane transport of therapeutic agents.\cite{schroeder09a, sundaram03a}
When combined with microbubbles, US is particularly effective: the oscillating microbubbles in the bloodstream induce oscillations in the vessel wall itself,\cite{chen11b} which in turn can enhance NP transport through the extracellular matrix.
These effects are typically attributed to acoustic cavitation of microbubbles. \cite{moradikashkooli23a,snipstad21a}

Interestingly, enhanced transport and penetration of liposomes in biofilms,\cite{ma15c} and of NPs in agarose gels,\cite{ma18a, karki22a} have also been observed under low-intensity US irradiation, insufficient to trigger bubble formation in the medium. 
These findings suggest that US may enhance NP thermal diffusion in biological hydrogels, such as the extracellular matrix, and other porous media. \cite{marshall16a,ma18a,marshall21a,karki22a}

Several {\it in vitro} studies have explored this hypothesis.
Wu and co-workers\cite{ma18a, karki22a} reported up to a eightfold increase in NP diffusion in agarose gels under US irradiation, with enhancement depending on NP size, US frequency, and amplitude.
In contrast, Davies and co-workers\cite{lovmo21a} found no improvement in NP penetration in collagen hydrogels, observing only mechanical deformation of the collagen matrix.
In a follow-up study using agarose gels and NPs of similar size to those in the experiments by Wu \etal.,\cite{einen23a} they reported a modest 0.3-fold increase in NP diffusion under US pulses with a 20\% duty cycle, and no enhancement at lower duty cycles.
These seemingly contradictory results underscore the need for a deeper mechanistic understanding of US-enhanced NP diffusion in hydrogels.

To our knowledge, only two analytical models have been proposed to describe oscillatory diffusion in porous media.
The two-state random walk (TSRW) model by Balakrishnan and Venkataraman,\cite{balakrishnan81a} describes particles alternating between thermal diffusion and oscillatory displacement.
More recently, Marshall and co-workers\cite{marshall16a, curran21a, marshall21a} developed a stochastic model incorporating US-driven oscillations and particle retention by the hydrogel matrix, predicting an effective acoustic diffusion coefficient as a function of US parameters.
While Marshall’s model qualitatively captures trends observed by Wu and co-workers,\cite{ma18a, karki22a} none
of the models explains the lack of enhancement reported by Davies and co-workers.\cite{lovmo21a, einen23a}

Computer simulations have been widely used to investigate NP diffusion in biological media, including ECM-like hydrogels\cite{hansing18a,hansing18b,motezakker24a} and crowded cytosolic environments.\cite{blanco17a,wang17a,blanco18b}
In these systems, NP diffusion is often anomalous due to hindered motion caused by macromolecular obstacles.
However, molecular-level simulations explicitly incorporating US to investigate nanoscale phenomena remain relatively  limited.\cite{okumura14a,man20a,araki21a,man21a,liang22a,price24a,lah25a}
To our knowledge, only one study has examined NP diffusion in hydrogels under US at the molecular level: Price \etal.\cite{price24a} reported anomalous sub-diffusion under US, consistent with experimental findings by Davies \etal.\cite{einen23a}
Yet, the molecular mechanisms governing US-enhanced NP diffusion in hydrogels and other biological tissues remain poorly understood.

In this work, we investigate the  conditions under which US can enhance NP diffusion in hydrogels using a coarse grained molecular model (see \reffig{fig:snapshot}). 
Our model features NPs diffusing throughout a periodic semi-flexible polymeric network and is studied using Langevin Dynamics (LD) with an external US-like force.
We adopt a bottom-up approach: starting with NP motion in an ideal dilute buffer solution, where the mean square displacement has an exact analytical solution, we validate our LD simulations under US.
We then extend our analysis to hydrogels with varying degrees of NP-network attraction.
Finally, we compare our results with experimental data to reconcile the divergent observations reported in the literature.

\section{\label{sec:theory} Theoretical background}
Molecular diffusion of NPs is traditionally quantified using the mean squared displacement
\begin{equation}
    \MSD \equiv \left< \left( \mathbf{x}(\tf) - \mathbf{x}(t_0)  \right)^2\right>,
    \label{eq:msd}
\end{equation}
where $\tau = \tf-t_0$ represents the lag time, defined as the difference between the final $\tf$ and the initial $t_0$ observation times.
The angle brackets $\left< \dots \right>$ indicate that the $\MSD$ is computed as an average over a statistical ensemble of multiple trajectories.
This averaging is essential because the trajectory of a single NP is inherently chaotic due to frequent collisions with solvent molecules.
Such random motion, known as Brownian motion, is a hallmark of NPs suspended in solution.

The foundational framework for interpreting the $\MSD$ of NPs is given by the
Einstein-Smoluchowski equation, which establishes a direct linear relationship between the $\MSD$ of a particle and the lag time
\begin{equation}
    \MSD = 
    2dD\tau. 
    \label{eq:einstein_smo}
\end{equation}
where $d$ is the dimensionality and $D$ is the diffusion coefficient of the particle.
For spherical NPs in a viscous medium, $D$ can be calculated from the Stokes-Einstein equation
\begin{equation}
    D_0 = \frac{\kT}{\gamma} = \frac{\kT}{3\pi\eta_0 d_{\rm H}},
    \label{eq:stokes_einstein}
\end{equation}
where $\gamma$ and $d_{\rm H}$ are the friction (or Stokes) coefficient and the hydrodynamic (or Stokes) diameter of the particle, respectively, $\eta_0$ is the viscosity of the solvent,  and $\kT$ is the thermal energy, defined by the product of the temperature $T$ and the Boltzmann constant $\kb$.
The subscript in $D_0$ emphasizes that \refeq{eq:stokes_einstein} is generally valid only in the ideal dilute solution limit.
When the concentration of macromolecules increases beyond this limit, molecular interactions become significant, and the diffusion of NPs may no longer be accurately described by Eqs. \ref{eq:einstein_smo} and \ref{eq:stokes_einstein}.
Under such conditions, molecular interactions cause the diffusion behavior of NPs to vary across different time scales, as discussed below. \cite{blanco18b,blanco17a}

At short time scales -- below the characteristic collision time of the NPs, typically on the order of a few nanoseconds -- molecular diffusion follows the Einstein-Smoluchowski equation (\refeq{eq:einstein_smo}).
However, the diffusion coefficient in this regime, denoted $\Dshort$, is generally smaller than the ideal dilute-limit value $D_0$, due to hydrodynamic interactions between NPs and surrounding macromolecules mediated by the solvent.\cite{tokuyama94a,tokuyama06a} 

At intermediate time scales, comparable to the collision time -- ranging from tens of nanoseconds to several microseconds -- NPs exhibit anomalous diffusion, which deviates from the linear behavior predicted by \refeq{eq:einstein_smo}. Instead, the $\MSD$ follows a power-law dependence on time\cite{saxton94a}
\begin{equation}
    \MSD = 
    2d \Gamma \tau^\alpha,
    \label{eq:anomalous_diffusion}
\end{equation}
where $\alpha$ is the so-called anomalous exponent and $\Gamma$ is a generalized transport coefficient.
A key implication of \refeq{eq:anomalous_diffusion} is that the diffusion coefficient, as defined by the Einstein-Smoluchowski relation (\refeq{eq:einstein_smo}), becomes time-dependent
\begin{equation}
    \Dt \equiv \frac{\MSD}{2d\tau}=\Gamma \tau^{\alpha-1}.
    \label{eq:time_dependent_D}
\end{equation}
This time dependence is sometimes overlooked in experimental studies of NP diffusion, potentially leading to inaccurate or ill-defined estimates of the diffusion coefficient.

At long, macroscopic time scales, standard diffusion is recovered, and the $\MSD$ once again follows \refeq{eq:einstein_smo} with a constant diffusion coefficient $\Dlong$.
In crowded systems, $\Dlong$ can be significantly smaller than both $D_0$ and $\Dshort$.\cite{ando13a,vilaseca11a,tokuyama06a,blanco18b,blanco17a}
This long-time diffusion coefficient can be interpreted as an effective diffusion coefficient that incorporates the macroscopic viscosity $\eta$  of the medium, including contributions from macromolecules, into the Stokes-Einstein equation (\refeq{eq:stokes_einstein}).
In practice, $\Dlong$ often corresponds to the diffusion coefficient measured in experiments, where lag times typically span milliseconds or longer.\cite{banks05a,pastor10a,ma18a,karki22a,einen23a}

\section{\label{sec:methods} Method}

\subsection{\label{sec:model} Molecular model for NP diffusion in soft tissue}
We have used a coarse-grained model to study the impact of an US-like oscillatory force in the diffusion of NPs in soft tissue. The model consists of a periodic, semi-flexible polymeric network and rigid NPs, as illustrated in the simulation snapshot shown in \reffig{fig:snapshot}. 
The solvent is treated implicitly through the equations of motion, as detailed in \refsec{sec:simulation_method}.
For brevity, we provide only a general overview of the model here and refer the reader to Appendix A for a complete technical description.

The hydrogel is a diamond lattice network composed of 32 polymer chains and 16 nodes, each with a functionality of 4, arranged in a diamond lattice structure.
Each polymer chain consists of 30 linearly connected monomer beads.
Bonded interactions within the hydrogel are modeled using Finite Extensible Non-linear Elastic (FENE) bonds and a harmonic bending potential, employing standard parameter values commonly used in similar hydrogel models.\cite{kosovan15a,beyer22a}
Steric repulsion between hydrogel beads is introduced \via~ the Weeks-Chandler-Andersen (WCA) potential, assuming that all beads have a diameter $\rbead$.

NPs are modelled as rigid spheres with diameter $\rNP$. 
Interactions between NPs, as well as between NPs and hydrogel beads, are governed by an augmented Lennard-Jones (LJ) potential.\cite{beyer24a}
For NP-NP interactions, the LJ potential simplifies to a repulsive WCA form with an offset.
The NP-hydrogel interactions are tuned by varying the depth of the potential well $\epsilon$ and the cut-off distance, allowing to explore networks with different degrees of stickiness. 

We have parametrized the model to replicate key features of NP diffusion in agarose hydrogels, aligning with reported experimental studies that track NP transport under US.\cite{ma18a,einen23a,karki22a}
NPs were modeled as rigid spheres with a diameter $\rNP = 10\, \rbead = \qty{100}{nm}$, consistent with the size of the polystyrene NPs used in the some of the above-mentioned experiments.\cite{ma18a,einen23a}
The hydrogel bead diameter was set to $\rbead = \qty{10}{nm}$, approximately matching the persistence length of an agarose fibril ($\sim\qty{9}{nm}$), as measured in the sol state by neutron scattering.\cite{guenet93}
The swelling equilibrium of the network was found to occur at a hydrogel volume fraction  $\phi \approx 0.86\%$ (\cf ~ \refsecSIswellingeq), which falls within the agarose concentration range used in the experimental studies discussed herein (approximately 0.5--0.9\%).\cite{ma18a,einen23a,karki22a} 
Consequently, $\phi=0.86\%$ provides an experimentally relevant network density for direct comparison between simulation and experiment, going beyond previous attempts using more generic models. \cite{price24a}

\subsection{\label{sec:simulation_method} Langevin Dynamics under an US field}
We simulate the motion of each particle in the model with time $t$  using Langevin Dynamics (LD) under an external field, given by
\cite{pathria96a,vangusteren81a}
\begin{equation}
    m \dvdt = 
    - \gamma \mathbf{v} 
    + \sqrt{2\gamma\kT}\bfxi
    - \nabla \mathbf{V}(\mathbf{x})
    + \Fext, 
    \label{eq:langevin}
\end{equation}
where $m$, $\mathbf{v}$ and $\gamma$ are the mass, velocity and the friction (or Stokes) coefficient of the particle, respectively.
$\bfxi$ is a Gaussian white noise term with zero mean and delta-correlated variance,  modeling the random thermal motion of the Brownian particle due to solvent collisions.
The potential gradient $\nabla \mathbf{V}$ at position $\mathbf{x}$, accounts for molecular interactions and is given by the force field of the molecular model, as detailed in Appendix A.

We describe the effect of a US wave traveling through the medium as the transfer of momentum from the implicit solvent to the explicit particles in the model using an external oscillatory force,
\begin{equation}
    \Fext =   A \cos(2\pi f t)\hat{\mathbf{n}} = \vmax \gamma \cos(2\pi f t)\hat{\mathbf{n}},
    \label{eq:F_osc}
\end{equation}
where $A$ is the amplitude of the oscillatory force, $f$ is the wave frequency, and $\hat{\mathbf{n}}$ is the unit vector in the direction of wave propagation.
We set the amplitude of the US force to $A = \vmax \gamma$, 
where $\vmax$ is the peak velocity in the oscillations of the fluid due to the US wave. 
Assuming linear acoustics, $\vmax$ can be directly related to the applied peak acoustic pressure\cite{kinsler00a}
\begin{equation}
    \Pmax = \rho c \vmax
    \label{eq:p_max}
\end{equation}
where $\rho$ is the density of the fluid and $c$ is the speed of sound. 
Although $\vmax$ is the direct input of our LD simulations, we report the results as a function of $\Pmax$ to enable direct comparison with \emph{in vitro} experiments measuring NP diffusion coefficient in agarose hydrogels\cite{ma18a,karki22a,einen23a} and drug delivery applications,\cite{sontum15a,snipstad2025a} which traditionally report $\Pmax$.
We calculate $\Pmax$ considering the density $\rho = \qty{997}{kg/m}^3$ of water at room temperature\cite{tanaka01d} and the speed of sound $c = \qty{1497}{m/s}$ in the same medium.\cite{bilaniuk93a}
We focus on the case of an US wave with frequency $f = \qty{1}{MHz}$ as it is representative of US waves typically used in drug delivery applications.\cite{einen23a,ma18a,sontum15a,snipstad2025a}

Preliminary computer simulations revealed that the specific way in which US irradiation is modeled to interact with the hydrogel network critically influences NP diffusion under US exposure. 
Inaccurate or overly simplistic representations can result in unphysical behaviors, as discussed in Appendix B.
For example, imposing a fixed oscillation amplitude independent of particle size, creates a nonphysical scenario where hydrogel particles oscillate faster than NPs, effectively dragging the NPs along.
While previous studies have addressed this issue by immobilizing the hydrogel network,\cite{price24a} our simulations indicate that this approach introduces caging effects even at very low hydrogel volume fractions (\cf~\refsecfixednetwork).
To avoid such artifacts, our model scales the US-induced force by particle size via $\gamma$ (\cf~\refeq{eq:stokes_einstein}), ensuring that both NPs and hydrogel particles respond proportionally to the US field.

We highlight the following simplifications in our method:

(i) \textbf{Spatial uniformity of the oscillatory field.}  
At our chosen frequency, the acoustic wavelength in soft tissue is $\lambda = c/f \approx1.5\,$mm, which is many orders of magnitude larger than both the particle size ($\rNP = \qty{100}{nm}$) and the simulation box length (\cf~ \refsec{sec:computational_details}).
Consequently, the imposed oscillatory force can be taken as spatially uniform across the simulation domain.

(ii) \textbf{Neglect of acoustic streaming.} Typical acoustic attenuation coefficients in soft tissues ($\sim$1\,dB/cm)\cite{menikou2017a} imply that spatial gradients and associated streaming velocities develop over length scales much larger than our simulation box length; for this reason, acoustic streaming is neglected in the present microscale simulations. 
We note that acoustic streaming can be important in macroscopic experimental setups and should be accounted for when interpreting bulk transport measurements.

(iii)  \textbf{Omission of Hydrodynamic Interactions (HI).} 
Our LD scheme accounts for solvent-mediated dissipation via a friction term but does not include HI. 
This choice was made to focus on the molecular-scale interaction between the NP and the hydrogel matrix and to keep the computational cost tractable. 
We acknowledge that HI can quantitatively affect relaxation times, contact durations, and diffusion coefficients, particularly at higher polymer densities or in closely confined geometries. \cite{ando13a,tokuyama06a,blanco17a}
However, the impact of HI is expected to be minimal at the low polymer volume fractions $\phi \le 2\%$ explored in this study.\cite{tokuyama94a} 
Moreover, HI contribute a constant scaling factor to the diffusion coefficient,\cite{tokuyama94a,tokuyama06a} which does not affect the qualitative trends investigated here.

(iv) \textbf{Continuous US exposure}.
In US-based drug delivery and \textit{in vitro} experiments, US is typically applied in pulses to minimize potential tissue damage.\cite{ma18a,karki22a,einen23a,sontum15a,snipstad2025a}
In contrast, our LD simulations employ continuous US exposure throughout the entire simulation time. 
Although our approach can be readily extended to include pulsed US, we intentionally restrict this study to continuous exposure to enable a clearer interpretation of the underlying mechanisms. 
The implications of pulsed versus continuous US are further discussed in Section~\ref{sec:comparison_exp}, where we compare our results with previous experimental studies performed under pulsed conditions.

\subsection{\label{sec:computational_details} Computational details}
For brevity, we provide here a general summary of the simulation setup and refer the reader to Appendix C for a complete technical description.
We performed LD simulations using the ESPResSo v4.2.1 software package,\cite{weeber24a} and constructed the hydrogel network with the pyMBE v1.0.0 molecule builder.\cite{beyer24a,pyMBEv1} 
The simulation box was a rectangular prism  of dimensions $L_\parallel \times L_\perp \times L_\perp$ subjected to periodic boundary conditions in all directions, as depicted in \reffig{fig:snapshot}.
The longitudinal dimension $L_\parallel$ was parallel to the wave propagation vector $\hat{\mathbf{n}}$, and chosen to  exceed the maximum particle displacement per wave cycle ($\sim \qty{300}{nm}$) in all studied systems. 
Box sizes ranged from $L_\parallel = \qty{687}{nm}$  to $L_\parallel = \qty{481}{nm}$, corresponding to polymer volume fractions  ranging from $\phi = 0.63 \%$ to $\phi = 1.84 \%$, respectively.
The polymer volume fraction was calculated as
\begin{equation}
    \phi = \frac{N_\mathrm{b}\pi \rbead^3}{6L_\parallel L^2_\perp}
    \label{eq:ex_volume}
\end{equation}
where $N_\mathrm{b} = 488$ is the number of hydrogel beads. 
For convenience, we set  $L_\parallel = 2L_\perp$ in all simulations.
The number of NPs ranged from 3 to 5, depending on system parameters, and was adjusted to minimize NP-NP interactions, particularly in simulations involving sticky networks. 

We performed LD production runs of typically $\qty{50}{ms}$, equivalent to $5 \cdot 10^8$ LD steps.
The temperature was maintained at $T = \qty{298.15}{K}$ by the Langevin thermostat.
The friction coefficient for each particle of diameter $d_{\rm NP}$ was set using the Stokes law $\gamma = 3\pi\eta_0 d_{\rm NP}$, with the dynamic viscosity of water at room temperature, $\eta_0 = \qty{0.89}{mPa \cdot s}$.\cite{kestin78b}
The mass of the NPs was estimated based on the density of polystyrene NPs (
$\rho_\mathrm{NP} = \qty{1.055}{g/mL}$) used in previous {\it in vitro} studies,\cite{ma18a,einen23a} as $m_\mathrm{NP}=\frac{1}{6}\pi \rho_\mathrm{NP}\rNP^3$. The mass of each hydrogel bead was arbitrarily set to $m_\mathrm{bead} = 0.1 m_\mathrm{NP}$.
As shown in \refsecmass~ in Appendix D, this choice  does not significantly affect NP diffusion. 

The key input parameter of our LD simulation is the peak velocity of the US field $\vmax$, used to calculate the amplitude of the US force (\cf~\refeq{eq:F_osc}).
For US to have a noticeable effect in NP motion, it needs to be, at least, comparable to the contribution from thermal motion to the NP velocity
$v_\mathrm{T} = \sqrt{3\kT/m_{\mathrm{NP}}} \sim 150$ nm/$\mus$. 
The range of considered $\vmax$ values approximately spans from  this lower threshold $v_\mathrm{T}$ to $10v_\mathrm{T}$.
This corresponds to acoustic peak pressures $\Pmax$ ranging from 0.18 to 2.18 MPa (\cf~\refeq{eq:p_max}), typically used in 
US-based drug delivery therapies and in-vitro experiments.\cite{ma18a,karki22a,einen23a,sontum15a,snipstad2025a}

During production runs, we measured the mean squared displacement $\MSD$, time-dependent diffusion coefficient $\Dt$, pore size $\rpore$, and the contact time  $\tc$ between particle pairs, defined as
\begin{equation}
    \tc = t(r > \rc) - t(r \leq \rc) \,,
    \label{eq:contact_time}
\end{equation}
where $t(r \leq \rc)$ is the first time a particle pair comes within contact distance, given by the cutoff in the Lennard-Jones potential $\rc$ (Eq. S1), and $t(r > \rc)$ is the time when contact is lost, \ie\ when the distance between the particle pair becomes $r > \rc$.
Contacts were monitored at every time step to capture transient fluctuations in $\tc$.
We calculated $\MSD$ using the algorithm described in Ref.\cite{ramirez10a}, $\Dt$ from \refeq{eq:time_dependent_D}, and $\rpore$ using the Monte Carlo algorithm described in Ref.\cite{trepte21a}
The accuracy in $\MSD$, $D(\tau)$, and $\tc$ were estimated as the standard error between average values measured across 10 independent LD runs with different random seeds.
The error in $r_\mathrm{pore}$ was estimated using the block analysis method.\cite{janke02c}
Typical LD production runs required approximately 5 days of computation per CPU (Intel Xeon Gold 6242) on the IDUN High Performance Computing cluster.\cite{sjalander19a}

\section{\label{sec:results}Results and Discussion}
\subsection{\label{sec:buffer}NP motion under US irradiation in buffer recovers thermal diffusion at long lag times}
We begin by examining the simplest case: a buffer solution of NPs in the ideal dilution limit.
In this regime, molecular interactions are negligible, and the  potential gradient in the Langevin equation of motion (\refeq{eq:langevin}) vanishes, {\it i.e.}, $\nabla \mathbf{V} = 0$.
As a result, the particle dynamics are governed solely by thermal fluctuations, solvent friction, and the oscillatory force induced by US irradiation.
Notably, the mean square displacement of the particles $\MSD$ has an analytical solution, obtained by integrating the Langevin equation in the so-called overdamped regime, 
\begin{equation}
    \MSD = 
    2dD_0\tau + \frac{1}{2}
    \left(  \frac{\Pmax}{\pi \rho c f} \right)^2
    \sin^2(\pi f \tau).
    \label{eq:msd_osc}
\end{equation}
A detailed derivation of this equation is provided in Appendix E.
We recall that the amplitude of the external force is related to $\Pmax$ via
\refeq{eq:p_max} as $A = \Pmax/(\rho c) = \vmax \gamma$.

\begin{figure*}[t]
    \centering
    \includegraphics[width=0.45\linewidth]{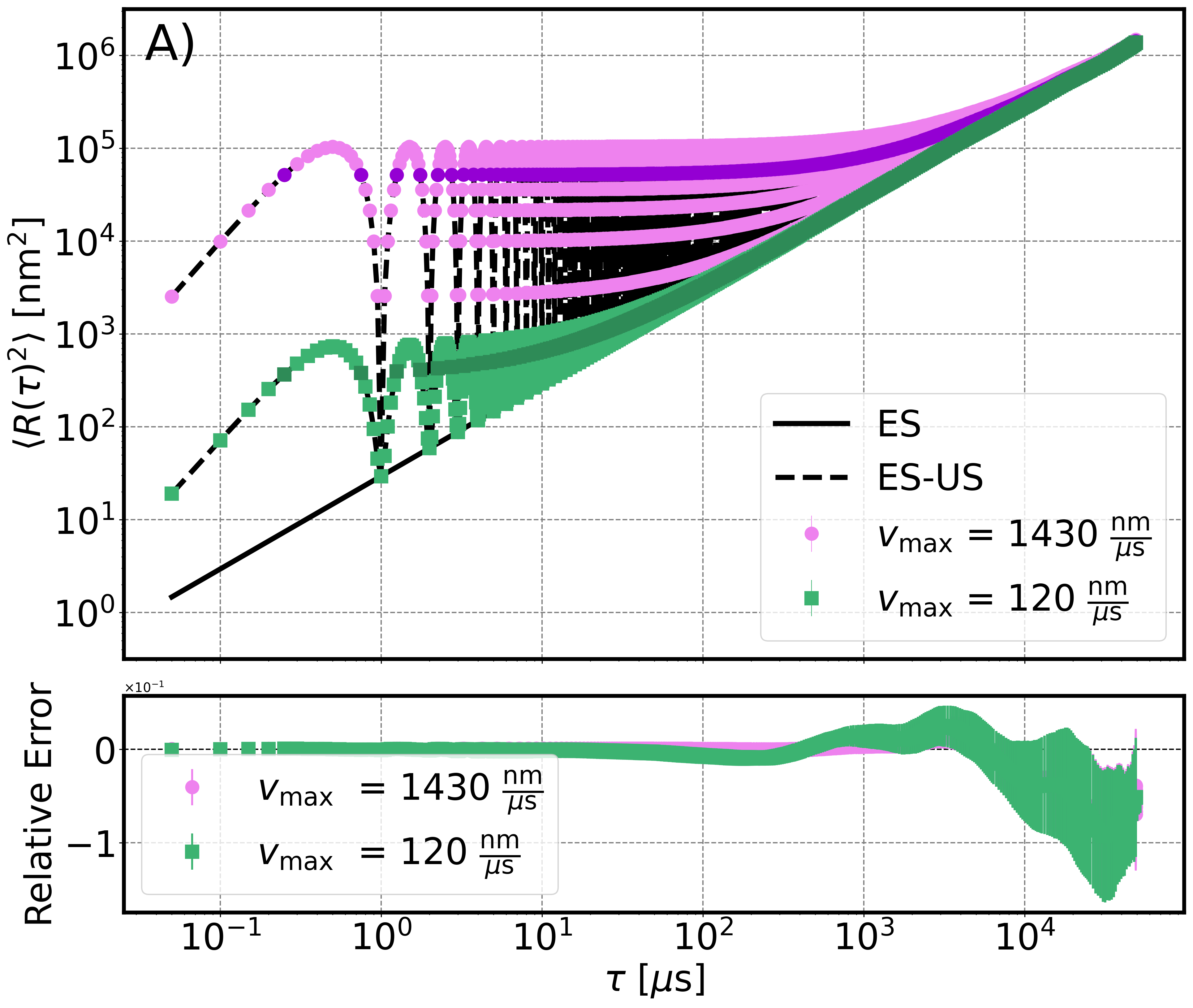}
    \includegraphics[width=0.45\linewidth]{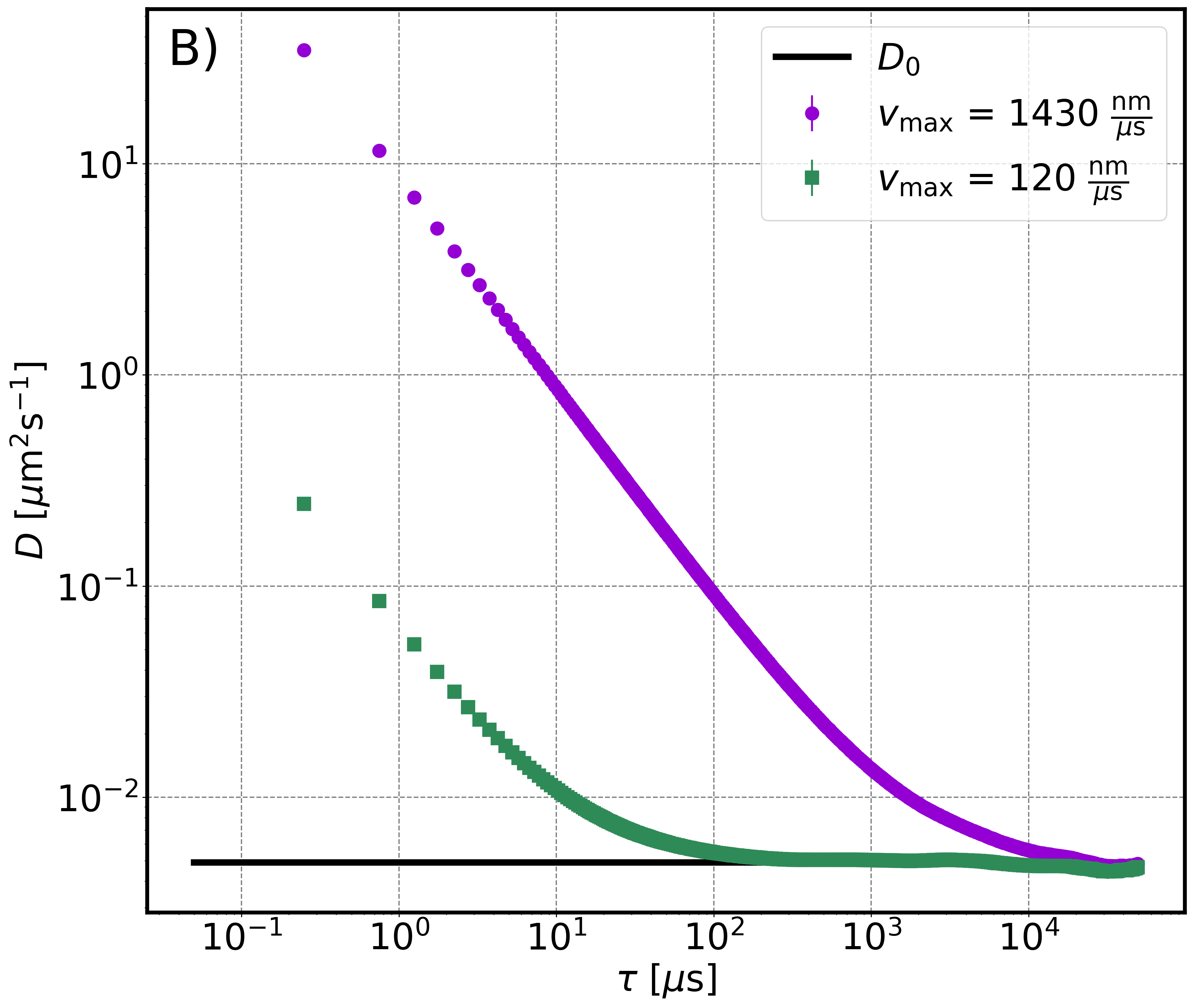}
    \caption{
    Mean squared displacement ($\MSD$, panel A) and diffusion coefficient at zero US amplitude ($D$, panel B) of NPs as a function of lag time $\tau$ obtained from LD simulations in a buffer (markers)  for two input US peak pressures $\Pmax$.
    Panel A includes the analytical solutions for $\MSD$ from the Einstein–Smoluchowski equation (\refeq{eq:einstein_smo}, labeled ES) and for the US-modified model (\refeq{eq:msd_osc}, labeled ES-US).
    The lower sub-panel of A shows the relative error between simulation results and the ES-US prediction, calculated as $(\MSD_{\textrm{LD}}-\MSD_{\textrm{ES-US}})/\MSD_{\textrm{ES-US}}$.
    In panel A, data points used to calculate $D$ are highlighted in darker shades.
    In panel B, the input Stokes–Einstein diffusion coefficient $D_0$ (\refeq{eq:stokes_einstein}) is shown as a reference (solid line).
    }
    \label{fig:buffer_msd}  
\end{figure*}

We begin by analyzing the general features of NP motion under US by solving  \refeq{eq:msd_osc} (ES-US), as shown in \reffig{fig:buffer_msd}.
At short lag times $\tau$, NP motion is dominated by the oscillatory component of the US field.
However, once per wave cycle, the displacement aligns with that expected from purely thermal motion, as described by the classical Einstein–Smoluchowski relation \refeq{eq:einstein_smo} (ES).
At longer $\tau$ values, thermal diffusion becomes the dominant contribution, and the influence of US diminishes, acting only as a small perturbation to the overall NP dynamics.

It is instructive to compare \refeq{eq:msd_osc} with earlier models of oscillatory diffusion.
The  two-state random walk (TSRW) model,\cite{balakrishnan81a} for instance, also predicts an $\MSD$ composed of
 two additive contributions: one arising from thermal motion, as described by the Einstein–Smoluchowski relation (\refeq{eq:einstein_smo}), and the other arising from the oscillatory motion.
At long times ($\tau \gg 1/f$), both models recover thermal diffusive behavior.
However, while \refeq{eq:msd_osc} yields a diffusion coefficient equal to $D_0$ (absence of US), the TSRW model predicts an effective diffusion coefficient that includes contributions from the oscillatory motion.
Similarly, Marshall's model\cite{marshall16a, curran21a, marshall21a}  predicts US-enhanced diffusion  proportional to the square of the wave amplitude, emphasizing the role of NP–hydrogel interactions as a key mechanism underlying US-enhanced diffusion. Such interactions are absent in  \refeq{eq:msd_osc} but the LD simulations overcome this limitation by explicitly including molecular interactions.

Despite its limitations, \refeq{eq:msd_osc} provides an exact  analytical benchmark in the ideal dilution regime, allowing to validate the LD simulations.
Indeed, the $\MSD$ obtained from LD simulations of NPs in buffer show excellent quantitative agreement with this expression, as shown in \reffig{fig:buffer_msd}A for the two limiting cases of US peak pressure considered here: $\Pmax = \qty{0.18}{MPa}$ and $\Pmax = \qty{2.18}{MPa}$.
Only a small relative error (less than $1\%$) is observed at lag times beyond $\qty{10}{ms}$, likely due to limited statistical sampling at these longer timescales.
At such long $\tau$ values, the time-dependent diffusion coefficient $D(\tau)$, computed from the slope of the $\MSD$ (\refeq{eq:time_dependent_D}), asymptotically approaches $D_0$ for both US amplitudes, as shown in \reffig{fig:buffer_msd}B.
For clarity, we report $D(\tau)$ only at time points corresponding to zero amplitude of the US wave, corresponding to the $\MSD$ values highlighted with a matching darker shade in \reffig{fig:buffer_msd}A.
At short $\tau$, $D$ exhibits oscillations reflecting the underlying US-induced motion (not shown in \reffig{fig:buffer_msd}B for clarity), but these oscillations decay at longer $\tau$ values, yielding a well-defined value of $D$.
We refer to such asymptotic value of $D$ as the diffusion coefficient at long times $\Dlong$.

Since \refeq{eq:msd_osc} is valid only in the absence of molecular interactions (ideal dilution limit), computer simulations are essential to explore NP dynamics in more complex environments.
In the following sections, we focus on NP diffusion in hydrogel networks, restricting our analysis to the results obtained from LD simulations.

\subsection{\label{sec:steric_network}Acoustically-enhanced diffusion of NPs can not be explained by steric caging}
\begin{figure}[t]
    \centering
    \includegraphics[width=0.99\linewidth]{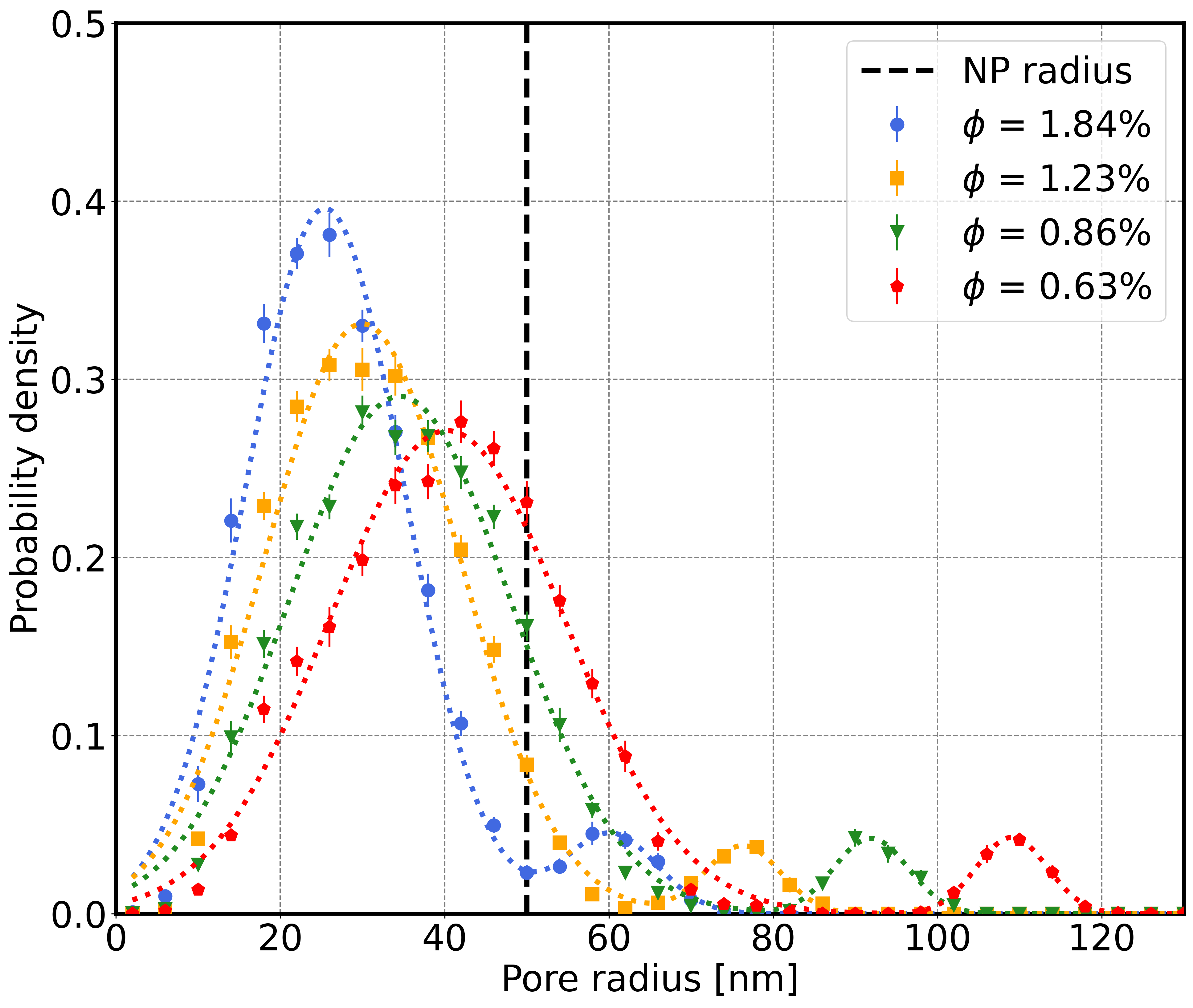}
    \caption{Probability density of pore radius in hydrogel networks with varying polymer volume fractions ($\phi$).
    The NP radius is indicated by a vertical line for reference.
    Dashed lines represent bimodal Gaussian fits to the distributions and are included as visual guides.}
    \label{fig:PSD}  
\end{figure}

\begin{figure*}[t]
    \centering
    \includegraphics[width=0.45\linewidth]{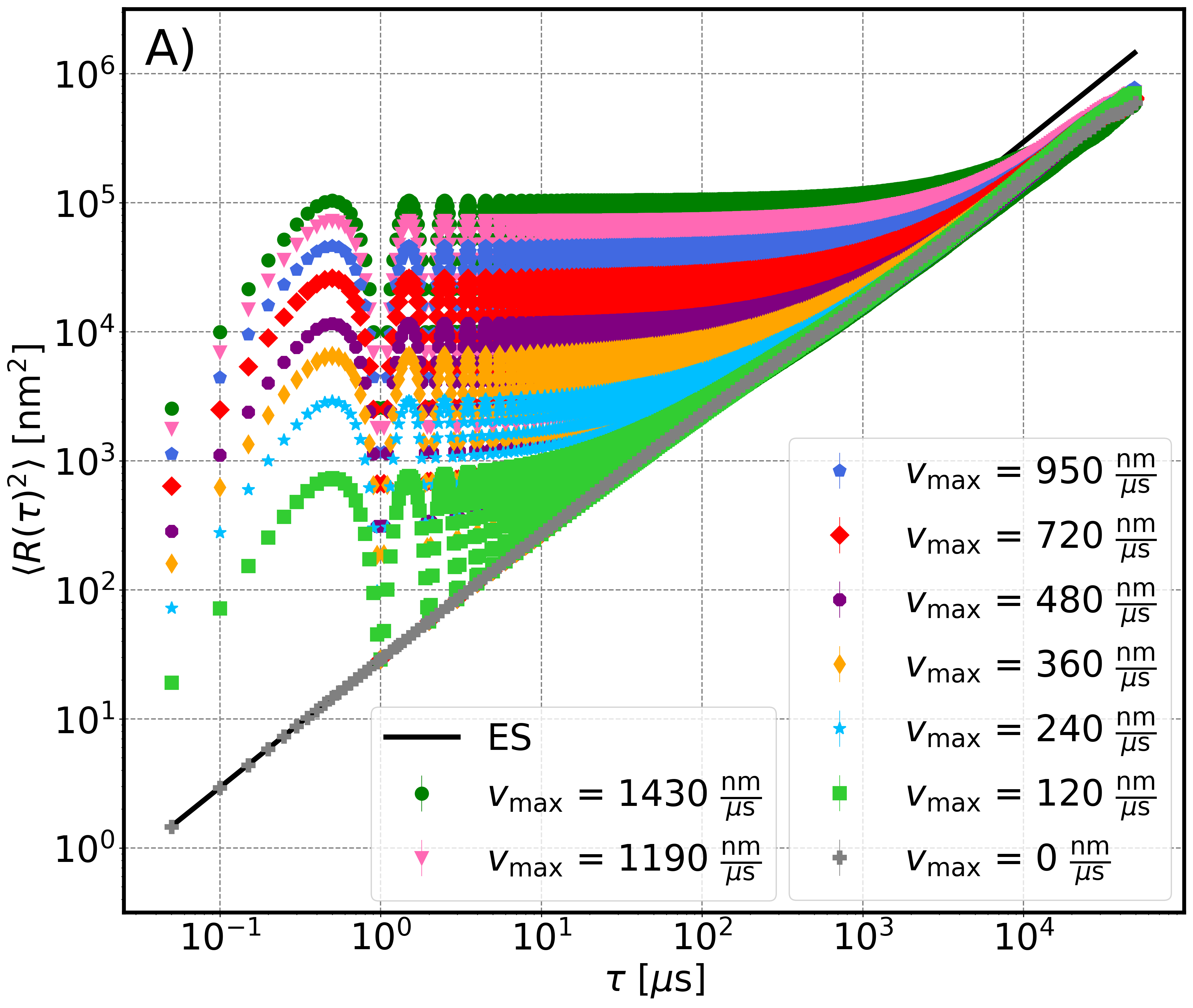}
    \includegraphics[width=0.45\linewidth]{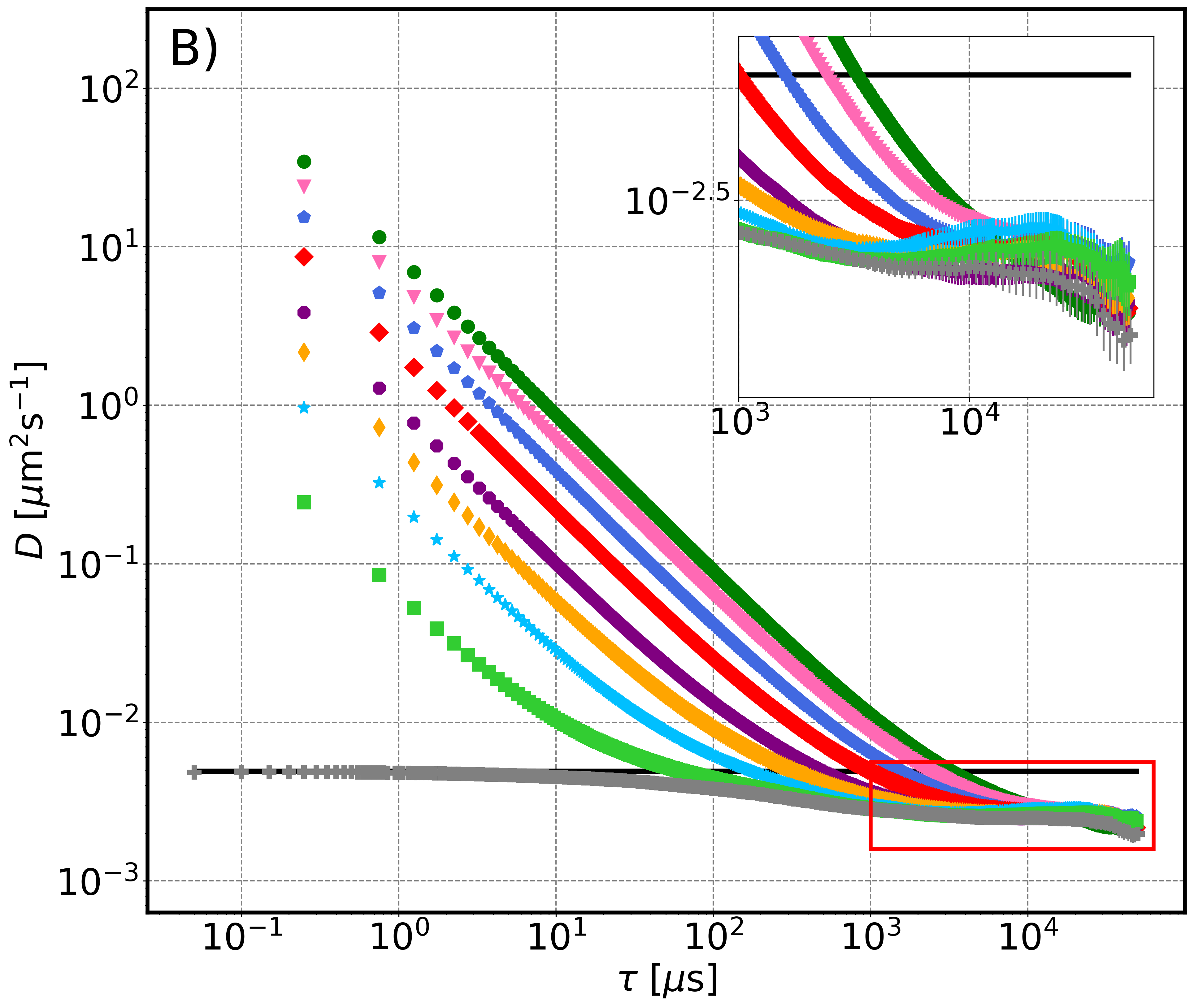} \\
    
     \caption{Mean squared displacement ($\MSD$, panel A) and diffusion coefficient at zero US amplitude ($D$, panel B) of NPs in a steric network of volume fraction $\phi = 0.86 \%$, under varying US peak pressures $\Pmax$.
     Panel A includes the analytical solution from the Einstein–Smoluchowski equation (\refeq{eq:einstein_smo}, labeled ES).
     In panel B, the input Stokes–Einstein diffusion coefficient $D_0$ (\refeq{eq:stokes_einstein}) is shown as a reference (solid line).} 
    \label{fig:steric_network_msd}
\end{figure*}

NP diffusion is known to be hindered in porous media, such as hydrogels and soft tissues, often resulting in sub-diffusive behavior. \cite{saxton94a,tabatabaei11a}
Theoretical models of US-enhanced NP diffusion in such environments typically attribute the enhancement to transient retention and release of NPs at pore boundaries.\cite{marshall16a}
Among the various mechanisms proposed, steric hindrance -- arising from size exclusion between NPs and the polymer network -- is commonly identified as the dominant factor limiting NP mobility. \cite{quesada-perez21a}
To assess whether US irradiation can overcome this steric confinement, we investigate NP diffusion in a hydrogel network modeled with purely steric repulsion, using the WCA potential (see Appendix A for technical details).
Here, we refer to this case as the "steric network" scenario.
 
In this scenario, the key parameter governing NP diffusion is the ratio between the NP size and the characteristic pore size of the hydrogel.
We investigated hydrogel networks with polymer volume fractions $\phi$ ranging from 0.63\% to 1.84\%, closely matching those of the agarose gels used in {\it in vitro} studies of US-enhanced NP diffusion.\cite{ma18a,einen23a,karki22a}
The resulting pore size distributions, shown in \reffig{fig:PSD}, exhibit two distinct pore populations, as evidenced by the bimodal Gaussian fits.
The smaller pores are located near network nodes, while larger pores occupy the interstitial regions.
In less dense networks ($\phi < 1\%$), the NP radius is comparable to the size of the smaller pores, allowing access to both pore populations.
In denser networks ($\phi > 1\%$), the NPs are excluded from the smaller pores and can only diffuse through the larger ones.
Increasing $\phi$ further causes the network pores to become smaller than the NP diameter, effectively trapping the particle and reproducing the caged dynamics observed in the fixed-network model described in Appendix B.1.
Consequently, steric hindrance is expected to be more pronounced at higher polymer volume fractions, although it remains significant even in the less dense networks studied here.

We begin our analysis of NP diffusion under US irradiation in the steric network with the intermediate case of $\phi = 0.86\%$, which corresponds to the swelling equilibrium of the network and closely matches the polymer volume fraction used in experimental studies.\cite{einen23a}
Additional results for other hydrogel volume fractions are provided in Appendix F.1.
As shown in \reffig{fig:steric_network_msd}, the $\MSD$ of NPs in this system exhibits similar qualitative features to the buffer case discussed in \refsec{sec:buffer}.
At short lag times $\tau$, NP motion is dominated by the oscillatory component of the US field.
At longer $\tau$, the system gradually transitions to a diffusive regime, converging towards the behavior observed in the absence of US ($\Pmax = 0$).
However, unlike the buffer case, the long-time $\MSD$ remains consistently below the (ideal) Einstein–Smoluchowski prediction (\refeq{eq:einstein_smo}), reflecting the persistent hindrance imposed by the hydrogel network.
Accordingly, the long-time diffusion coefficient, $\Dlong$, extracted from the slope of $\MSD$ at large $\tau$, is consistently smaller than the Stokes–Einstein value $D_0$ for all US peak pressures $\Pmax$, as shown in \reffig{fig:steric_network_msd}B.
While increasing $\Pmax$ delays the onset of thermal diffusion, it does not appreciably alter the asymptotic value of $\Dlong$, as highlighted in the inset of \reffig{fig:steric_network_msd}B.

\begin{figure}[h]
    \centering
    \includegraphics[width=0.95\linewidth]{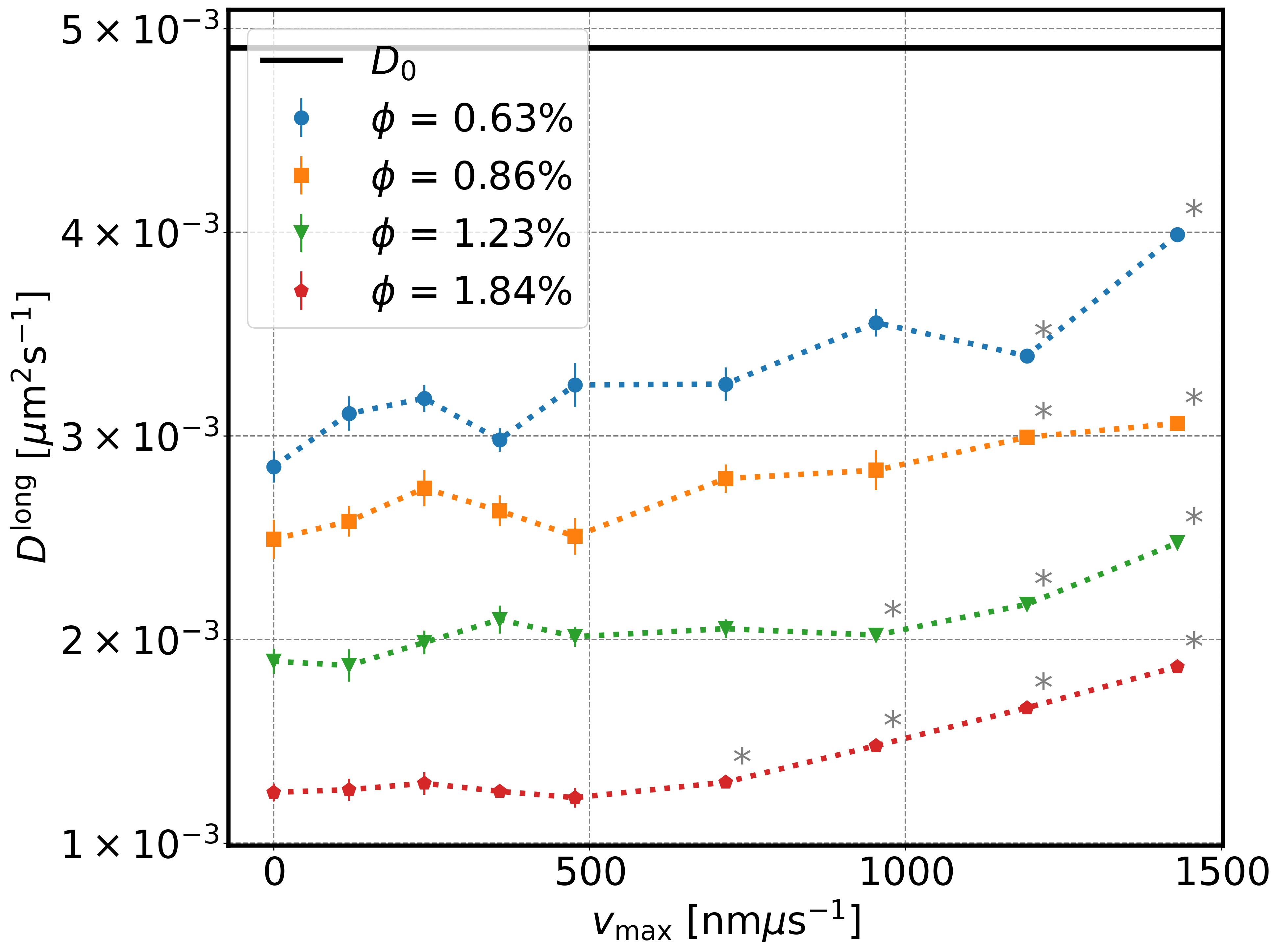}
    \caption{Diffusion coefficient at long lag times ($\Dlong$) as a function of US peak pressure ($\Pmax$) for steric networks with varying polymer volume fractions ($\phi$).
    The Stokes–Einstein diffusion coefficient $D_0$ (see \refeq{eq:stokes_einstein}) is shown as a reference (solid line).
    }
    \label{fig:steric_network_D_v}
    
\end{figure}

\begin{figure*}[t]
    \centering
    \includegraphics[width=0.45\linewidth]{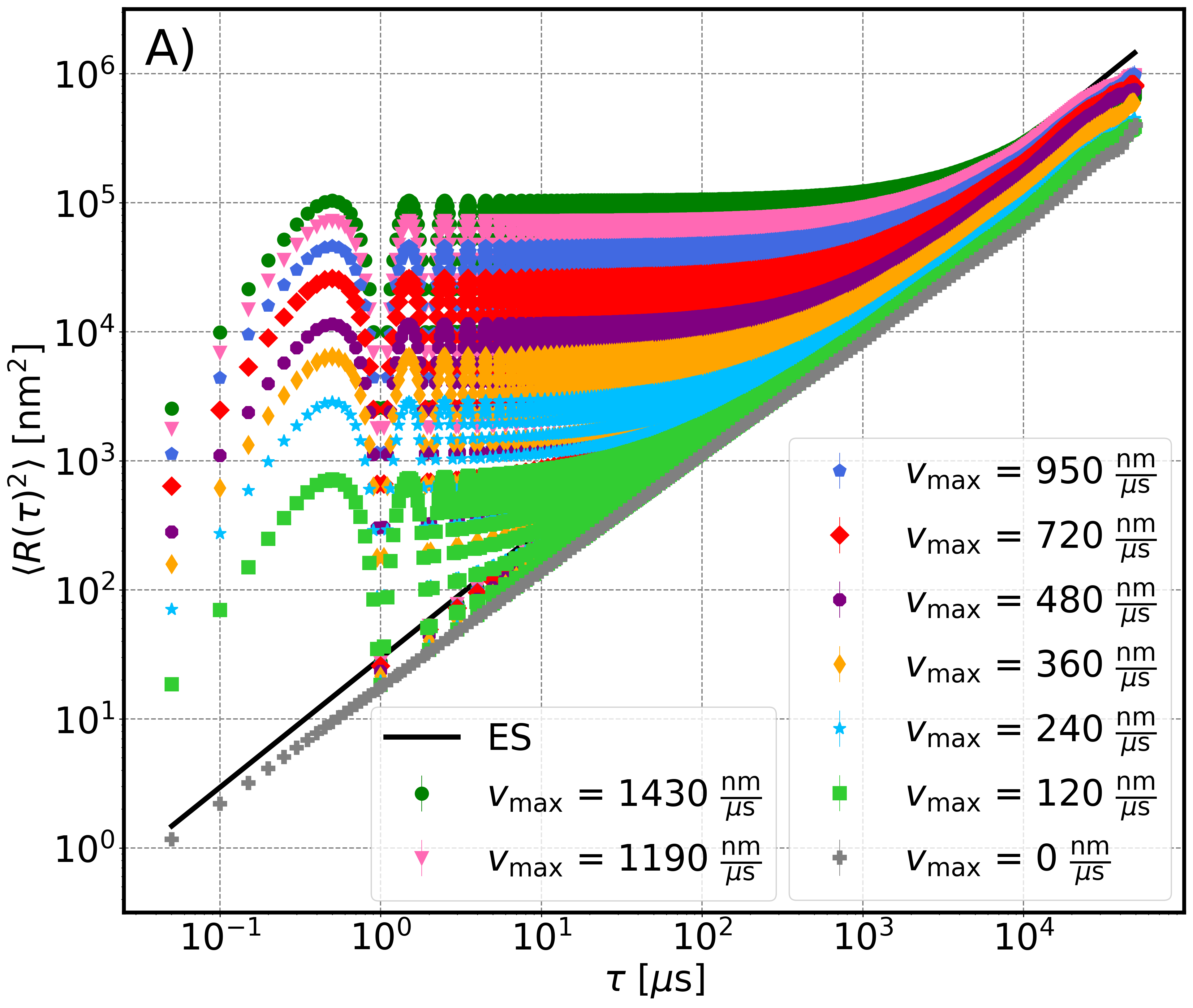}
    \includegraphics[width=0.45\linewidth]{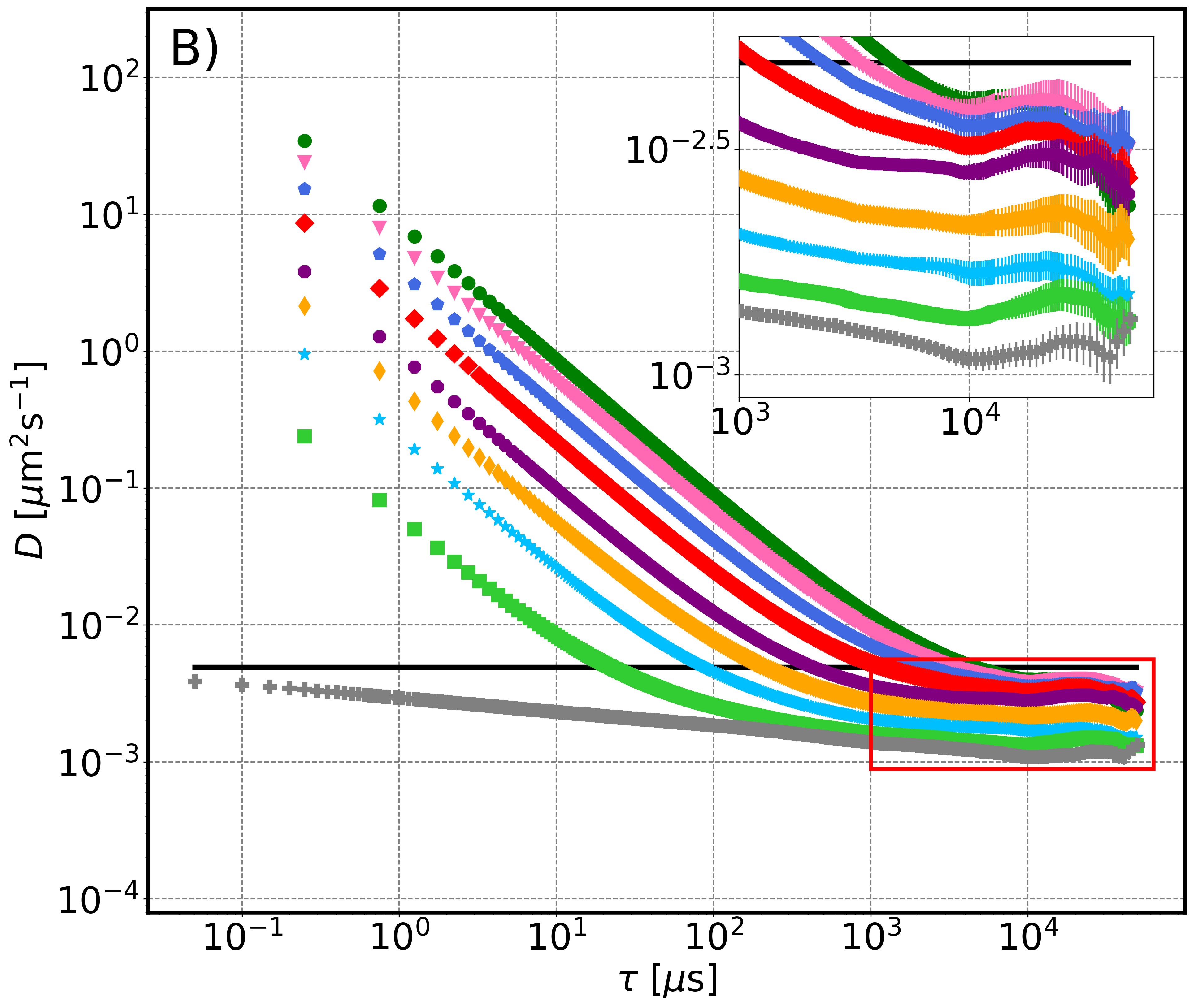} \\
    
     \caption{Mean squared displacement ($\MSD$, panel A) and diffusion coefficient at zero US amplitude ($D$, panel B) of NPs in a sticky network with polymer volume fraction $\phi = 0.86 \%$ and Lennard-Jones interaction depth $\epsilon = 1\kT$, under varying US peak pressures $\Pmax$.
     Panel A includes the analytical solution from the Einstein–Smoluchowski equation (\refeq{eq:einstein_smo}, labeled ES).
     In panel B, the Stokes–Einstein diffusion coefficient $D_0$ (\refeq{eq:stokes_einstein}) is shown as a reference (solid line).}
    \label{fig:sticky_network_msd}
\end{figure*}

A broader overview of $\Dlong$ calculated across all steric network systems is presented in \reffig{fig:steric_network_D_v}.
Asterisks indicate cases where a clear diffusive plateau was not reached within the simulation time.
Across the range of US peak velocity $\Pmax$ studied, we observe no significant enhancement of $\Dlong$ in the steric network, except in those flagged cases where $\Dlong$ could not be estimated accurately.
In such cases, the simulations were not sufficiently long to reach the long-time diffusive regime, resulting in overestimated $\Dlong$ values.
However, the inset of \reffig{fig:steric_network_msd}B suggests that the overestimated $\Dlong$ values would approach those obtained in the accurately sampled cases if the simulations were run for a longer time.
In steric networks, the NP–hydrogel contact times are found to be very short, below the US wave period ($\tc < 1 \mus$, \cf~Appendix F.3), and thus the US field does not substantially perturb the transient NP–hydrogel interactions.
These findings suggest that steric hindrance alone can not account for the US-induced enhancement of NP diffusion observed experimentally.\cite{ma18a,karki22a,einen23a} 

\subsection{US irradiation enhances diffusion of NPs with sufficient attraction to the hydrogel}
Beyond steric effects, various molecular interactions (\eg, hydrophobic interactions, specific binding, and electrostatic forces) can contribute to NP retention within hydrogel networks.
These attractive interactions can increase NP residence time and modulate the impact of US on NP diffusion.

To explore this hypothesis, we introduce an attractive Lennard-Jones (LJ) potential between NPs and hydrogel beads, modeling generic adhesive interactions. 
We refer to this scenario as the "sticky network".
This simplified model allows systematic control of the interaction strength via the LJ potential well depth $\epsilon$.
We again focus on a hydrogel network with volume fraction of $\phi = 0.86 \%$.

We begin with the case $\epsilon = 1k_\mathrm{B}T$, representing attractive interaction comparable to the thermal energy, shown in \reffig{fig:sticky_network_msd}A.
Results for other interactions strengths are provided in Appendix F.2.
The $\MSD$ of NPs in the sticky network (\reffig{fig:sticky_network_msd}A) is consistently lower than in the steric case (\reffig{fig:steric_network_msd}A) when US is absent ($\Pmax = 0$, gray crosses), indicating stronger NP retention due to attractive interactions.
Under US irradiation ($\Pmax > 0$), all systems exhibit similar $\MSD$ behavior at short lag times ($\tau < 10^3\mus$), dominated by oscillatory motion.
However, at longer lag times ($\tau > 10^3\mus$), the sticky network shows a clear monotonic increase in $\MSD$ with increasing $\Pmax$, indicating US-enhanced NP mobility.
This trend is reflected in the diffusion coefficient $D$ at long lag times, which increases with $\Pmax$ in the sticky network, as shown in \reffig{fig:sticky_network_msd}B.

\begin{figure}[t]
    \centering
    \includegraphics[width=0.9\linewidth]{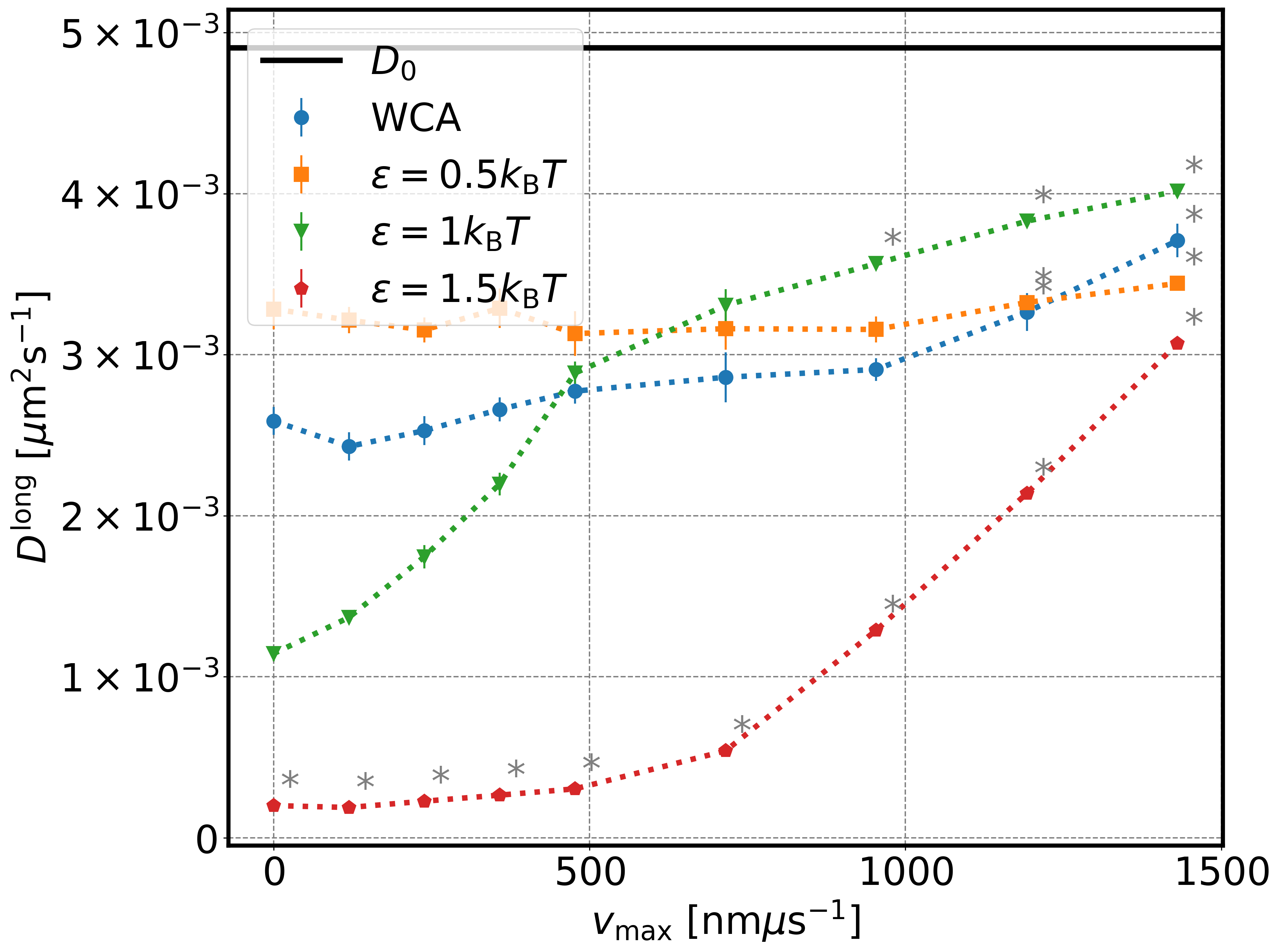}
    \caption{Diffusion coefficient at long lag times ($\Dlong$) as a function of US peak pressure ($\Pmax$) for sticky networks with polymer volume fractions $\phi = 0.86\%$ and varying Lennard-Jones interaction depths $\epsilon$.
    The Stokes–Einstein diffusion coefficient $D_0$ (\refeq{eq:stokes_einstein}) is shown as a reference (solid line).}
    \label{fig:sticky_network_D_v}
    
\end{figure}

A comprehensive comparison of $\Dlong$ across interactions strengths is shown in \reffig{fig:sticky_network_D_v}, for networks with a hydrogel volume fraction of $\phi = 0.86\%$.
For weak attraction ($\epsilon = 0.5\kT$), the behavior of $\Dlong$ as a function of $\Pmax$ resembles that of the steric network (labeled WCA), with no significant enhancement.
Interestingly, $\Dlong$ is systematically higher in this weakly sticky network than in the steric case, suggesting that moderate attractive interaction may facilitate NP diffusion.
We attribute this effect to a NP hopping mechanism, as described in Ref.\cite{tabatabaei11a}, although a full mechanistic analysis is beyond the scope of the present work.
For stronger interactions ($\epsilon \geq 1\kT$), a clear monotonic increase in $\Dlong$ with $\Pmax$ is observed, confirming US-enhanced diffusion at long lag times.
We note that the long-time plateau of $D$ was not fully reached within the simulation time for the strongest interaction case ($\epsilon = 1.5k_\mathrm{B}T$). This leads to a systematic overestimation of $\Dlong$, which does not alter the qualitative trend depicted in \reffig{fig:sticky_network_D_v}.
These findings indicate that US can enhance NP diffusion at long timescales, provided sufficient attractive interaction with the hydrogel network exists.

\begin{figure}[h]
    \centering
    \includegraphics[width=0.9\linewidth]{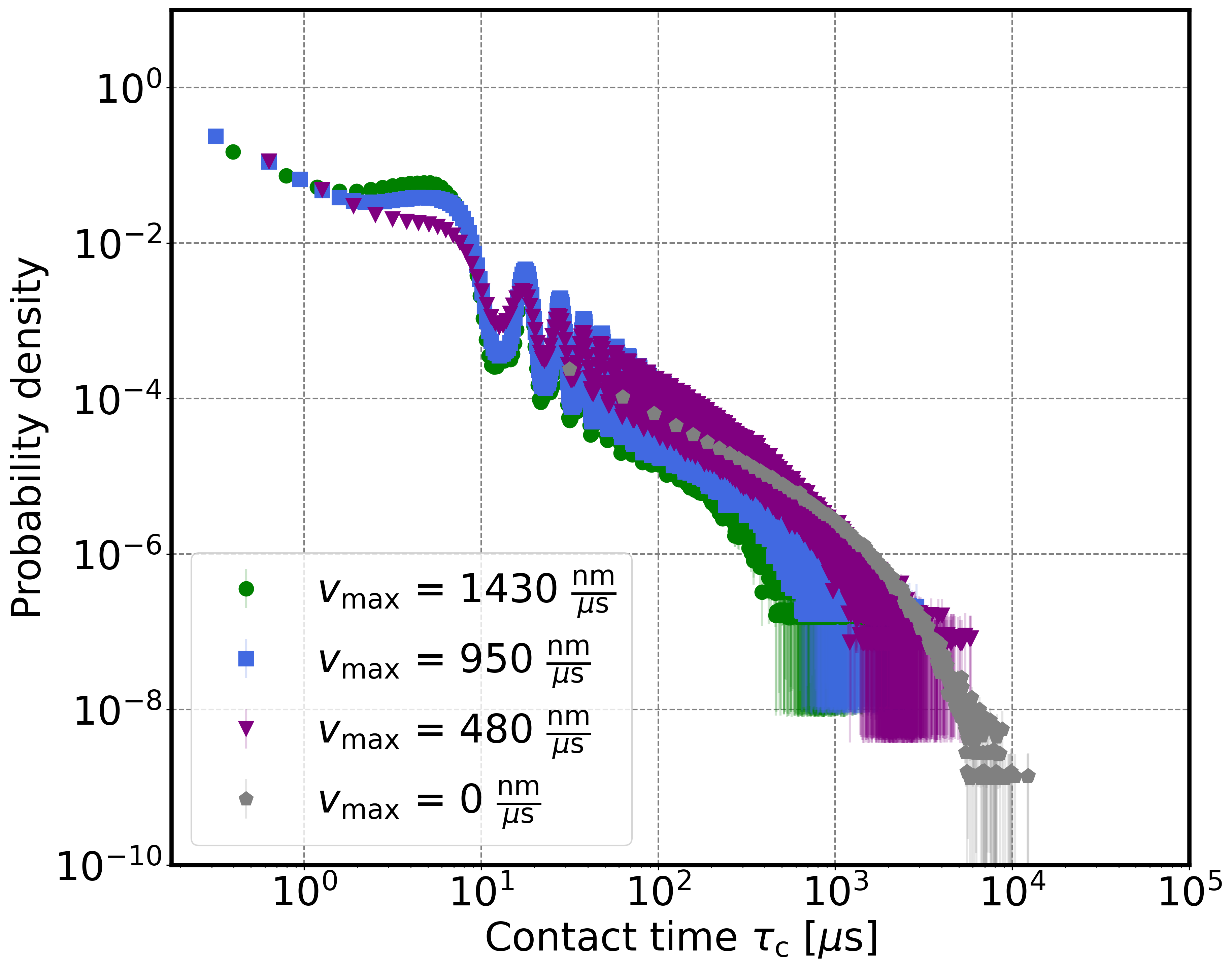}
    \caption{Probability density of the hydrogel–NP contact times in a sticky network with interaction strength $\epsilon = 1\kT$, measured at various US peak pressures $\Pmax$.
    Each contact time corresponds to a specific NP-hydrogel particle pair and quantifies the duration of individual adhesive interactions.
    }
    \label{fig:sticky_network_tc}

\end{figure}

To investigate the underlying mechanism, we analyzed the contact time $\tc$ (defined in \refeq{eq:contact_time}) between NP–hydrogel particle pairs.
\reffig{fig:sticky_network_tc} shows the probability density of $\tc$ for the sticky network with $\epsilon = 1 \kT$; results for other interaction strengths are provided in Appendix F.3.
Without US ($\Pmax = 0$), the probability density of $\tc$ decays monotonically, consistent with random thermal escape from transient contacts.
Under US irradiation ($\Pmax > 0$), the contact time distribution becomes oscillatory, reflecting the interplay between US-induced motion and the slower thermal relaxation of the hydrogel network relative to the NPs.
Visual inspection of the trajectories shows that, during the phases of coherent motion, NPs and hydrogel beads oscillate synchronously with the acoustic field, leading to extended sticking events.
Once per acoustic cycle, however, the oscillatory force reverses direction, and NP motion becomes dominated by diffusion, breaking NP–hydrogel contacts and producing release events.
Averaged over many US cycles, this periodic stick-and-release process reduces the overall contact time $\tc$ and facilitates NP escape from attractive regions of the network.
We refer to this as the “stick-and-release” mechanism, schematically illustrated in Appendix G, which explains the systematic decrease in $\tc$ and the corresponding enhancement of NP diffusion with increasing $\Pmax$ (\reffig{fig:sticky_network_D_v}).

\subsection{Comparison with previous experiments in the literature \label{sec:comparison_exp}}
In this section, we contextualize our simulation results by comparing them with previous experimental studies on NP diffusion in hydrogels under US exposure.
To enable a direct comparison, all key parameters from both simulations and experiments are summarized in Table 1.

\begin{table*}[t]
\centering
\caption{Comparison between experimental and simulation studies reporting NP diffusion in hydrogels under US exposure.
For each study, we summarize the NP diameter ($\rNP$), polymer volume fraction ($\phi$), US frequency ($f$), US peak pressure ($\Pmax$), duty cycle ($DC$), US exposure time per pulse ($\tus$, \cf~\refeq{eq:DC}), and the relative acoustic diffusion enhancement ($\Da / \Dm$).
For the experimental data,\cite{ma18a,karki22a,einen23a} $\phi$ was estimated from the reported agarose concentration ($\Cp$) and the density of dry agarose\cite{laurent67a} ($\np = \qty{1.64}{g/mL}$) as $\phi = \Cp/\np$.
For our data, $\phi$ is calculated from \refeq{eq:ex_volume}.
Following the nomenclature in Refs.\cite{ma18a,karki22a}, $\Dm$ denotes the diffusion coefficient measured in the absence of US and $\Da = D - \Dm$ is the acoustic contribution to diffusion coefficient $D$.
For each study, we report the minimum and maximum values of $\Da / \Dm$ corresponding to the $\Da$ value measured at the lowest and highest $DC$ or $\Pmax$ values.
The value marked with an asterisk ($^*$) indicates the relative diffusion enhancement obtained from simulations in which $\Dlong$ is overestimated.}
\begin{tabular}{|c|c|c|c|c|c|c|c|}
\hline
Reference & $\rNP$ / nm & $\phi$ & $f$ / MHz & $\Pmax$ / MPa & $DC$ & $\tus$ / $\mus$ & $\Da / \Dm$ \\
\hline
Ma \etal\cite{ma18a} & 20 & $0.49\%$ & 1 &  Not reported. & $20\%$ & Not reported. & 1.7 \\
\hline
Ma \etal\cite{ma18a} & 100 & $0.49\%$ & 1 & Not reported. & $20\%$ & Not reported. & 2.3 \\
\hline
Karki \etal\cite{karki22a} & 20 & $0.49\%$ & 1 & $0.1-0.18$ & $10\%$ & Not reported. & $2.4-7.9$ \\
\hline
Karki \etal\cite{karki22a} & 40 & $0.49\%$ & 1 & $0.1-0.18$ & $10\%$ & Not reported. & $3.0-6.3$ \\
\hline
Einen \etal\cite{einen23a} & 100 & $0.91\%$ & 1 & 1 & $0.1 - 20\%$ & $100-2000$ & $0-0.3$ \\
\hline
This work & 100 & $0.63 - 1.84\%$ & 1 & $0.18-2.18$ & $100\%$ & 50000 & $0 - 1.9(14.4^*)$ \\
\hline
\end{tabular}
\end{table*}

Wu and co-workers (Ma \etal\cite{ma18a} and Karki \etal\cite{karki22a}) conducted a series of experiments in which fluorescent NPs were introduced at one end of an agarose gel along the direction of US propagation. 
They monitored the NP concentration profiles over time and proposed a phenomenological model in which the effective diffusion coefficient of the NPs is expressed as the sum of two components:
\begin{equation}
D = \Dm + \Da \,.
\label{eq:Dm_Da}
\end{equation}
Here, $\Dm$ represents the thermal (molecular) diffusion coefficient, while $\Da$ accounts for an effective acoustic contribution to the diffusion coefficient induced by US.
In their approach, $\Dm$ was determined by fitting concentration profiles from control experiments (without US) to Fick's second law of diffusion, and $\Da$ was extracted by fitting the US-exposed data to the diffusion–advection equation.

Our LD simulations support the physical plausibility of this additive model, but only under conditions where NPs experience significant retention by the hydrogel network, such as in our sufficiently sticky network scenarios (see \reffig{fig:sticky_network_D_v}).
In these cases, US can facilitate the release of NPs escape from adhesive interactions, resulting in a genuine enhancement of long-time diffusion.
In contrast, for systems with purely steric or weakly attractive, our simulations show no systematic increase in the long-time diffusion coefficient with increasing US amplitude. 
Instead, the diffusion coefficient asymptotically approaches the thermal value observed in the absence of US.
These findings suggest that strong NP–network interactions are a necessary condition for US-enhanced diffusion at long times, consistent with the assumptions underlying Wu and co-workers' model.

In their experiments, Ma \etal\cite{ma18a} reported an US-induced diffusion enhancement of approximately $\Da/\Dm \approx 2.3$ for NPs with a diameter of $\rNP \approx \qty{100}{nm}$, the same size used in our simulations.
Our model reproduces a comparable enhancement for the sticky network case with $\epsilon = 1\kT$, yielding an acoustic enhancement of 
$(\Dlong_{\qty{1.42}{MPa}} - \Dlong_{0}) / \Dlong_{0} = \Da/\Dm = 1.9$, where the subscript in $\Dlong$ indicates the applied $\Pmax$ value.
Ma \etal\ also observed a smaller enhancement ($\Da/\Dm \approx 1.7$) for smaller NPs ($\rNP \approx \qty{20}{nm}$), which our Langevin dynamics simulations rationalize by noting that smaller particles experience weaker attractive interactions, owing to their reduced surface area, and thus benefit less from US exposure.

In a follow-up study, Karki \etal\cite{karki22a} found that NPs with $\rNP \approx \qty{20}{nm}$ exhibited similar $\Da/\Dm$ values to those of intermediate-sized NPs ($\rNP \approx \qty{40}{nm}$).
They reported substantial acoustic diffusion enhancements, up to $\Da/\Dm = 7.9$.
Our simulations with the most sticky network ($\epsilon = 1.5,\kT$) also yield large enhancement factors, reaching $(\Dlong_{\qty{2.13}{MPa}} - \Dlong_{0}) / \Dlong_{0} = \Da/\Dm = 14.4$, albeit at higher $\Pmax$ values than those used by Karki \etal.
It is worth noting that our simulated NPs ($\rNP = \qty{100}{nm}$) are larger than those employed in their experiments (20–40 nm), making it plausible that lower acoustic pressures are sufficient to release smaller NPs from the hydrogel network. 

More recently, Einen \etal \cite{einen23a} measured the $\MSD$ of NPs in agarose gels under US, using a condensed milk/aqueous solution mixture as the solvent. 
They fitted their data to the generalized Einstein–Smoluchowski relation for anomalous diffusion (\refeq{eq:anomalous_diffusion}) and measured the generalized transport coefficient $\Gamma$  
and the anomalous scaling exponent $\alpha$ of the NPs in the hydrogel.
They found  pronounced sub-diffusive behavior for NPs with radius $\rNP \approx \qty{100}{nm}$, with  $\alpha \sim 0.6$.
This value is close to the Rouse model prediction for polymer chain dynamics ($\alpha = 0.5$),\cite{rubinstein03a} suggesting that the NPs were strongly associated with the agarose polymer network and moving in tandem with the network’s thermal fluctuations. 
However, Einen \etal~ observed only a modest increase in NP diffusion under US, in contrast to the larger enhancements reported by Wu and co-workers. 

A key difference between our simulations and the experimental studies is that, in the latter, US was applied in pulses, whereas our simulations assume continuous exposure.
In experiments, the pulse duration is typically reported using the so-called duty cycle ($DC$), defined as the fraction of time during which the US field is active within each pulse period:
\begin{equation}
DC = \frac{\tus}{\tpulse},
\label{eq:DC}
\end{equation}
where $\tus$ is the duration of continuous US exposure within a single pulse and $\tpulse$ is the pulse length.
Because $DC$ is a relative quantity, pulses with the same duty cycle can correspond to different absolute exposure times.
For example, a pulse with $\tus = \qty{10}{ms}$ and $\tpulse = \qty{2}{ms}$ and another with $\tus = \qty{5}{ms}$ and $\tpulse = \qty{1}{ms}$ both correspond to $DC = 20\%$, even though they represent different US conditions.

Einen \etal~ used pulses with $\tpulse = 10000 \mus$ and varied $DC$ from 0.1\% to 20\%, corresponding to $\tus$ durations of 100$\mus$ to 2000$\mus$ per pulse, respectively. 
They observed only a modest accoustic enhancement of $(\Gamma_{20\%}-\Gamma_{0})/\Gamma_{0} = \Da /\Dm \approx 0.3$ at the highest $DC$, where the subscript in $\Gamma$ indicates the $DC$, with no significant acoustic enhancement at lower $DC$s.
This is consistent with our results, which show that US reduces the NP–hydrogel contact time $\tc$ only over lag times of at least $\tau \gtrsim 1000\mus$, requiring multiple oscillation cycles. 
At low $DC$s, where US is on for less than this critical time length, the impact on $\tc$ and, consequently, on NP diffusion is minimal. 
This provides a plausible mechanistic explanation for the lack of enhancement observed by Einen \etal~ at low $DC$s.

Unfortunately, Wu and co-workers \cite{ma18a,karki22a} reported only the duty cycle, not the pulse length, making it impossible to determine whether the larger effect of US-enhanced NP diffusion was due to longer pulse durations.
We emphasize the importance of reporting both duty cycles and absolute pulse duration in such studies, as both parameters may critically influence the extent to which US can enhance NP diffusion in hydrogels.

\section{Conclusions\label{sec:conclusions}}
In this work, we investigated the molecular mechanisms underlying US-enhanced NP diffusion in hydrogels using coarse-grained LD simulations. 
Our model captures the essential physics of NP motion within a polymeric network under US exposure, enabling a systematic investigation of how network retention and US amplitude influence diffusion behavior.

The simulations reproduce key experimental trends reported in the literature. 
Specifically, we find that US can significantly enhance NP diffusion following an "stick-and-release" mechanism, in which NPs  stick transiently to the network and are periodically released by the US oscillations.
The key conditions under which the “stick-and-release” mechanism emerges are sufficient attractive interactions between NPs and the hydrogel network and US intensities high enough to overcome local adhesion without disrupting the network structure. 
In these cases, US reduces the particle-network contact time and promotes escape events, leading to an increased long-time diffusion coefficient. 
This mechanism aligns with experimental observations of effective acoustic diffusion in agarose gels at high-duty-cycle US exposure.

In contrast, for weakly retained NPs, US has minimal impact on long-time diffusion, consistent with studies reporting no enhancement. 
Our results also suggest that the duration of US exposure is critical: enhanced diffusion emerges only when pulse lengths span multiple oscillation cycles, providing a mechanistic explanation for the lack of enhancement observed in low duty-cycle experiments.
Finally, our results emphasize that the specific model used to describe how US couples to the hydrogel is critical in determining both the extent and nature of NP diffusion enhancement.

While our coarse-grained approach enables a controlled, bottom-up investigation, it necessarily involves several simplifications. 
We represent the ECM as a homogeneous, periodic polymer network, thereby neglecting structural heterogeneities, crosslinking variability, and patchy interactions that may be critical in biological tissues. 
Incorporating such complexity in future models will be essential to evaluate how spatial inhomogeneities affect NP transport under US. 
Additionally, hydrodynamic interactions are treated at the level of Langevin friction, neglecting long-range coupling effects between the NP, network, and the surrounding fluid. 
Further studies that explicitly include hydrodynamic interactions will be important to assess their impact on NP mobility and on the dynamic response of the hydrogel network to US.
We also plan to extend our model to incorporate \textit{pulsed} US exposure, allowing us to systematically investigate how the duty cycle and the length of continuous US exposure per pulse modulates the stick-and-release mechanism and the resulting diffusion enhancement.

Overall, our findings identify NP-hydrogel interactions and US exposure parameters as key factors governing acoustically-enhanced NP diffusion. 
Our coarse-grained molecular model captures the qualitative trends expected for such systems, demonstrating that the "stick-and-release" mechanism can lead to acoustically-enhanced NP diffusion.
The quantitative magnitude of this effect is modulated by the effective stickiness between the NPs and the hydrogel matrix, which in turn depends on the specific physicochemical properties of both components.
These insights reconcile seemingly contradictory experimental results and offer a mechanistic framework to inform the design of US-mediated delivery strategies in biological hydrogels and tumor ECM.

\section*{Acknowledgements}
P.M.B., H.H.R. and R.S.D  acknowledge the funding from the European
Union’s Horizon Europe research and innovation program under
the Marie Sklodowska-Curie Grant Agreement No. 101062456 (ModEMUS).
We acknowledge the computational resources from the High Performance Computing center IDUN from the Norwegian University of Science and Technology.
We thank the group of Prof. C. Holm, and specially Dr. J.-N. Grad, for technical assistance to setup the US wave in ESPResSo and Dr. P. Košovan for insightful discussions. 
We thank the group of Prof. Davies for interesting discussions on US-enhanced drug delivery using NPs as carriers.
We thank C. B. Hunskår, M. S. Otervik, C. R. Amundsen and S. B. Johnsen for their valuable preliminary research work within the ModEMUS project during their studies at NTNU.

\bibliographystyle{elsarticle-num} 

\bibliography{icp,our_bibliography}

\appendix
\begin{figure*}
    \centering
    \includegraphics[width=1\linewidth]{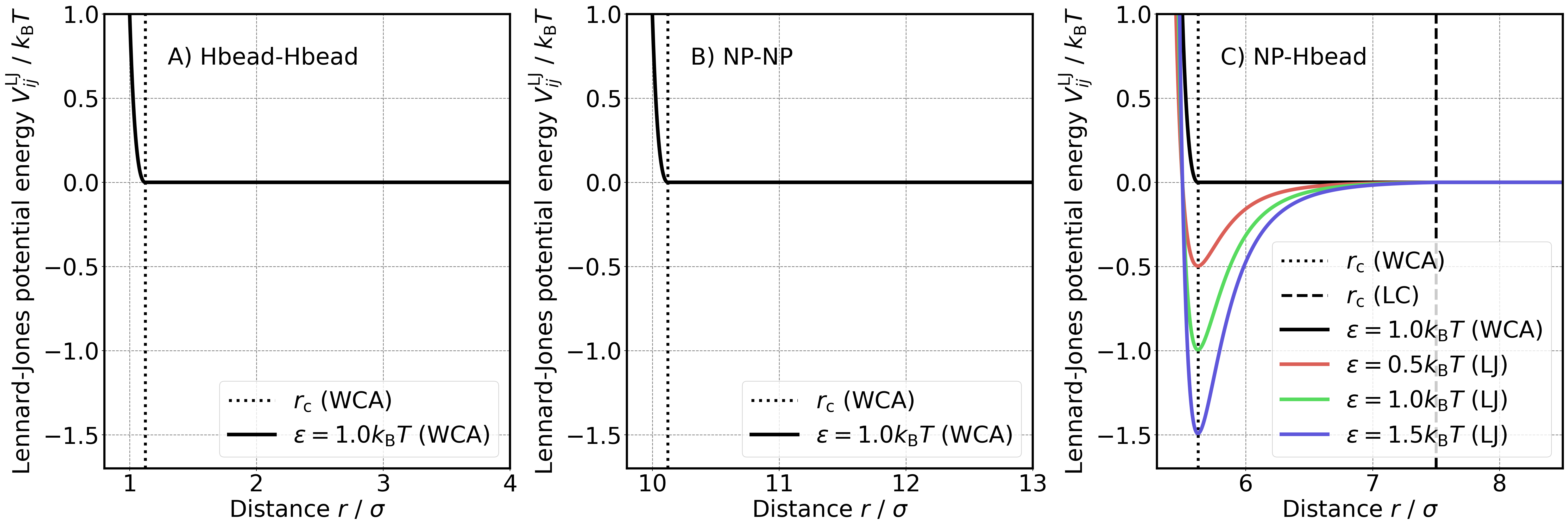}
    \caption{Lennard-Jones (LJ) potential $V^{\mathrm{LJ}}$ between particle pairs as a function of their center-to-center distance $r$,  calculated using Eqs. \ref{eq:lj} and \ref{eq:offset}.
    Panel A: hydrogel bead (Hbead) pairs interact through the purely steric Weeks-Chandler-Andersen (WCA) potential with $\roff = 0$ and $\rc = 2^{1/6}\sigma$.
    Panel B: nanoparticle (NP) pairs interact through the WCA potential with $\roff = 9\sigma$ and and $\rc = 2^{1/6}\sigma+\roff$.
    Panel C: two different interactions are considered for NP-Hbead interactions, both with $\roff = 4.5\sigma$.
    For the steric network, the WCA potential is used with $\rc = 2^{1/6}\sigma + \roff$.
    For the sticky network, the LJ potential is applied with
     $\rc = 3\sigma + \roff$ and $\epsilon$ values ranging from $0.5\kT$ to $1.5\kT$.
     In all panels, dotted lines indicate the value of $\rc$ for the WCA potential, while dashed lines represent the corresponding value for the LJ potential. 
     }
    \label{fig:rcut}
\end{figure*}
\section{Additional information about the simulation model \label{sec:additional_info_model}}
In this section, we expand upon the general description of the molecular model provided in the main text. 
\subsection{Non-bonded interactions}
Non-bonded interactions between particles are modeled using the following pairwise augmented Lennard-Jones (LJ) potential\cite{beyer24a}
\begin{equation}
\begin{split}\label{eq:lj}
  V^{\mathrm{LJ}}_{ij}(r) =&
    \begin{cases}
      \infty & \text{for}\ r < \roff\\
      4 \epsilon \left[ \left(\displaystyle\frac{\sigma}{r-\roff}\right)^{12}
      - \left(\displaystyle\frac{\sigma}{r-\roff}\right)^6\right] - C
      & \text{for}\ \roff < r < \rc \\
      0 & \text{for}\ r > \rc
    \end{cases},\end{split}
\end{equation}
where $r$ is the center-to-center distance between particles $i$ and $j$, $\sigma$ sets the effective range of repulsion, $\epsilon$ is the depth of the attractive well, $\rc$ is the cutoff distance beyond which interactions vanish and $\roff$ is an offset that shifts the potential.
The constant $C = 4\epsilon( (\sigma / \rcut)^{12} - (\sigma / \rcut)^{6} )$ ensures continuity at the cutoff, where $\rcut = \rc - \roff$.
This augmented LJ potential reduces to the classical LJ form when $\roff = 0$ and $\rcut \rightarrow \infty$.

We define the hydrogel bead diameter as $\rbead = \qty{10}{nm}$, which serves as the unit of length in our reduced unit system ($\rbead = 1$).
We set $\sigma = \rbead = 1$ for all particle pairs, making the shape of the LJ potential independent of particle size.
Instead, we use a size-dependent offset $\roff$ based on the diameters $d_i$ and $d_j$ of the interacting particles: 
\begin{equation}
    \roff = \frac{1}{2}(d_i+d_j)-\sigma\,.
    \label{eq:offset}
\end{equation}
This offset ensures that the potential reflects the physical contact distance between particles. 
For $\rc= 2^{1/6}\sigma + \roff$ and $\epsilon = \kT =1$, the potential (\refeq{eq:lj}) reduces to the Weeks-Chandler-Andersen (WCA) form, regardless of the particle size. 

As illustrated in \reffig{fig:rcut}, non-bonded interactions are modeled using the steric WCA potential for (i) hydrogel bead (Hbead) pairs (\reffig{fig:rcut}A) and (ii) nanoparticle (NP) pairs (\reffig{fig:rcut}B). 
For NP–Hbead pair interactions (\reffig{fig:rcut}C), we consider two scenarios: (i) the `steric network' (black line), where only steric hindrance is present, modeled with the WCA potential; and (ii) the `sticky network' (colored lines), which include attractive interactions modeled using the LJ potential with $\epsilon$ values ranging from $0.5\kT$ to $1.5\kT$. 
For the sticky network, we set $\rc = 3\sigma + \roff = 7.5\sigma$, as $V_{ij}^{\mathrm{LJ}}(r=7.5\sigma)\approx 0$ for all considered $\epsilon$ values.

\subsection{Bonded interactions}

Hydrogel beads are connected via the Finite Extensible Non-linear Elastic (FENE) potential:
\begin{equation}
    \label{eq:fene}
  V^{\mathrm{FENE}}_{ij}(r) = - \frac{1}{2} \kfene \rmax^2 \ln \left(1-\left( \frac{r}{\rmax} \right)^2\right)\,, 
\end{equation}
where $\kfene = 30 \kT / \sigma^2$ is the spring constant and  $\rmax = 1.5\sigma$ is the maximum stretching length.
These values are standard for Kremer-Grest models of linear polymer chains\cite{everaers20a}, and have been used in similar hydrogel models in the literature.
\cite{kosovan15a,beyer22a}

To control hydrogel stiffness, we apply a harmonic bending potential:
\begin{equation}
    \label{eq:bending}
  V^{\mathrm{bend}}_{ijk}(\theta) =  \frac{1}{2} \kbend (\theta-\theta_0)^2 \,,
\end{equation}
where $\theta$ is the bond angle between beads $i$, $j$ and $k$, $\theta_0 = \pi \, \mathrm{rad}$ is the equilibrium bond angle, and $\kbend = 6\kT/\mathrm{rad}^2$ is the bending constant. 
The latter was tuned to achieve a hydrogel volume fraction at swelling equilibrium consistent with experimental agarose gels.

\subsection{Swelling equilibrium}
Swelling equilibrium of a hydrogel is reached when the pressure in the simulation box (or the `system') $\Psys$  equals that of the supernatant (or the `reservoir') $\Pres$. 
Within the used model, the supernatant is a polymer-free aqueous buffer solution and the system is the hydrogel phase.
In line with the implicit solvent description in the model, we assume equal solvent contribution to the total pressure of both the system $\Psys_{\mathrm{sol}}$ and the reservoir $\Pres_{\mathrm{sol}}$. The pressure difference is
\begin{equation}
    \Delta P = \Psys - \Pres = \Psys_{\mathrm{pol}} + \Psys_{\mathrm{sol}} - \Pres_{\mathrm{sol}} \approx \Psys_{\mathrm{pol}},
    \label{eq:DeltaP}
\end{equation}
where $\Psys_{\mathrm{pol}}$ is the hydrogel polymer contribution to the system pressure. 

\begin{figure}
    \centering
    \includegraphics[width=0.9\linewidth]{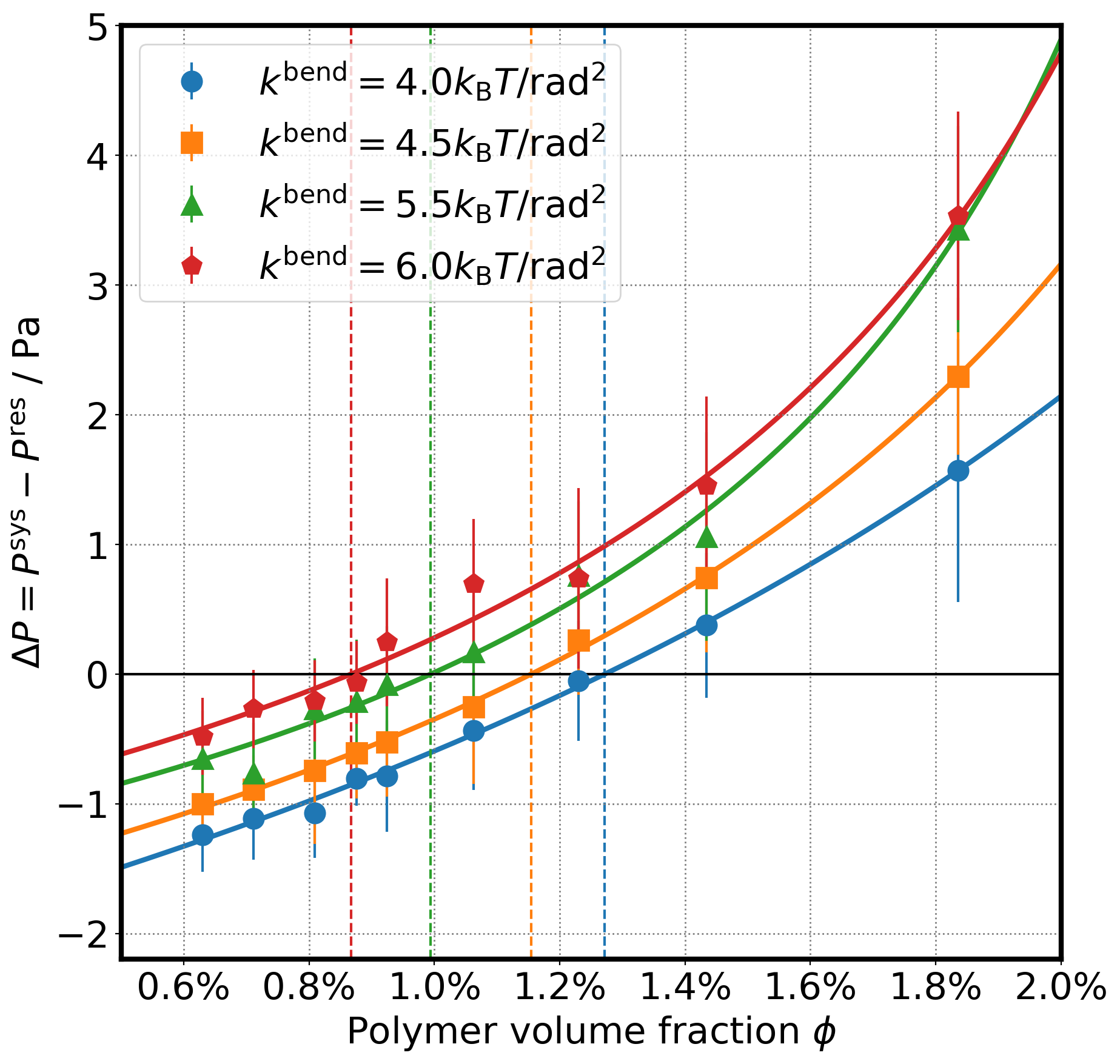}
    \caption{Pressure difference $\Delta P$ (\refeq{eq:DeltaP}) between the hydrogel system and the supernatant reservoir as a function of hydrogel volume fraction $\phi$ (\refeq{eq:phi}) for hydrogels with varying stiffness, controlled by the bending constant $\kbend$ (\refeq{eq:bending}).
    The vertical dashed lines mark the $\phi$ value of swelling equilibrium $\Delta P = 0$ for each system.}
    \label{fig:pv_bending_parametrization}
\end{figure}

Following the so-called PV protocol,\cite{kosovan15a,beyer22a} we measured $P_{\mathrm{sys}}$ at different hydrogel volume fractions, according to
\begin{equation}
    \phi = \frac{N_\mathrm{b}\pi \rbead^3}{6V},
    \label{eq:phi}
\end{equation}
where $\Nbead = 488$ is the total number of hydrogel beads and $V$ is the simulation box volume. For $\kbend = 6\kT/\mathrm{rad}^2$, the model reaches swelling equilibrium at $\phi \approx 0.86\%$, as shown in  \reffig{fig:pv_bending_parametrization}, consistent with experimental  agarose gels values ($\phi \sim 0.5-0.9 \%$).\cite{ma18a,einen23a} 

\newpage

\section{Effect of hydrogel response to ultrasound on nanoparticle diffusion}
The mechanical response of a macroscopic hydrogel network, such as agarose, to nanoscale perturbations induced by ultrasound (US) irradiation is not physically intuitive, making it challenging to incorporate accurately into molecular-scale models.
Importantly, the dynamics of the hydrogel network and its coupling to US can directly influence NP diffusion through the matrix.

To investigate this, we evaluated four distinct modeling approaches:

\begin{enumerate}
  \renewcommand{\labelenumi}{\roman{enumi}.}
  \item \emph{Fixed network}: All hydrogel beads are immobilized throughout the simulation.
  \item \emph{Flexible network unaffected by US}: Hydrogel beads undergo thermal motion but are not influenced by US forces.
  \item \emph{Flexible network subjected to the same US force as the NPs}: All particles, regardless of size, experienced the same US force.
  \item \emph{Flexible network under size-scaled US force}: The US force is scaled by particle size, ensuring equal contribution to the particle velocity.
\end{enumerate}

Model (iv), which we consider the most physically consistent, is employed and discussed throughout the main text. 
The implications and outcomes of all four approaches are analyzed in detail below.

\subsection{Fixed Network}
\label{subsec:fixed_network}

\begin{figure}
    \centering
    \includegraphics[width=0.9\linewidth]{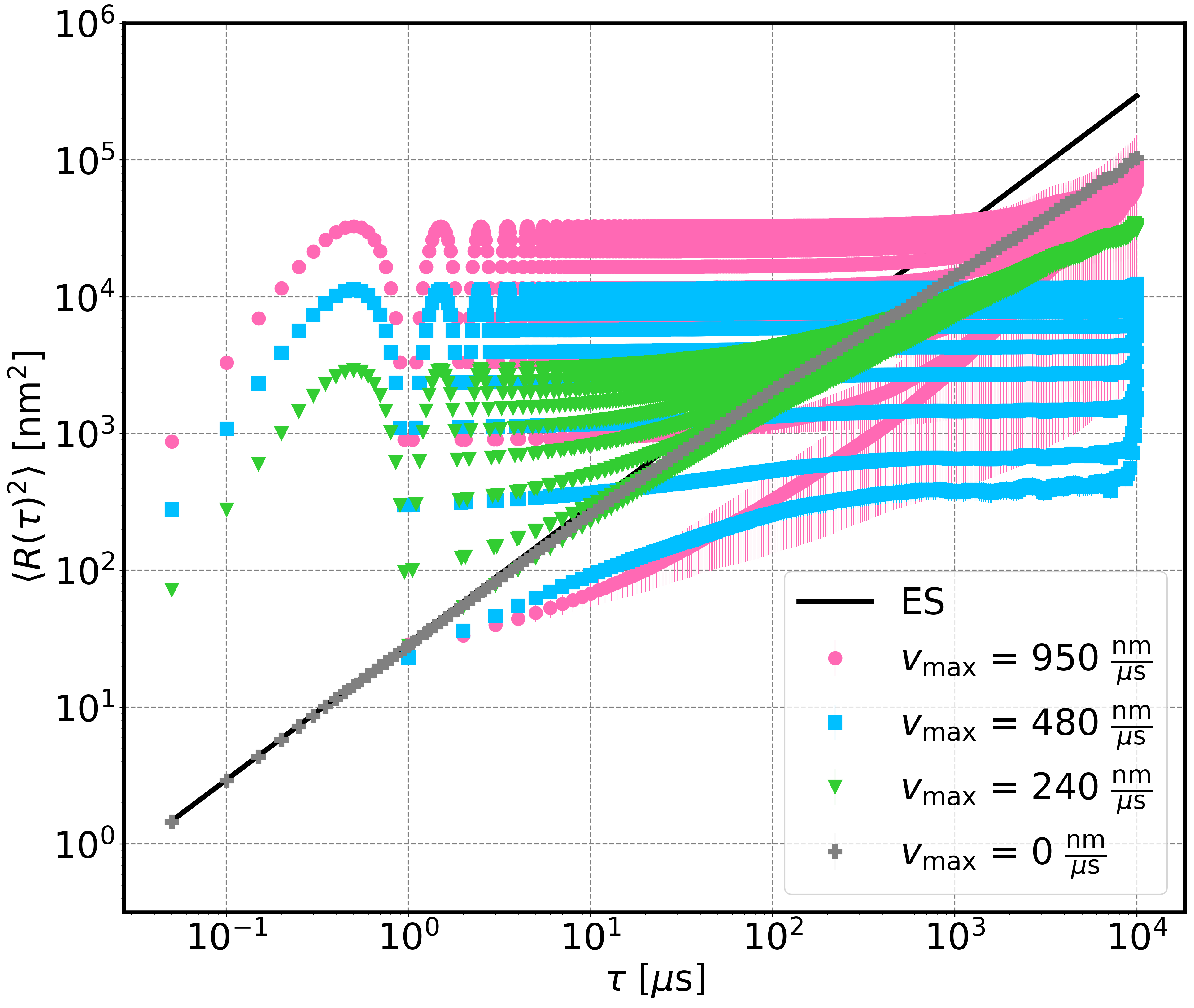}
    \caption{Mean squared displacement, $\MSD$, of NPs in a fixed hydrogel network (polymer volume fraction $\phi = 0.86\%$), shown as a function of lag time $\tau$, for different US peak pressures $\Pmax$. 
    A \emph{fixed network} is assumed, in which all hydrogel beads remain immobilized throughout the simulation.
    The Einstein-Smoluchowski (ES) analytical solution (\refeq{eq:einstein_smo}) is shown in black for reference.}
    \label{fig:MSD_fixed_network}
\end{figure}

In this model, after relaxing the hydrogel network to its swelling equilibrium, hydrogel beads are held fixed in space in an arbitrary configuration, and only NPs are subjected to the oscillatory US force. 
\reffig{fig:MSD_fixed_network} shows the $\MSD$ of the NPs in a fixed network with a polymer volume fraction of $\phi = 0.86\%$, corresponding to swelling equilibrium. 
A pronounced caging effect emerges, where NPs repeatedly collide with the static network and return to their original positions after each oscillation cycle. 
This is most evident at $\Pmax = \qty{0.72}{MPa}$ (blue squares), where the net  $\MSD$ remains nearly constant. 
At the lower amplitudes ({\it e.g.}, $\Pmax = \qty{0.36}{MPa}$, green triangles), the effect is less severe but still noticeable compared to the control system without US (gray crosses). 
Interestingly, the caging effect appears reduced at the highest peak velocity ($\Pmax = \qty{1.42}{MPa}$, pink circles), possibly
due to the force overcoming confinement. 
However, large error bars prevent definitive conclusions in this case. 

A similar semi-fixed network model was employed by Price \textit{et al.}\cite{price24a}, in which hydrogel beads were tethered to fixed lattice positions via harmonic springs, thereby restricting their motion with a rigid spring constant.
Caging effects were not explicitly reported, but the relatively low hydrogel density and large reported pore size ($100-320$ nm) - significantly greater than the diameters of the NPs used in the study (30 nm and $\sim$ 90 nm) - likely prevented their emergence.
Furthermore, only very low US amplitudes ($0.04-0.13$ MPa) - below the minimum US peak pressure ($\approx$ 0.18 MPa, \cf~ Section 3.3) required for the acoustic contribution to NP motion to exceed that of thermal diffusion - were examined for the larger NPs ($\sim$ 90 nm), where steric hindrance from the hydrogel would have been more pronounced, further mitigating any potential caging effects.

In summary, the fixed network model induces strong NP confinement and was therefore deemed unsuitable for further analysis.

\subsection{Flexible network unaffected by US}
\label{subsec:flex_NO_US}
To mitigate the pronounced caging effect observed in the fixed network model, we next allowed hydrogel beads to undergo thermal motion while excluding them from the applied US force.
In this model, only the NPs respond to US, while the hydrogel network remains passive and unaffected by the oscillatory field.

\begin{figure} 
    \centering
    \includegraphics[width=0.9\linewidth]{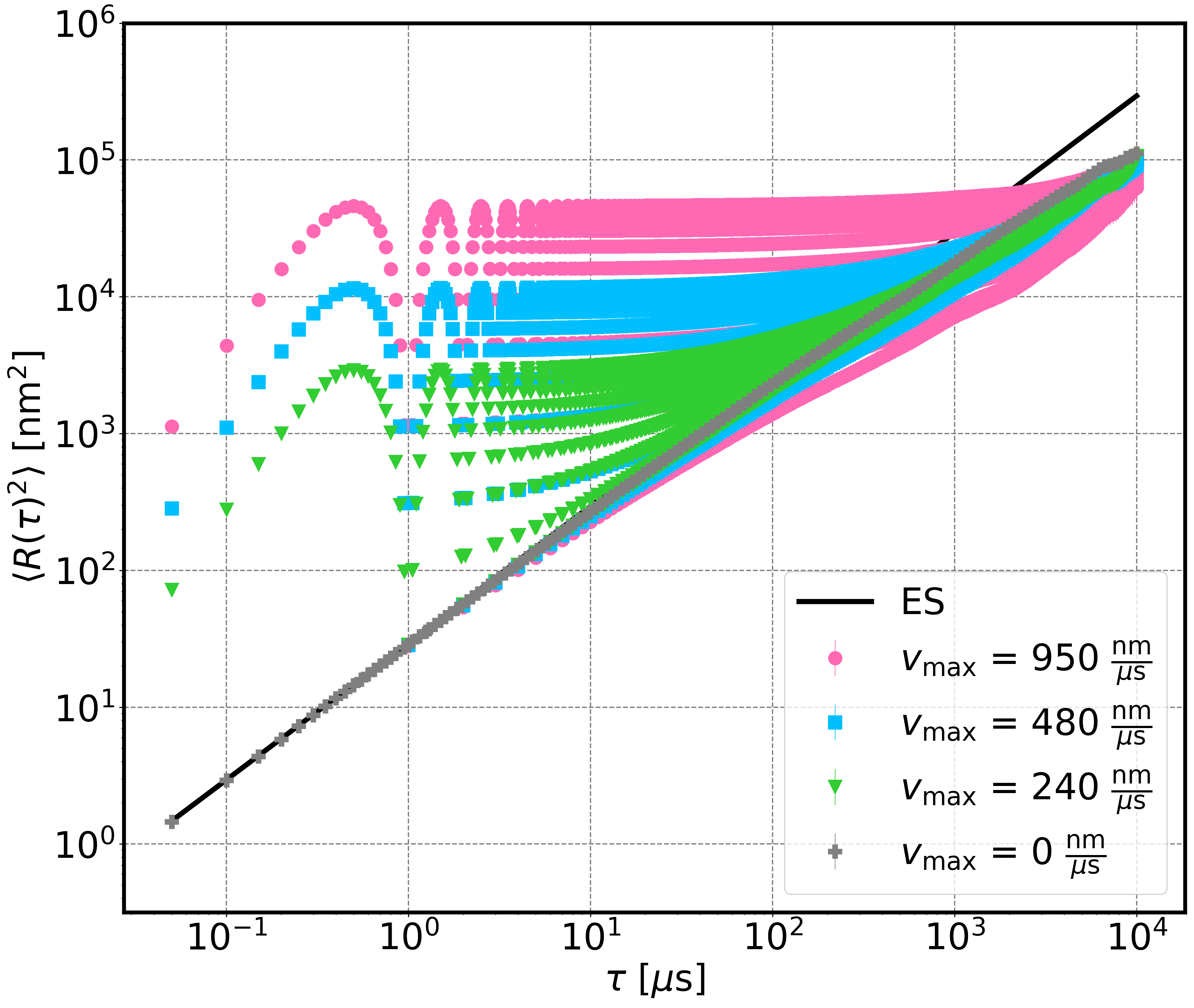}
    \caption[$\MSD$ of NPs in a Flexible, Non-oscillating Network]{Mean squared displacement, $\MSD$, of NPs in a flexible, non-oscillating hydrogel network with polymer volume fraction $\phi = 0.86\%$, shown as a function of lag time $\tau$ for various US peak pressures, $\Pmax$. 
    A \emph{flexible network unaffected by US} is assumed, where hydrogel beads undergo thermal motion but are not influenced by US forces.
    The Einstein-Smoluchowski (ES) analytical solution (\refeq{eq:einstein_smo}) is shown in black for reference.}
    \label{fig:MSD_flex_no_US}
\end{figure}

As shown in \reffig{fig:MSD_flex_no_US}, the thermal motion of the network beads diminishes the caging effect compared to the fixed network.
However, at high US amplitudes ({\it e.g.}, $\Pmax = \qty{1.42}{MPa}$, pink circles), NPs still experience confinement due to repeated collisions with the passive hydrogel network.
This residual caging limits the realism of the model.
Since hydrogels are composed of flexible polymer chains that respond to mechanical perturbations, neglecting their interaction with US is physically inconsistent.
For this reason, this model was also deemed inadequate for further analysis.

\subsection{Flexible network affected by US}
\label{subsec:flex_with_US}
The final two models incorporate a flexible hydrogel network in which the polymer beads are also subjected to the US force $\Fext$.
We explored two variants of this setup:
(i) a uniform US force model, where all particles - regardless of size - experience the same oscillatory force; and (ii) a size-scaled US force model, where the applied US force is scaled according to particle size, ensuring an equal contribution of US to the particle velocity.

As discussed in \refsec{sec:analytical_ES_US}, for an ideally dilute system, the velocity $\mathbf{v}$ of a particle in our model is given by
\begin{equation}
\mathbf{v} =
\frac{{\rm d} \mathbf{x}}{{\rm d} t} =
\sqrt{\frac{2k_{\rm B}T}{\gamma}}\boldsymbol{\xi}
+\frac{\Fext}{\gamma} =
\sqrt{\frac{2k_{\rm B}T}{\gamma}}\boldsymbol{\xi}
+\frac{A}{\gamma}\cos(2\pi f t)\hat{\mathbf{n}},
\label{eq:ideal_v}
\end{equation}
where $A$ is the amplitude of the US force, $\gamma$ is the friction coefficient of the particle, and the remaining symbols (not relevant for the discussion here) are defined in \refsec{sec:analytical_ES_US}.
The first term describes the contribution from thermal fluctuations, while the second term accounts for the deterministic oscillatory motion induced by the US wave.

In the uniform US force model, we assume that all particles experience the same force amplitude $A$, regardless of their size.
According to \refeq{eq:ideal_v}, the US-induced velocity component is inversely proportional to $\gamma$, and thus to particle size (\cf~Eq. 3 in the main text).
Consequently, smaller particles exhibit larger velocity amplitudes and oscillate more rapidly than larger ones under US, leading to size-dependent US dynamics.

To ensure consistent US motion across particle sizes, in the size-scaled US force model, we scaled the US force amplitude proportionally to each particle’s friction coefficient, such that $A = \gamma \vmax$, where $\vmax$ is the peak velocity of the US field.
This results in an equal contribution of the US field to the velocity of all particles, regardless of their size:
\begin{equation}
\mathbf{v} =
\sqrt{\frac{2k_{\rm B}T}{\gamma}}\boldsymbol{\xi} + 
\vmax \cos(2\pi f t)\hat{\mathbf{n}}.
\label{eq:our_v}
\end{equation}

\begin{figure} 
    \centering
    \includegraphics[width=0.9\linewidth]{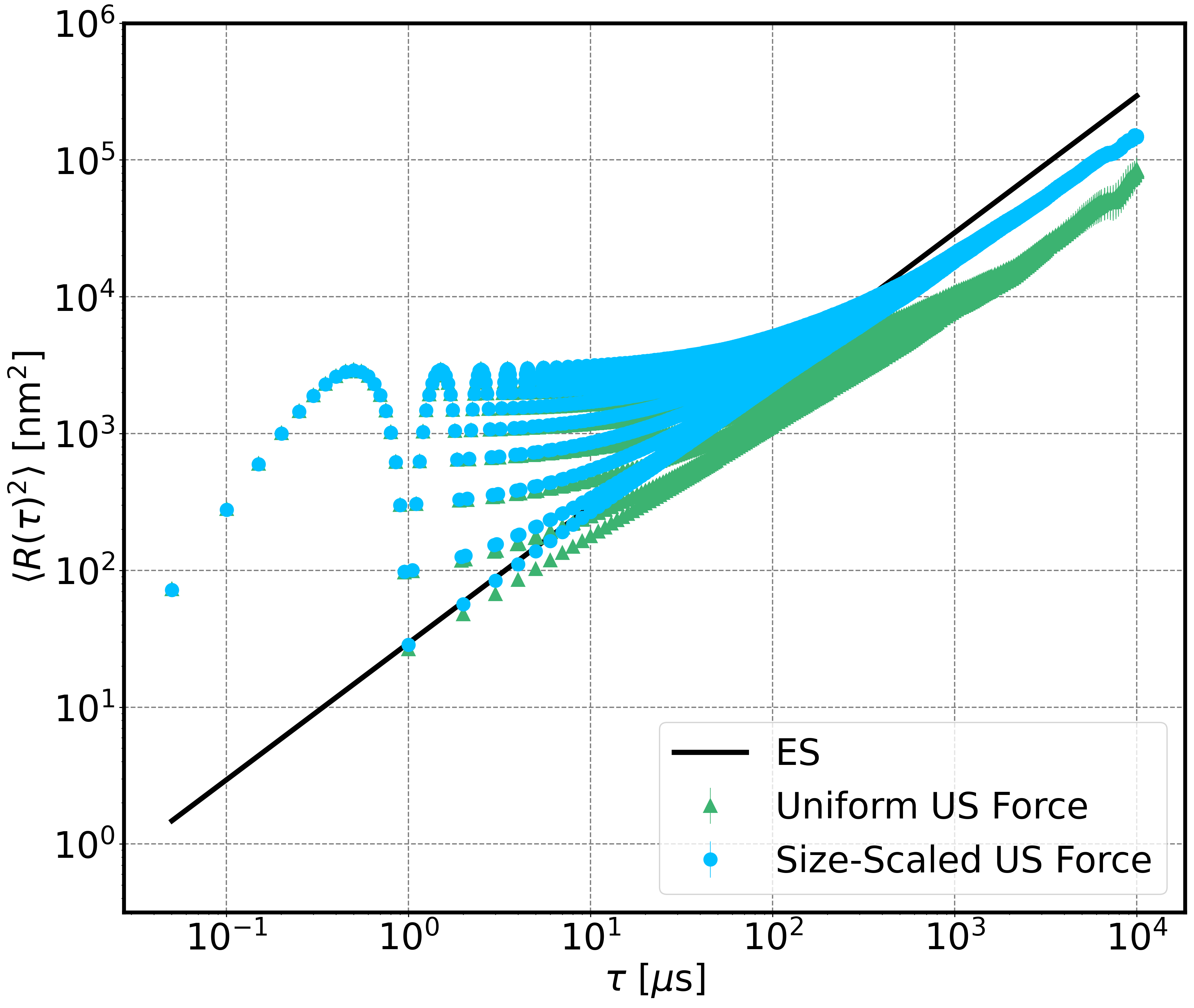}
    \caption{Mean squared displacement, $\MSD$, of NPs in a flexible hydrogel network (polymer volume fraction $\phi = 0.86\%$) under US irradiation with $\vmax = 240$ nm/$\mu$s. 
    Two models are compared: \emph{the uniform US force model} (green triangles), where the same force is applied to all particles, and the \emph{size-scaled US force model} (blue circles), where the force is scaled by particle size.
    The Einstein-Smoluchowski (ES) analytical solution (\refeq{eq:einstein_smo}) is shown in black for reference.
    }
    \label{fig:MSD_flex_with_US}
\end{figure}

\reffig{fig:MSD_flex_with_US} compares the $\MSD$ of NPs in the two models.
In the uniform US force model (green triangles), all particles experience the same force amplitude: $A_{\text{bead}} = A_{\text{NP}} = \gamma_{\mathrm{NP}} \vmax$.
In the size-scaled US force model (blue circles), the US force amplitude is scaled with each particle's friction coefficient: $A_{\text{NP}} = \gamma_{\mathrm{NP}}\vmax$ and $A_{\text{bead}} = \gamma_{\mathrm{bead}}\vmax$.
In both cases, the peak velocity of the US field was fixed at $\vmax = 240$ nm/$\mu$s.

As shown in \reffig{fig:MSD_flex_with_US}, the uniform US force model results in reduced NP diffusion at long times compared to the scaled model.
Since the force on the NPs is identical in both cases, this difference arises from the more rapidly oscillating network beads in the uniform model, which causes frequent collisions and increases macromolecular hindrance.
In contrast, by ensuring all components oscillate at the same velocity, the size-scaled model minimizes relative motion between components, which results in enhanced NP transport, in relation to the uniform US force model.

Moreover, in the uniform US force model, the large US force applied to the low-drag network beads was sufficient to disrupt inter-bead bonds at $\vmax > 240$ nm/$\mu$s, compromising network integrity.
These findings underscore the importance of applying a size-scaled US force to preserve realistic hydrogel mechanics and reliable diffusion behavior.
Accordingly, the size-scaled US force model was adopted for all subsequent simulations in this study.

\begin{figure}
    \centering
    \includegraphics[width=0.9\linewidth]{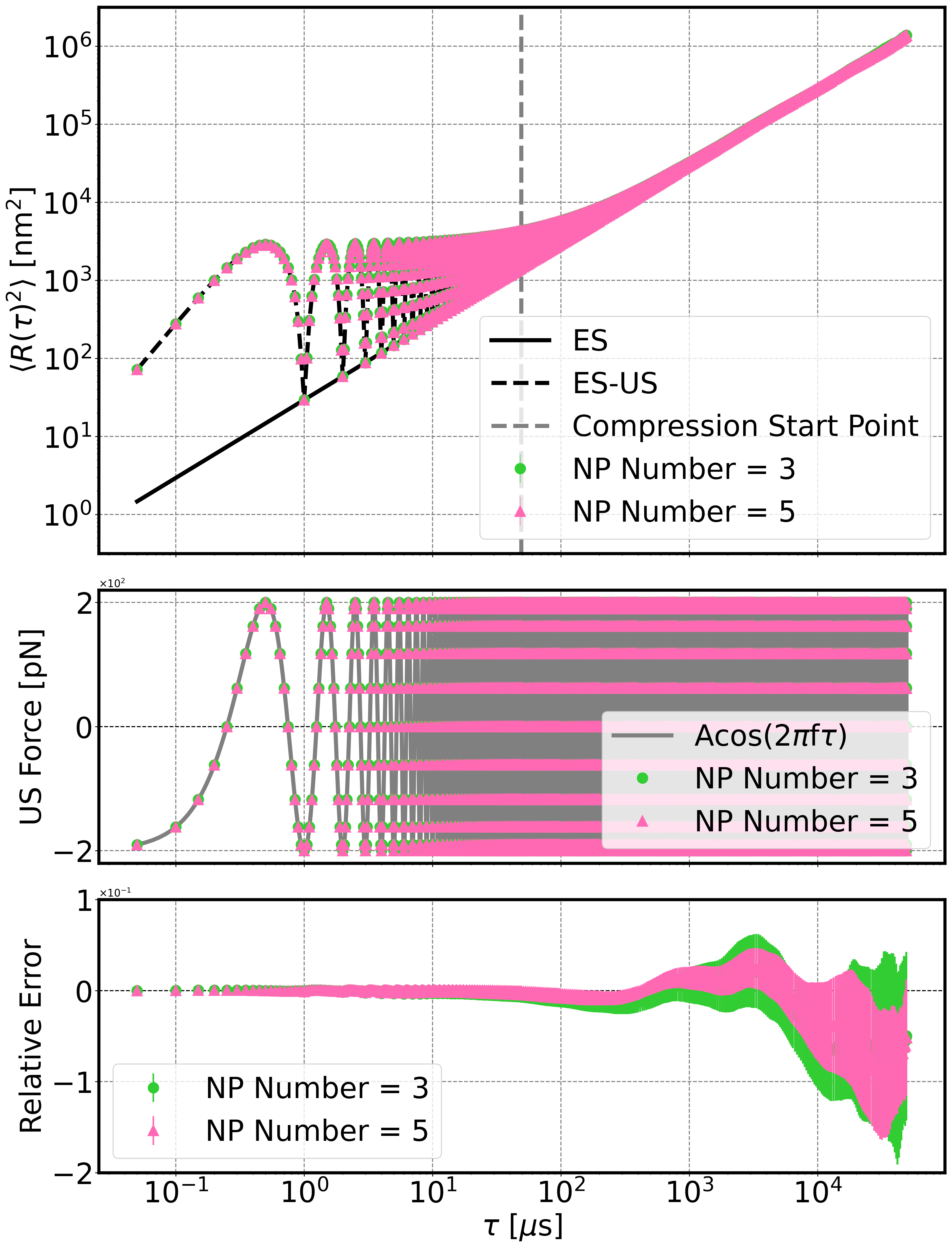}
    \caption{
Top panel: MSD of NPs as a function of lag time $\tau$ in a buffer without hydrogel, under US with peak velocity $\vmax = \qty{240}{nm}/\mus$. 
Simulation results (markers) are compared with the exact analytical solution from \refeq{eq:msd_osc_final} (ES-US). 
For reference, the thermal motion in the absence of US is shown using \refeq{eq:einstein_smo} (ES). 
The vertical line at $p = 40\mus$ marks the onset of non-linear sampling in the multiple-tau correlation algorithm implementation\cite{ramirez10a}.
Middle panel: External force from the US field as a function of $\tau$, given by \refeq{eq:F_osc} (solid line). Markers indicate the $\tau$ values at which MSD was sampled.
Bottom panel: Relative residuals between simulations and the analytical solution from \refeq{eq:msd_osc_final}, defined as
$(\langle R^2 \rangle_{\mathrm{LD}} - \langle R^2 \rangle_{\mathrm{ES-US}}) / \langle R^2 \rangle_{\mathrm{ES-US}}$.
    }
    \label{fig:sampling_benchmark}
\end{figure}

\section{Additional computational details\label{sec:additional_computational_details}}
To improve stability in the numerical integration of the LD simulations, we employ an internal system of reduced units.
The smaller characteristic length in the model is the hydrogel bead diameter $\rbead = \qty{10}{nm}$, which defines the unit of length $\sigma^* = \rbead$.
In reduced units, this sets the hydrogel bead diameter to $\rbead^* = 1 $ and the NP diameter to $\rNP^* = 10$.
We define the unit of time $t^*$ as one-tenth of the US wave period, $t^* = 1/(10f)=\qty{100}{ns}$. 
This choice balances the need for a small time step for stable numerical integration with the ability to reach long-time scales necessary for measuring the long-time diffusion coefficient of the NPs.
The unit of energy $\epsilon^*$ is set to the thermal energy at room temperature $\epsilon^*=\kT$, where $\kb$ is the Boltzmann constant and $T = \qty{298.15}{K}$ is the temperature.
This fixes the unit of mass as $m^*= \epsilon^*t^{*2}/\sigma^{*2}$.

The simulation protocol consisted of two stages: (i) network initialization and relaxation, where firstly we generated a fully stretched diamond hydrogel network and gradually compressed the simulation box to reach the target polymer volume fraction $\phi$. The network was then relaxed (in the absence of the US force) for $10^6$ LD steps, and the final configuration was used as the starting structure for (ii) production LD simulations. Here NPs were introduced, and their motion tracked over $5 \cdot 10^8$ LD steps.
LD integration was performed using the Velocity Verlet algorithm with a time step $\mathrm{d}t = 0.001t^* = \qty{0.1}{ns}$. 

The key observable in our study was the mean squared displacement $\MSD$ of the NPs.
To improve the accuracy of $\MSD$, we segmented each particle trajectory into $N_\mathrm{seg}$ intervals of equal lag time $\tau$ and averaged over them, according to
\begin{equation}
    \MSD = 
    \frac{1}{N_\mathrm{seg}}\sum_{i = 0}^{N_\mathrm{seg}-1}(\mathbf{x}((i+1)\tau) - \mathbf{x}(i\tau))^{2},
    \label{eq:msd_tau}
\end{equation}
where $N_\mathrm{seg} = t_\mathrm{sim}/\tau$ and $t_\mathrm{sim}$ was the total simulation time. 
As $\tau$ increases, $N_\mathrm{seg}$ decreases, reducing $\MSD$ accuracy at long times.
To mitigate this, we averaged over all particles of the same type and across multiple independent LD  production runs with different random seeds.

We used the multiple-tau correlation algorithm\cite{ramirez10a} to efficiently compute the averages in \refeq{eq:msd_tau} on the fly.
We validated our implementation of the algorithm for an ideal system consisting of NPs in a buffer solution, which has an analytical solution given by \refeq{eq:msd_osc_final} (ES-US) as shown in \reffig{fig:sampling_benchmark}.
This algorithm samples $\MSD$ linearly for short lag times (below a threshold time $p$) and compresses data for longer $\tau$, reducing memory usage and runtime. 
At the first compression level, $\MSD$ is sampled every $\tau = 1/(20f) = \qty{50}{ns}$, corresponding to 20 points per US cycle (see \reffig{fig:sampling_benchmark}, middle panel).
Beyond $p = 40\mus$, corresponding to 40 wave periods, data are grouped into compression levels $n$ with intervals $\tau \in [2^{n - 1} p,2^{n}(p-1)]$.
At each compression level, the data from the lower level is grouped to produce a single data point, preserving sampling across different points of the wave period, as can be observed in the middle panel of \reffig{fig:sampling_benchmark}.
For both NPs numbers $N_\mathrm{NP}$ used in this study, the computed $\MSD$ agrees well with the analytical solution from \refeq{eq:msd_osc_final} (ES-US) within statistical accuracy for $\tau \lesssim 2\cdot 10^4\mus$. 
At longer $\tau$ values, small systematic deviations appear, and we excluded these data points from the calculation of the long-time diffusion coefficient $\Dlong$.

\section{Choice of mass for hydrogel beads}
We sat the mass of the NPs to $m_\mathrm{NP}=\frac{1}{6}\rho_\mathrm{NP}\pi\rNP^3 \approx \qty{330}{MDa}$, where 
$\rho_\mathrm{NP} = \qty{1.055}{g/mL}$ is the density 
of the polystyrene nanoparticles (F8801 FluoSpheres\textsuperscript{\textregistered}, ThermoFisher, USA) used in previous {\it in vitro} experiments.\cite{ma18a,einen23a} 
In our reduced unit system, this corresponds to a NP mass of $\mNP^*= 1.3$.
We arbitrarily assigned the hydrogel bead mass as $\mbead^* = 0.1 \mNP = 0.13$.
This value is larger than the estimated mass of an agarose bead, $\mbead=\frac{1}{6}\np\pi\rbead^3 \approx \qty{517}{kDa}$ or $\mbead=0.002\mNP$, where $\np = \qty{1.64}{g/mL}$ is the density of dry agarose.\cite{laurent67a}

Using the physically accurate bead mass would require a significantly smaller time step to maintain numerical stability, which would drastically increase the computational cost of sampling long-time NP diffusion.
In practice, the particle mass only affects the inertia term in the Langevin equation (\refeq{eq:langevin}), which becomes negligible at sufficiently long times.

\begin{figure}
    \centering
    \includegraphics[width=0.9\linewidth]{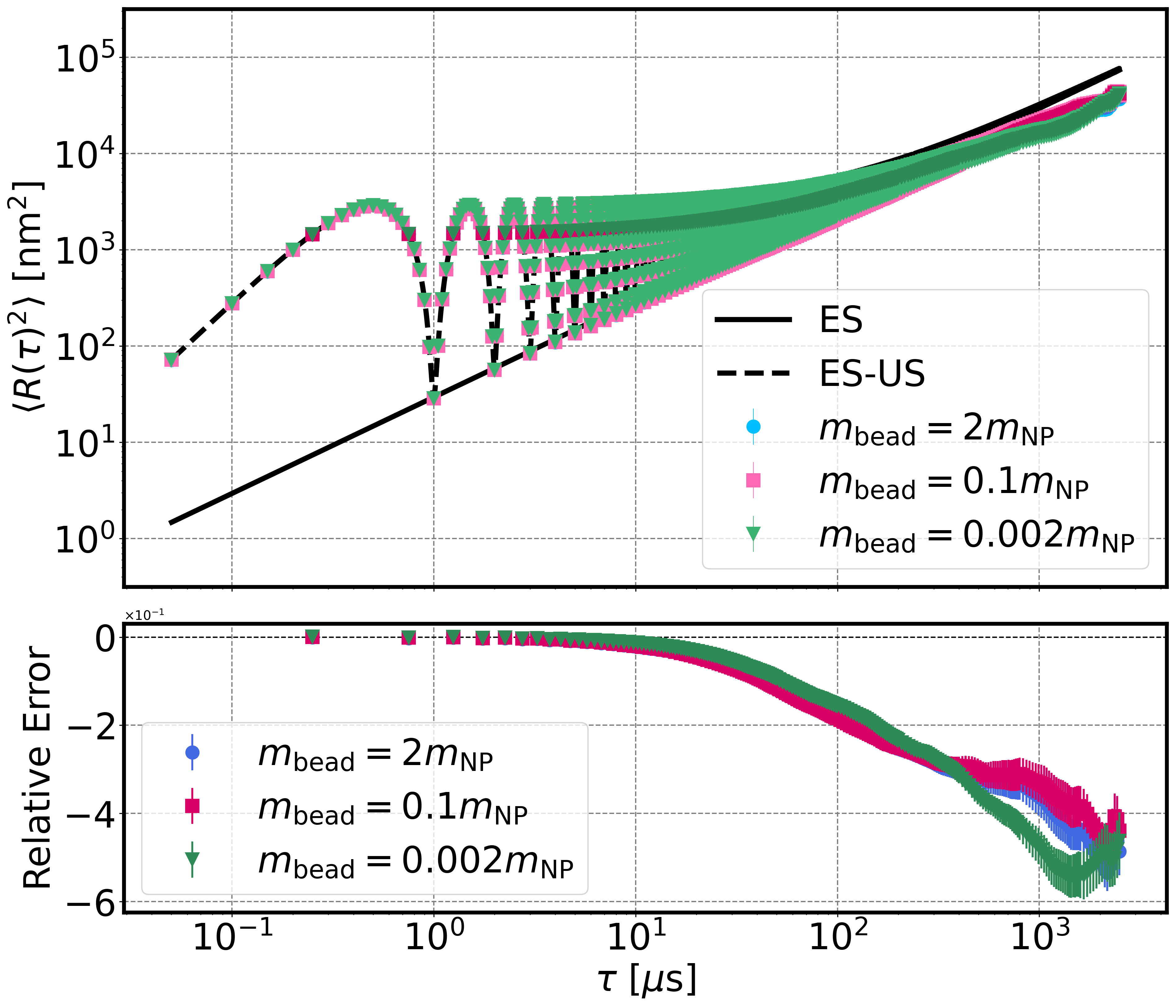}
    \caption{Mean square displacement $\MSD$ of the NPs as a function of lag time $\tau$ in a steric hydrogel network with polymer volume fraction $\phi = 0.86 \%$, for different hydrogel bead masses. 
    Simulations were performed under US irradiation with peak velocities $\Pmax = \qty{0.36}{MPa}$.
    The analytical benchmarks from the Einstein-Smoluchowski (ES) equation (\refeq{eq:einstein_smo}) and its US-extended form (ES-US, Eq. \ref{eq:msd_osc_final}) are shown for reference.
    The lower panel shows the relative error of the simulations respect to the ES baseline, defined as $(\MSD_{\textrm{LD}}-\MSD_{\textrm{ES-US}})/\MSD_{\textrm{ES-US}}$.
    }
    \label{fig:mass_test}
\end{figure}

To assess the impact of $\mbead$ on NP diffusion, we performed LD simulations with varying $\mbead$, as shown in \reffig{fig:mass_test}.
We observe that the $\MSD$ curves for all tested $\mbead$ are nearly identical, with only minor differences at long lag times. 
To examine these differences more closely, we computed the relative deviations from the analytical benchmark (\refeq{eq:msd_osc_final}), shown in the lower panel of \reffig{fig:mass_test}.
The relative deviations are small and consistent across bead masses, with slight discrepancies at larger $\tau$ values where statistical sampling is poorer.
These results justify our choice of $\mbead = 0.1 \mNP$, which enabled efficient simulations without compromising the qualitative conclusions of our study.

\section{Analytical solution for the mean squared displacement of a Brownian particle under oscillatory stress in dilute solution\label{sec:analytical_ES_US}}
We derive the mean square displacement  of a Brownian particle subjected to an oscillatory external force in a dilute solution, starting from the Langevin equation:\cite{pathria96a,vangusteren81a} 
\begin{equation}
    m \frac{{\rm d} \mathbf{v}}{{\rm d} t}= 
    - \gamma \mathbf{v} 
    + \sqrt{2\gamma\kT}\bfxi
    - \nabla \mathbf{V}(\mathbf{x})
    + \Fext \,. 
    \label{eq:langevin}
\end{equation}
Here $m$, $\mathbf{v}$ and $\gamma$ are the particle mass, velocity and the friction (Stokes) coefficient, respectively.
$\kT$ is the thermal energy, which depends on the temperature $T$ and the Boltzmann constant $\kb$.
$\bfxi$ is a white noise with the following ensemble properties
\begin{align}
    \left< \bfxi(t) \right> &= 0, \label{eq:xi_mean} \\
    \left< \bfxi(t) \cdot \bfxi(t')\right> &= d \delta(t-t'), \label{eq:xi_var}
\end{align}
where $\delta$ is the Dirac delta function, $d$ is the dimensionality of the system and the brackets $\left< \dots \right>$ denote the ensemble time average.
This white noise term mimics the random thermal motion of the Brownian particle due to its collisions with the solvent.
The term $\nabla \mathbf{V}$ is the gradient of the potential energy at particle position $\mathbf{x}$ due to molecular interactions, and 
$\Fext$ is an external force acting on the particle. 

In the so-called overdamped limit, when the inertial effects are negligible compared to the friction force $(m \frac{{\rm d} \mathbf{v}}{{\rm d} t} \ll \gamma \mathbf{v})$, \refeq{eq:langevin} simplifies to
\begin{equation}
    0 = 
    - \gamma \mathbf{v} 
    + \sqrt{2\gamma\kT}\bfxi
    - \nabla \mathbf{V}(\mathbf{x})
    + \Fext \,. 
    \label{eq:langevin_overdamped}
\end{equation}
This approximation is valid for NPs in viscous media such as water, where overdamped dynamics dominate at lag times beyond a few nanoseconds. 
In the ideal dilution limit, molecular interactions vanish ($\nabla \mathbf{V}(\mathbf{x}) = 0$) and \refeq{eq:langevin_overdamped} can be further simplified
\begin{equation}
    0 = 
    - \gamma \mathbf{v} 
    + \sqrt{2\gamma\kT}\bfxi
    + \Fext. 
    \label{eq:langevin_ideal}
\end{equation}

We now consider the particular case of a Brownian particle  under oscillatory stress due to an external field 
\begin{equation}
    \Fext =   A\cos(2\pi f t)\hat{\mathbf{n}} \,,
    \label{eq:F_osc}
\end{equation}
where $A$ is the peak force amplitude, $f$ is the frequency, and 
$\hat{\mathbf{n}}$ is the unit vector in the direction of wave propagation.
Such external field could be, for example, an acoustic oscillation due to US irradiation.
Substituting \refeq{eq:F_osc} into \refeq{eq:langevin_ideal} and solving for $\mathbf{v}$, yields the particle velocity:
\begin{equation}
    \mathbf{v} = 
    \frac{{\rm d} \mathbf{x}}{{\rm d} t} =
    \sqrt{\frac{2\kT}{\gamma}}\boldsymbol{\xi}
    +\frac{A}{\gamma}\cos(2\pi f t)\hat{\mathbf{n}} \,.
    \label{eq:langevin_ideal_osc}
\end{equation}

As a first step to calculate the mean square displacement, we integrate the equation of motion \refeq{eq:langevin_ideal_osc}, 
\begin{align}
    \int \textrm{d}{\mathbf{x}} =&
     \int_{t_0}^\tf \sqrt{\frac{2\kT} {\gamma}}\boldsymbol{\xi}\dt +
     \int_{t_0}^\tf \frac{A}{\gamma}\cos(2\pi f t)\hat{\mathbf{n}}\dt,
    \label{eq:v_integral} \\
    \mathbf{x}(\tf) - \mathbf{x}(t_0) =& 
    \int_{t_0}^\tf \sqrt{\frac{2\kT} {\gamma}}\boldsymbol{\xi}\dt \nonumber  \\  
     &+
      \frac{A}{2\pi f\gamma} 
    \left(  \sin(2\pi f \tf) - \sin(2\pi f t_0)\right) \hat{\mathbf{n}}\,, \label{eq:diff_x}
\end{align}
where $t_0$ and $\tf$ are the initial and the final sampling times, respectively.
Defining the lag time $\tau = \tf-t_0$, the MSD is:
\begin{equation}
    \MSD \equiv \left< \left( \mathbf{x}(\tf) - \mathbf{x}(t_0)  \right)^2\right> \,.
    \label{eq:msd}
\end{equation}
Combining Eqs. \ref{eq:msd} and \ref{eq:diff_x}, it follows that
\begin{align}
    & \MSD =  \notag \\ 
    & 
    < ( \int_{t_0}^\tf \sqrt{\frac{2\kT} {\gamma}}\boldsymbol{\xi}\dt \notag \\
    & +      \frac{A}{2\pi f\gamma} 
    \left(  \sin(2\pi f \tf) - \sin(2\pi f t_0)\right) \hat{\mathbf{n}} )^2 > = \notag \\
   & \left< \left( \int_{t_0}^\tf \sqrt{\frac{2\kT} {\gamma}}\boldsymbol{\xi}\dt \right)^2 \right> \notag \\
    &+
     \left< \left( \frac{A}{2\pi f\gamma} 
    \left(  \sin(2\pi f \tf) - \sin(2\pi f t_0)\right) \hat{\mathbf{n}} \right)^2 \right> \notag \\
    &+ \left<  \frac{A}{\pi f\gamma} \left(  \sin(2\pi f \tf) - \sin(2\pi f t_0)\right) \hat{\mathbf{n}} \int_{t_0}^\tf \sqrt{\frac{2\kT} {\gamma}}\boldsymbol{\xi}\dt \right>.
    \label{eq:msd_lengthy}
\end{align}
For clarity, let us simplify the right-hand side of \refeq{eq:msd_lengthy} term-by-term. 
The first term yields the classical Brownian contribution using the dissipation-fluctuation theorem\cite{howard93a} and the white noise property (\refeq{eq:xi_var}):
\begin{align}
    &\left< \left( \int_{t_0}^\tf \sqrt{\frac{2\kT} {\gamma}}\boldsymbol{\xi}\dt \right)^2 \right> = \notag \\
    &
    \frac{2\kT} {\gamma} \int_{t_0}^\tf \int_{t_0}^\tf  
    \left< \boldsymbol{\xi}(t) \cdot \boldsymbol{\xi}(t') \right>
    \textrm{d}t\textrm{d}t' =
     \notag \\
    & \frac{2\kT} {\gamma} \int_{t_0}^\tf \textrm{d}t
    \int_{t_0}^\tf  d \delta(t-t')  \textrm{d}t'
    = \frac{2d\kT}{\gamma}\tau\,.
    \label{eq:msd_int_dissipation_fluctuation}
\end{align}
To simplify the second term, we use  the trigonometric identity
\begin{align}
    \sin(2\pi f \tf) - \sin(2\pi f t_0) &=
    2\sin\left( \frac{2\pi f (\tf -t_0)}{2} \right)\cos\left(  \frac{2\pi f (\tf + t_0)}{2}\right) \notag \\
    &= 2\sin\left( \pi f \tau \right)\cos\left(  \pi f (\tf + t_0)\right)\,.
    \label{eq:trig_sin_diff}
\end{align}
Subtitution \refeq{eq:trig_sin_diff} into the second term of \refeq{eq:msd_lengthy}  yields:
\begin{align}
    & \left< \left( \frac{A}{2\pi f\gamma} 
    \left(  \sin(2\pi f \tf) - \sin(2\pi f t_0)\right) \hat{\mathbf{n}} \right)^2 \right> = \notag \\ 
    &    \left< \left(  \frac{A}{\pi f\gamma} \right)^2
    \sin^2 \left( \pi f \tau \right) \cos^2 \left(  \pi f (\tf + t_0)\right) \right> = \notag \\
    & \left(  \frac{A}{\pi f\gamma} \right)^2
    \sin^2 \left( \pi f \tau \right) \left< \cos^2 \left(  \pi f (\tf + t_0)\right) \right> \,,  
    \label{eq:sin_cos_av}
\end{align}
where we used that $\hat{\mathbf{n}}$ is a unit vector and therefore $\hat{\mathbf{n}}^2 = 1$. 

The specific choice of initial ($t_0$) and final ($t_f$) sampling times is critical when evaluating the average in the cosine term of \refeq{eq:sin_cos_av}, as this term depends on the sum $(t_0 + t_f)$ rather than the difference $(\tau)$.
Certain sampling procedures, such as selecting $t_0$ and $t_f$ at equivalent phases within the wave period, can introduce systematic biases and lead to incorrect estimates of $\MSD$.
To avoid such artifacts, we average the $\MSD$ obtained from the computer simulations over multiple combinations of $t_0$ and $t_f$ uniformly distributed across the wave period (see \refsec{sec:additional_computational_details}).

Assuming $t_0$ and $\tf$ values are evenly distributed across the wave period,  the average in  \refeq{eq:sin_cos_av} simplifies to
\begin{equation}
    \left(  \frac{A}{\pi f\gamma} \right)^2
    \sin^2 \left( \pi f \tau \right) \left< \cos^2 \left(  \pi f (\tf + t_0)\right) \right> 
    = \frac{1}{2} \left(  \frac{A}{\pi f\gamma} \right)^2
    \sin^2 \left( \pi f \tau \right)  
    \label{eq:sin_cos_av_sol}
\end{equation}
where we used that the average of the cosine-squared function over different angles $\theta$ is $\left < \cos^2 \theta\right> = 1/2$.

Lastly, the third term in the right-hand side of \refeq{eq:msd_lengthy} vanishes due to the zero mean of the noise (\refeq{eq:xi_mean}):
\begin{align}
    \left<  \frac{A}{\pi f\gamma} \left(  \sin(2\pi f \tf) - \sin(2\pi f t_0)\right) \hat{\mathbf{n}} \int_{t_0}^\tf \sqrt{\frac{2\kT} {\gamma}}\boldsymbol{\xi}\dt \right> &= \notag \\   
    \frac{2A}{\pi f\gamma}  \sin\left( \pi f \tau \right)\left<\cos\left(  \pi f (\tf + t_0) \right) \right>  \hat{\mathbf{n}} \int_{t_0}^\tf \sqrt{\frac{2\kT} {\gamma}}\left<\boldsymbol{\xi}\right>\dt 
    &= 0 \,.
\end{align}

Combining all terms, \refeq{eq:msd_lengthy} can be written as
\begin{equation}
    \MSD = 2dD\tau + \frac{1}{2}
    \left(  \frac{A}{\pi f\gamma} \right)^2
      \sin^2(\pi f \tau) \,,
    \label{eq:msd_osc}
\end{equation}
where we have used  the Einstein relation
\begin{equation}
    D = \frac{\kT}{\gamma}.
    \label{eq:einstein_relation}
\end{equation}
In this work, we set the amplitude of the wave to $A = \vmax \gamma$, where $\vmax$ is the peak velocity of the wave,
\begin{equation}
    \MSD = 
    2dD\tau + \frac{1}{2} 
    \left(  \frac{v_{\textrm{max}}}{\pi f} \right)^2
    \sin^2(\pi f \tau) \,.
    \label{eq:msd_osc_final_vmax}
\end{equation}
Assuming linear acoustics, $\vmax$ can be directly related to the applied peak acoustic pressure\cite{kinsler00a}
\begin{equation}
    \vmax = \frac{\Pmax}{\rho c}
    \label{eq:p_max}
\end{equation}
where $\rho$ is the density of the fluid and $c$ is the speed of sound. 
Substituting \refeq{eq:p_max} into \refeq{eq:msd_osc_final_vmax} yields the analytical solution displayed in the main text:
\begin{equation}
    \MSD = 
    2dD_0\tau + \frac{1}{2}
    \left(  \frac{\Pmax}{\pi\rho \,c f} \right)^2
    \sin^2(\pi f \tau).
    \label{eq:msd_osc_final}
\end{equation}
Here we stress that the average in $\MSD$ must be done over $\tau$ values uniformly sampled across the wave period for Eqs. \ref{eq:msd_osc},  \ref{eq:msd_osc_final_vmax} and \ref{eq:msd_osc_final} to be valid.
For small oscillation forces, the last term becomes negligible, and Eq. \ref{eq:msd_osc_final} reduces to the classical Einstein-Smoluchowski equation\cite{howard93a}
\begin{equation}
    \MSD = 
    2dD\tau. 
    \label{eq:einstein_smo}
\end{equation}

\section{Additional results for steric and sticky networks}
\subsection{Steric networks}

\begin{figure}[h!]
    \centering
    \includegraphics[width=0.9\linewidth]{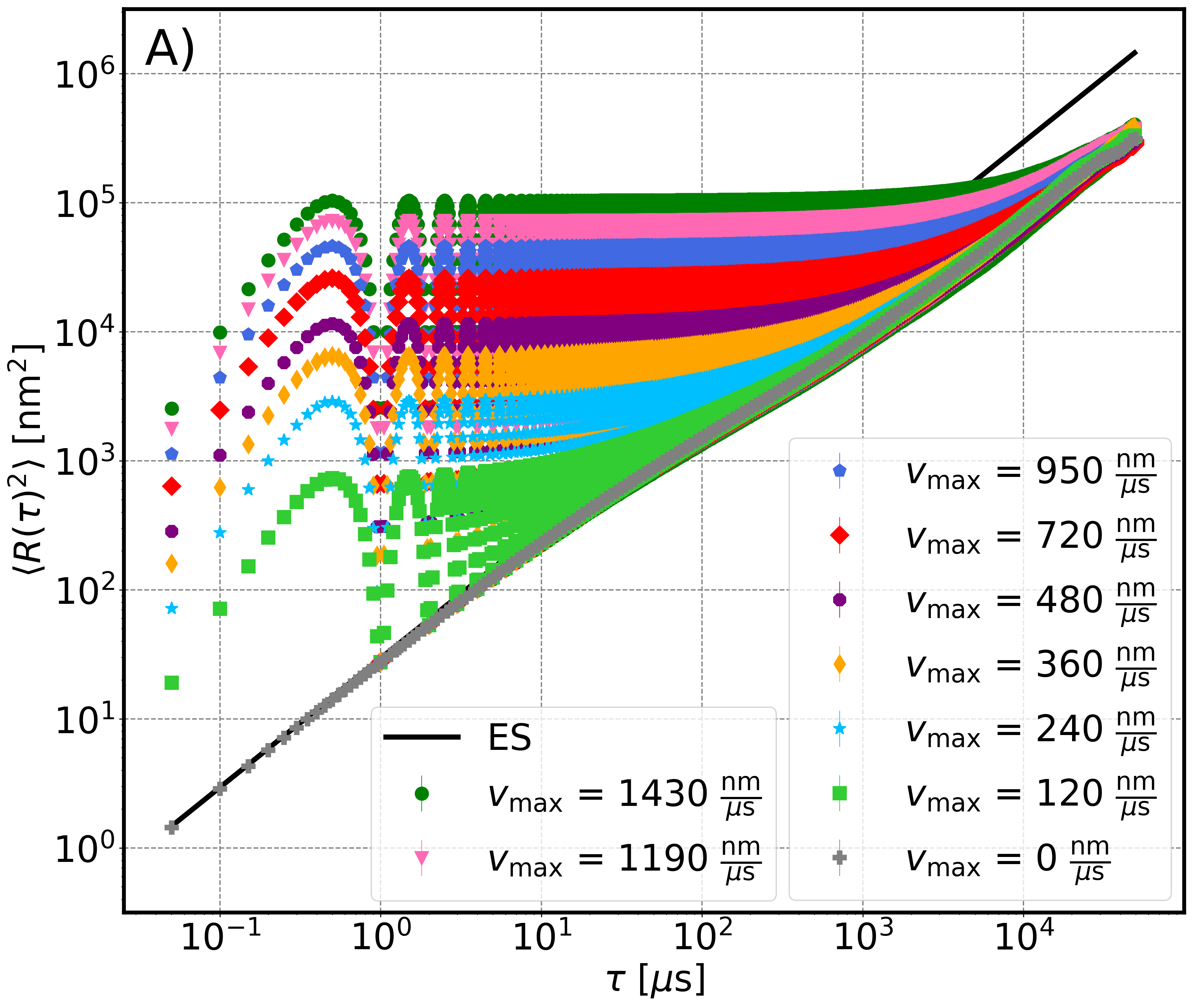}
    \includegraphics[width=0.9\linewidth]{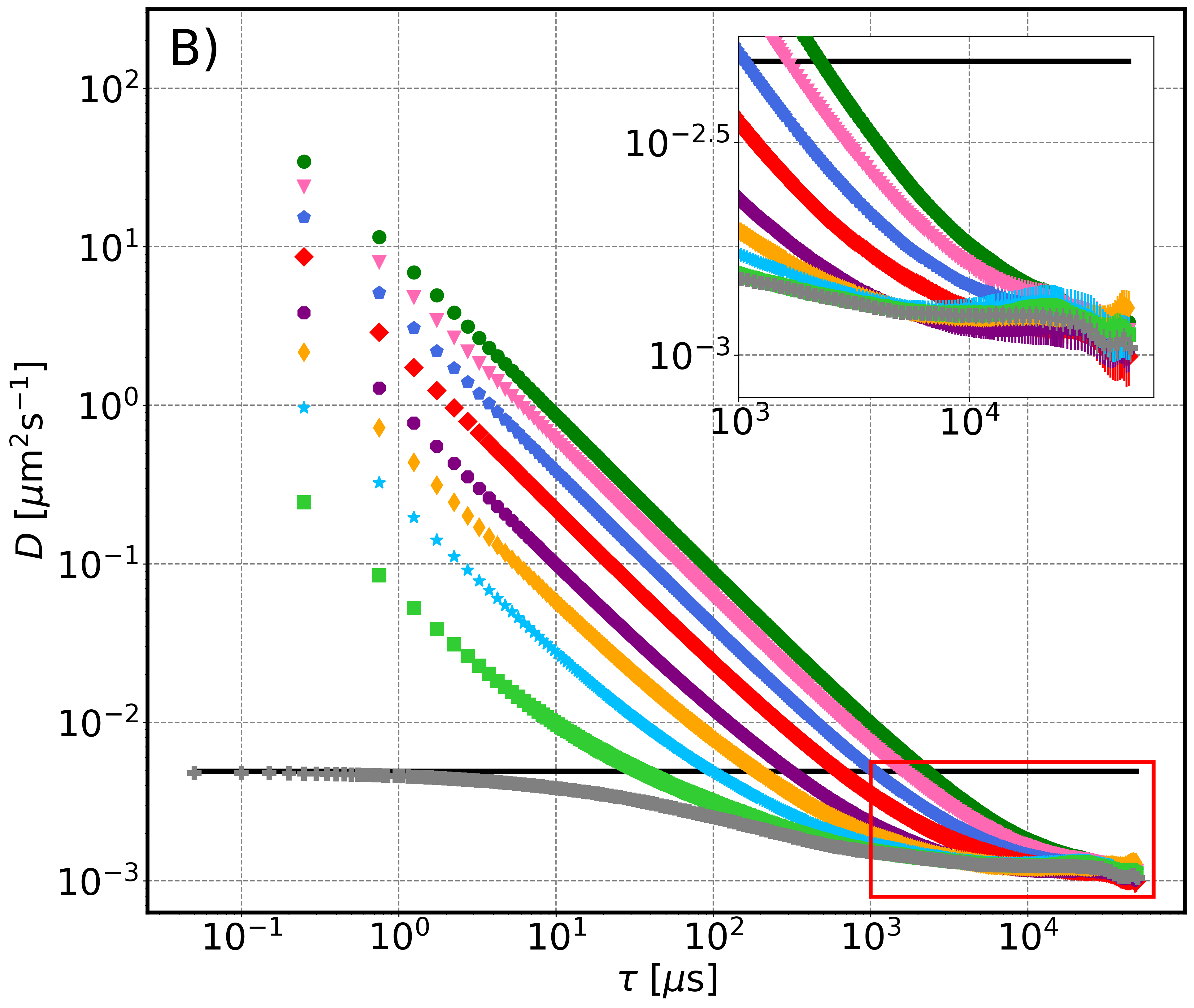} \\
    
     \caption{Mean squared displacement ($\MSD$, panel A) and diffusion coefficient ($D$, panel B) of NPs in a steric hydrogel network with polymer volume fraction $\phi = 1.84 \%$, under varying US peak pressures $\Pmax$.
     Panel A includes the analytical $\MSD$ solution from the Einstein–Smoluchowski equation (\refeq{eq:einstein_smo}, labeled ES).
     In panel B, the input Stokes–Einstein diffusion coefficient $D_0$ (\refeqD) is shown as a reference (solid line).} 
    \label{fig:steric_network_msd_VF_1.84}
\end{figure}

\begin{figure}[h!]
    \centering
    \includegraphics[width=0.9\linewidth]{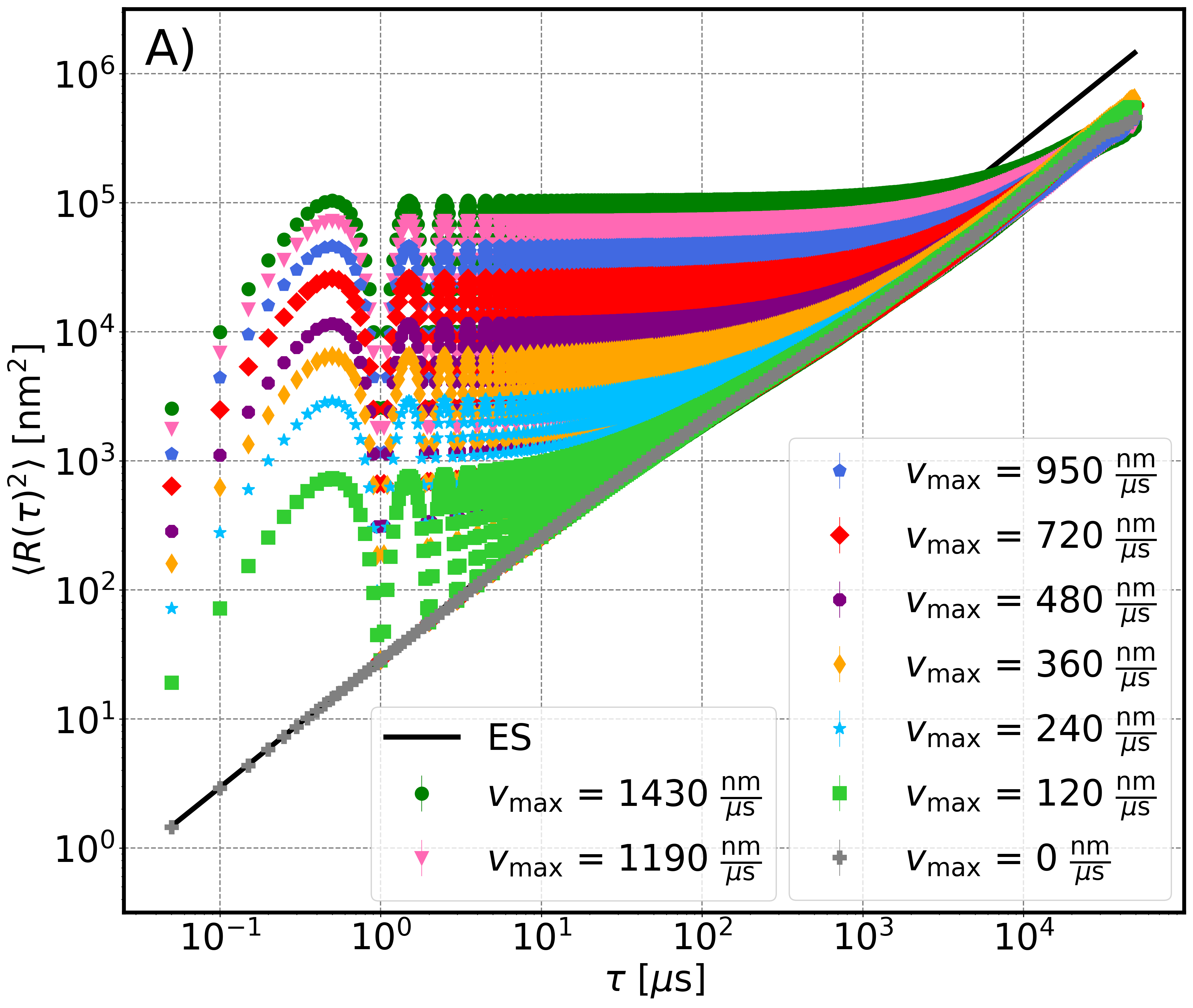}
    \includegraphics[width=0.9\linewidth]{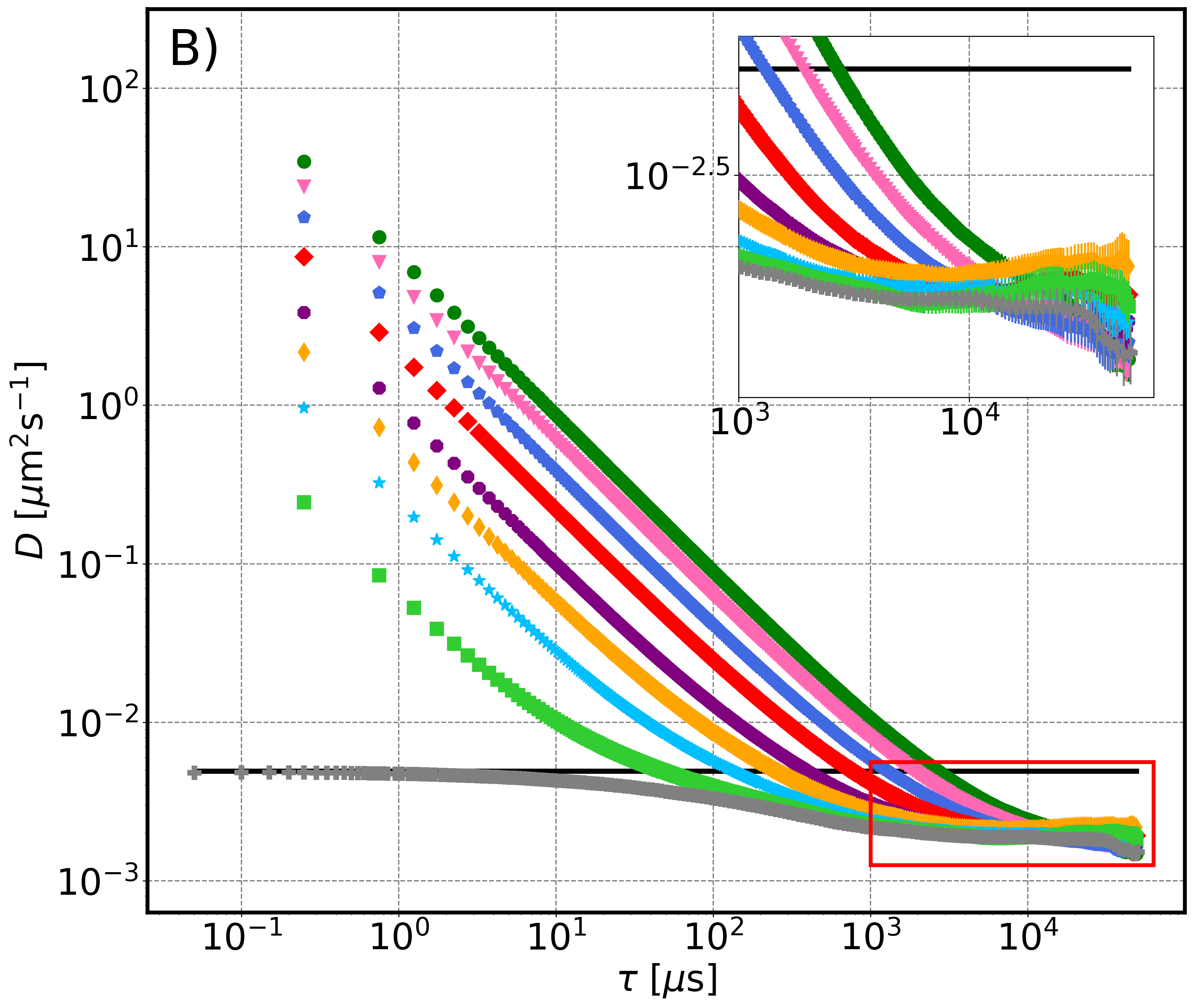} \\
    
     \caption{Mean squared displacement ($\MSD$, panel A) and diffusion coefficient ($D$, panel B) of NPs in a steric hydrogel network with polymer volume fraction $\phi = 1.23 \%$, under varying US peak pressures $\Pmax$.
     Panel A includes the analytical $\MSD$ solution from the Einstein–Smoluchowski equation (\refeq{eq:einstein_smo}, labeled ES).
     In panel B, the input Stokes–Einstein diffusion coefficient $D_0$ (\refeqD) is shown as a reference (solid line).} 
    \label{fig:steric_network_msd_VF_1.23}
\end{figure}

\begin{figure}[h!]
    \centering
    \includegraphics[width=0.9\linewidth]{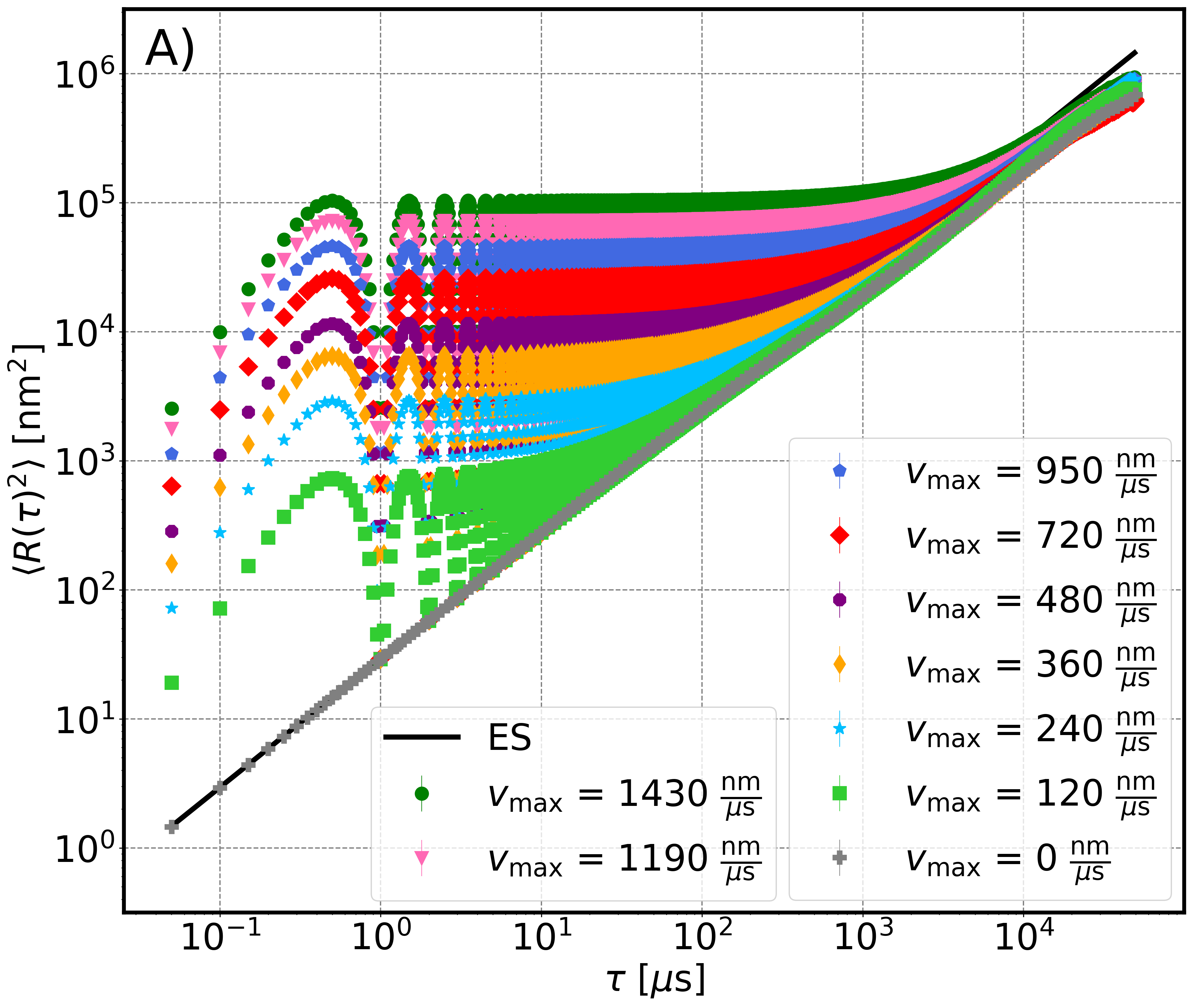}
    \includegraphics[width=0.9\linewidth]{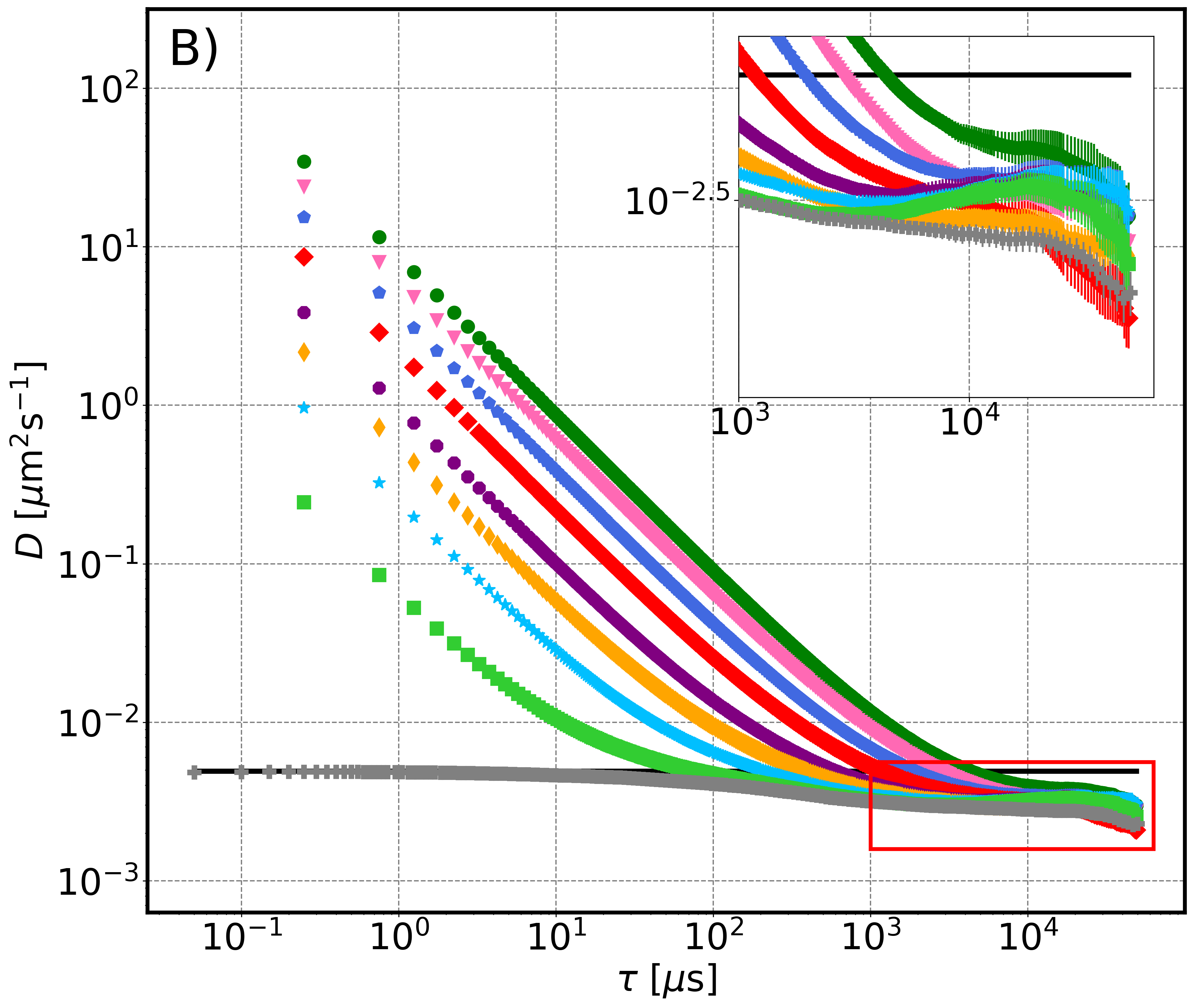} \\
    
     \caption{Mean squared displacement ($\MSD$, panel A) and diffusion coefficient ($D$, panel B) of NPs in a steric hydrogel network with polymer volume fraction $\phi = 0.63 \%$, under varying US peak pressures $\Pmax$.
     Panel A includes the analytical $\MSD$ solution from the Einstein–Smoluchowski equation (\refeq{eq:einstein_smo}, labeled ES).
     In panel B, the input Stokes–Einstein diffusion coefficient $D_0$ (\refeqD) is shown as a reference (continuous line).} 
    \label{fig:steric_network_msd_VF_0.63}
\end{figure}

\newpage
\clearpage
\subsection{Sticky networks}

\begin{figure}[h!]
    \centering
    \includegraphics[width=0.9\linewidth]{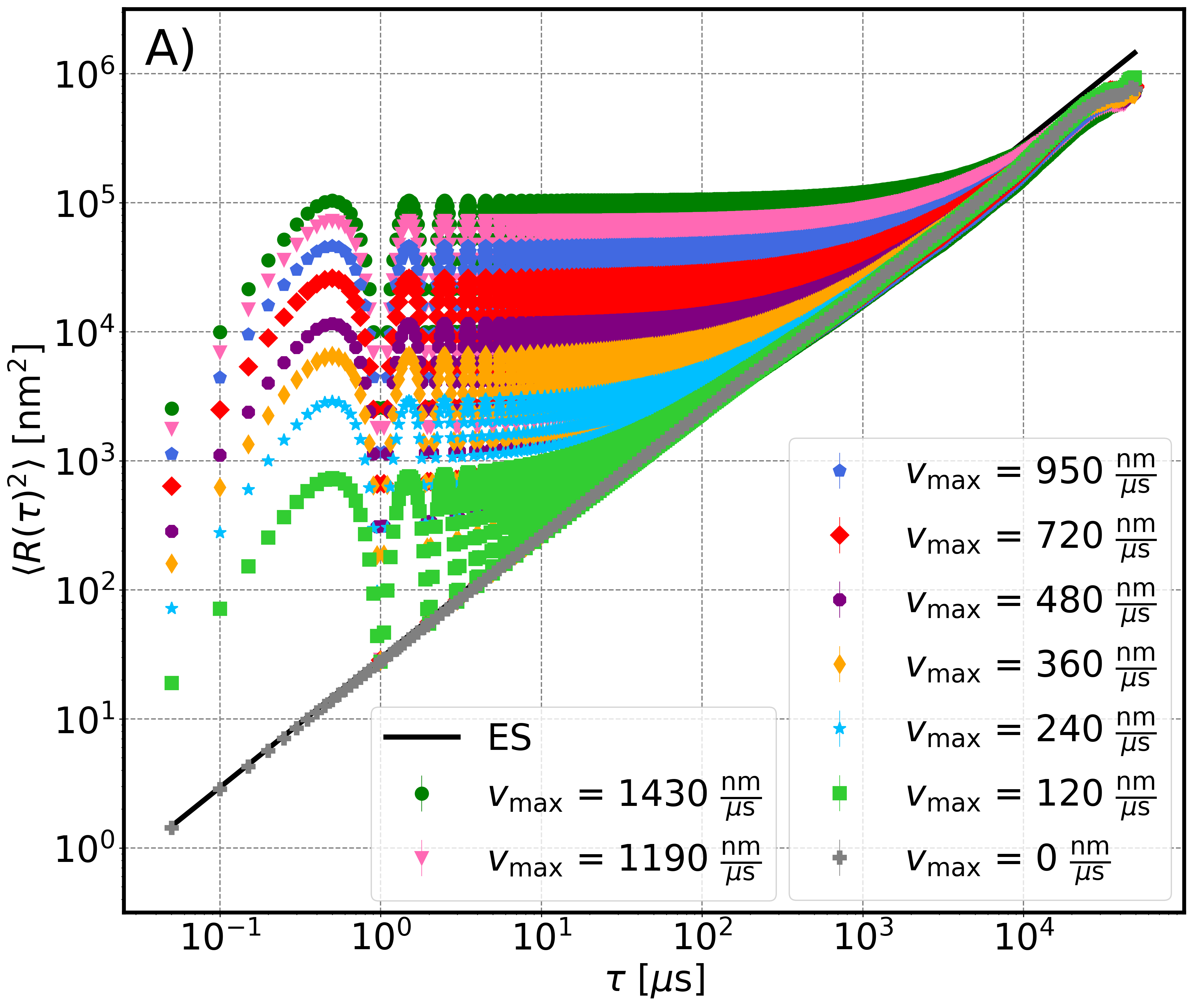}
    \includegraphics[width=0.9\linewidth]{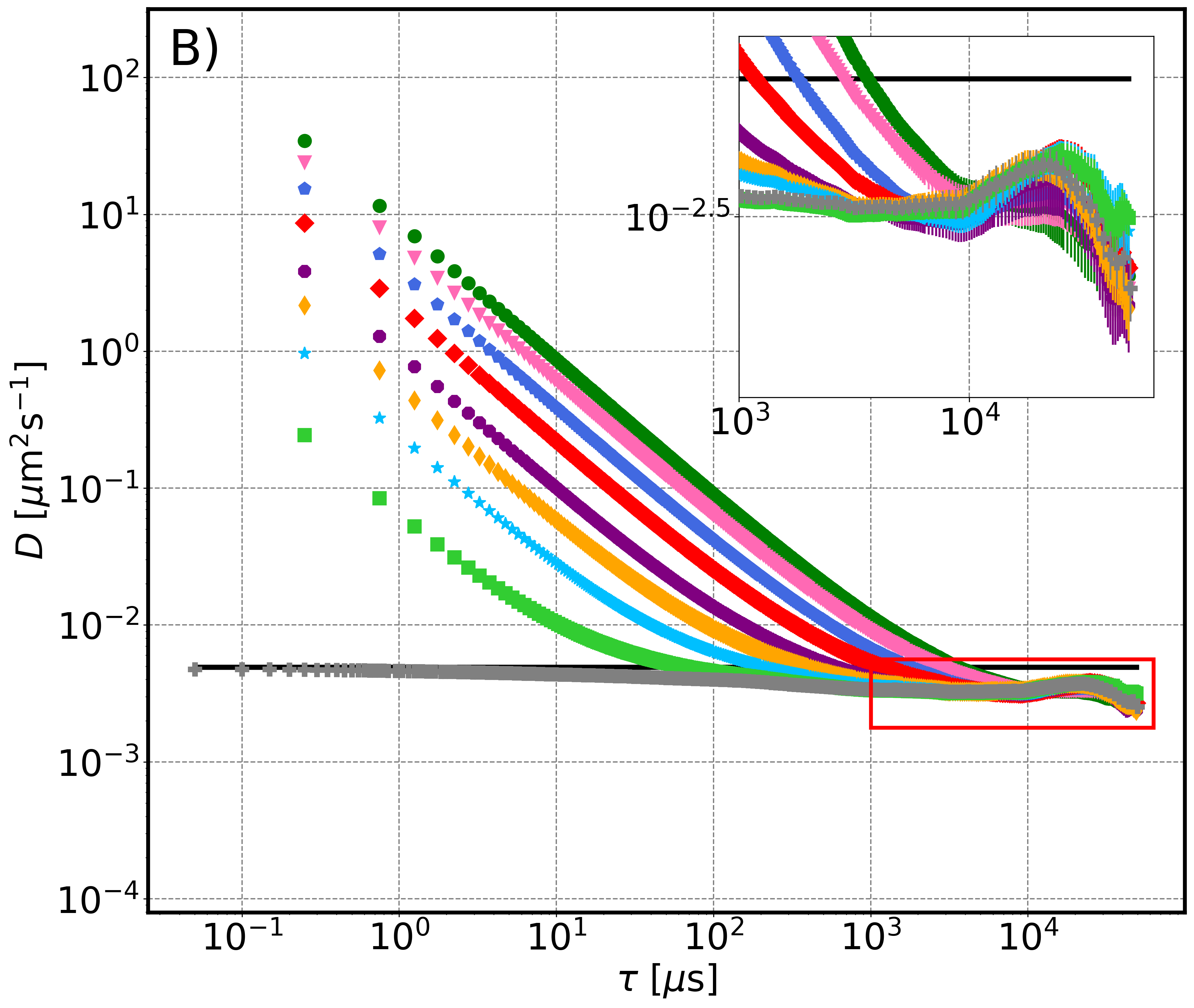} \\
    
     \caption{Mean squared displacement ($\MSD$, panel A) and diffusion coefficient ($D$, panel B) of NPs in a sticky network with polymer volume fraction $\phi = 0.86 \%$ and Lennard-Jones interaction strength $\epsilon = 0.5\kT$, under varying US peak pressures $\Pmax$.
     Panel A includes the analytical $\MSD$ solution from the Einstein–Smoluchowski equation (\refeq{eq:einstein_smo}, labeled ES).
     In panel B, the input Stokes–Einstein diffusion coefficient $D_0$ (\refeqD) is shown as a reference (solid line).}
    \label{fig:sticky_network_msd_epsilon_0.5}
\end{figure}

\begin{figure}[h!]
    \centering
    \includegraphics[width=0.9\linewidth]{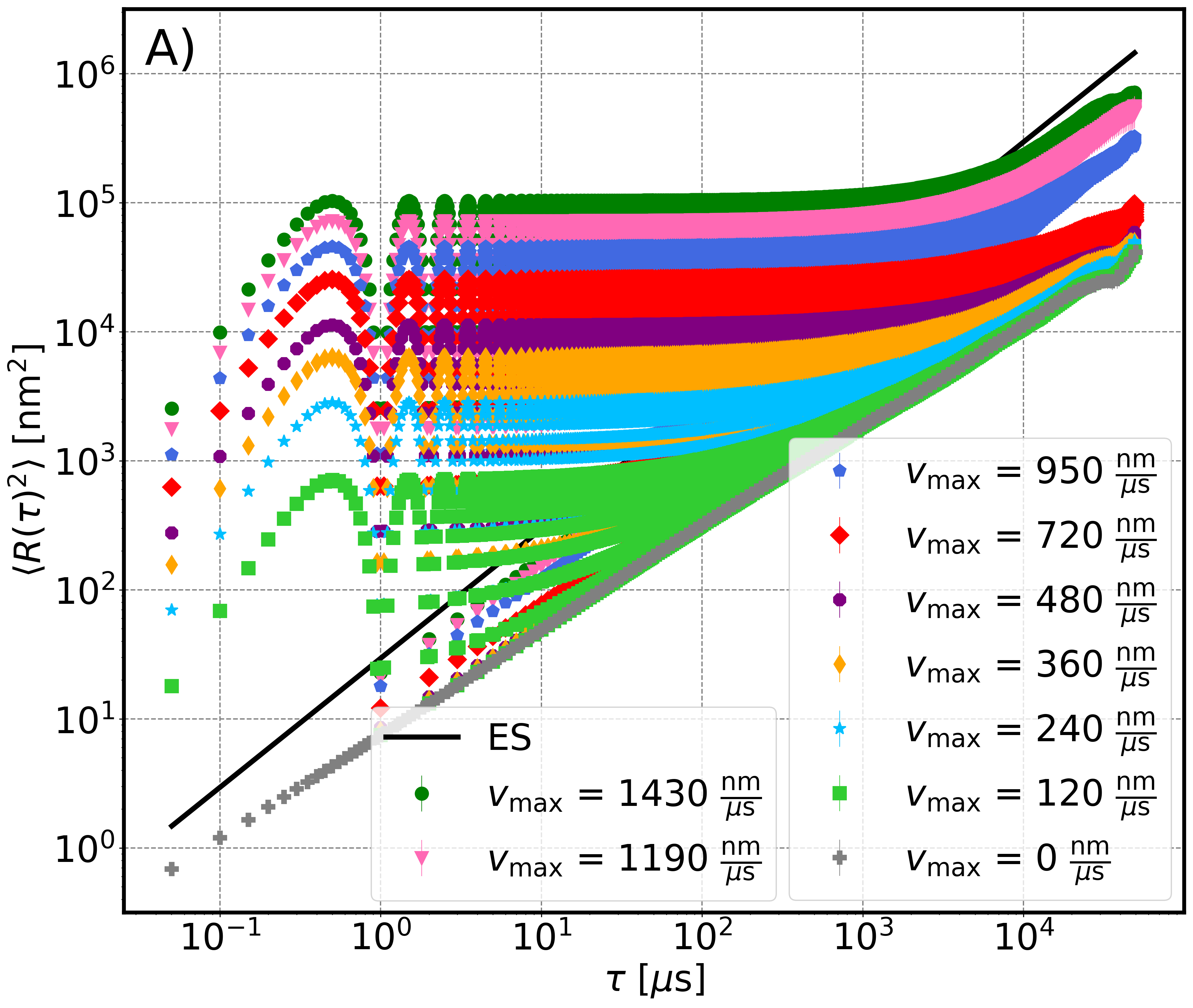}
    \includegraphics[width=0.9\linewidth]{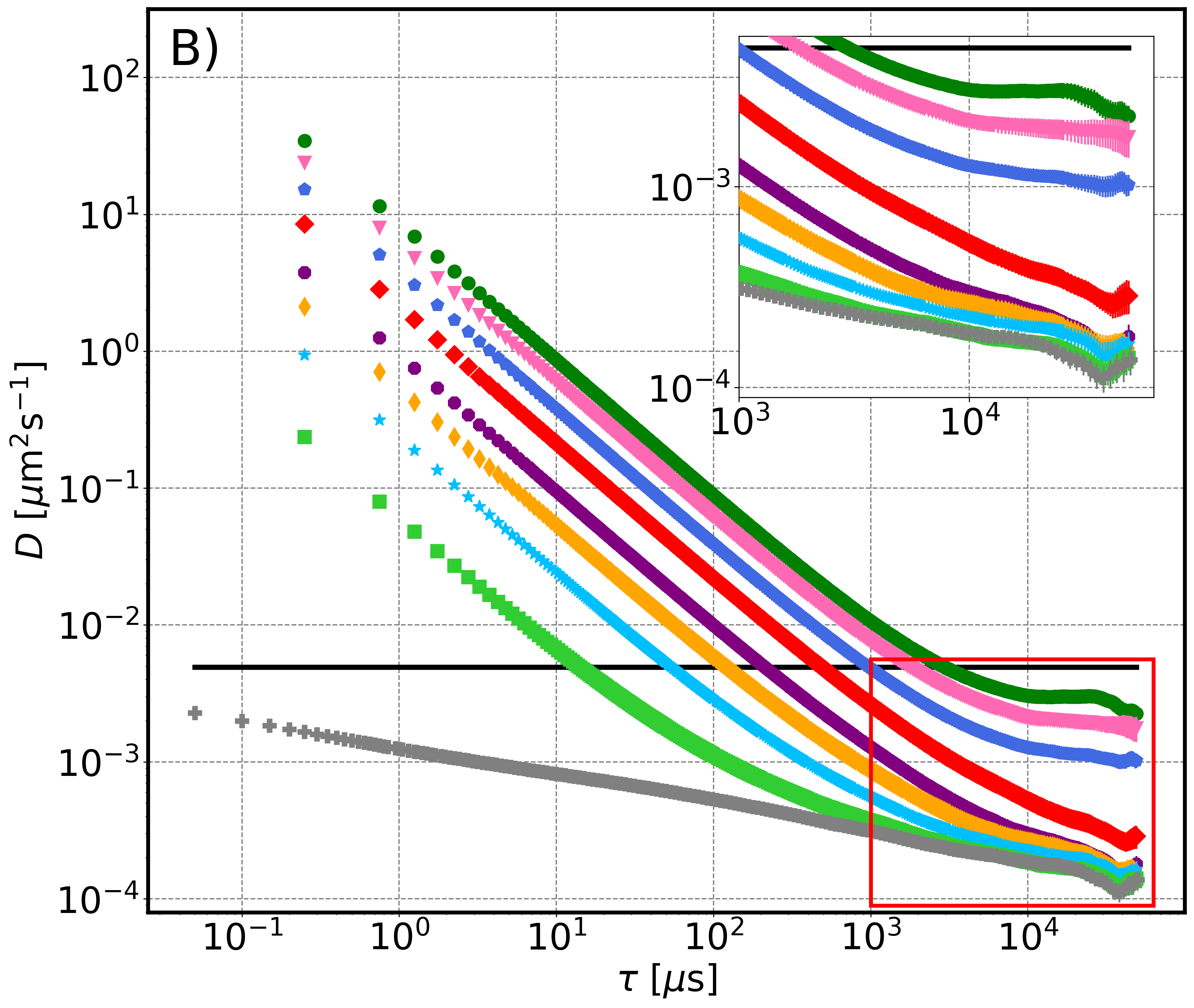} \\
    
     \caption{Mean squared displacement ($\MSD$, panel A) and diffusion coefficient ($D$, panel B) of NPs in a sticky network with polymer volume fraction $\phi = 0.86 \%$ and Lennard-Jones interaction strength $\epsilon = 1.5\kT$, under varying US peak pressures $\Pmax$.
     Panel A includes the analytical $\MSD$ solution from the Einstein–Smoluchowski equation (\refeq{eq:einstein_smo}, labeled ES).
     In panel B, the input Stokes–Einstein diffusion coefficient $D_0$ (\refeqD) is shown as a reference (solid line).}
    \label{fig:sticky_network_msd_epsilon_1.5}
\end{figure}

\newpage
\clearpage

\subsection{Contact time distributions in both networks}

\begin{figure}[h!]
    \centering
    \includegraphics[width=0.9\linewidth]{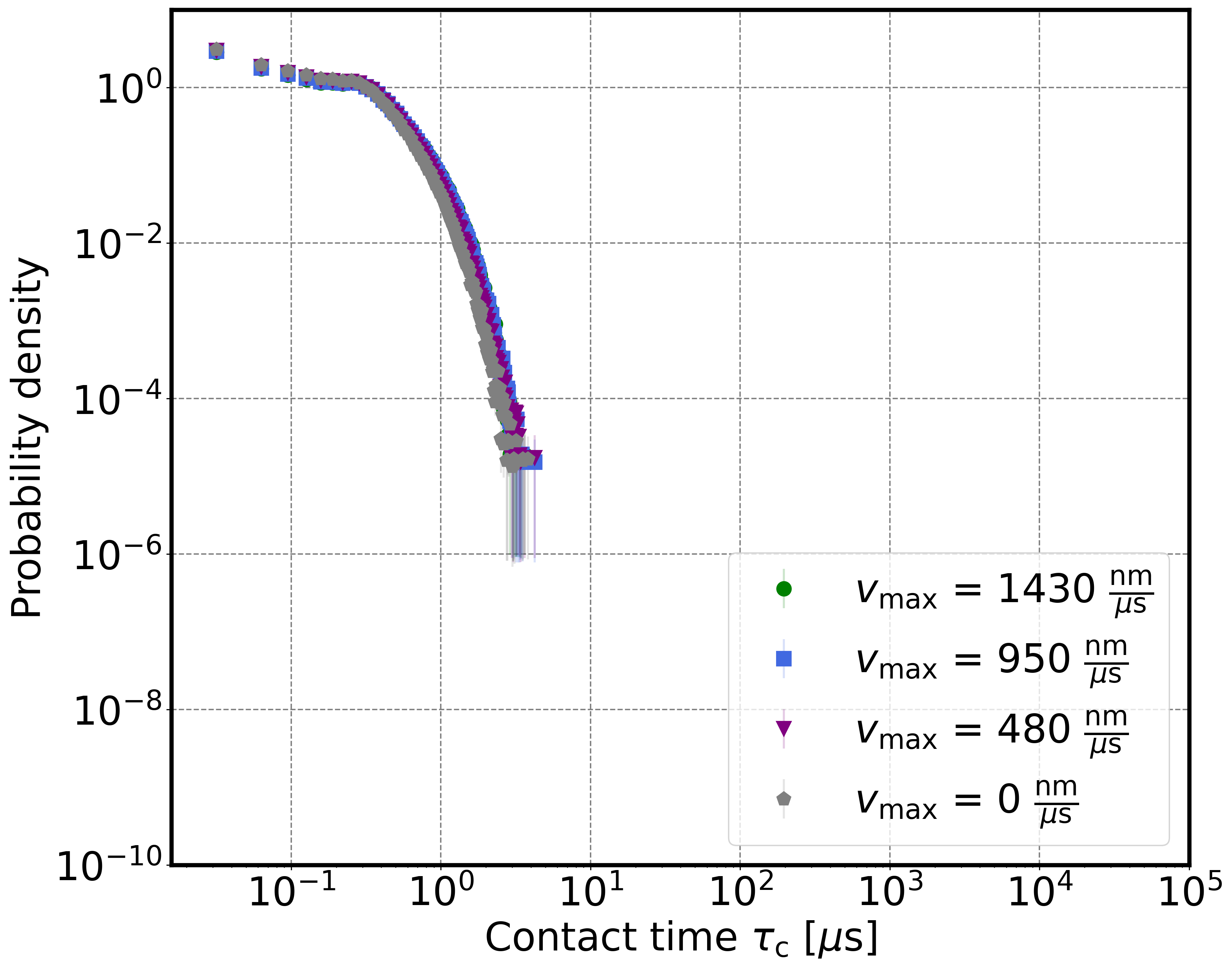}
    \caption{Probability density of hydrogel–NP contact times in a steric hydrogel network with polymer volume fraction $\phi = 0.86\%$, measured at various US peak pressures $\Pmax$.
    Each contact time corresponds to a specific NP-hydrogel bead pair and quantifies the duration of individual steric interactions.
    }
    \label{fig:steric_network_tc}
\end{figure}

\begin{figure}[h!]
    \centering
    \includegraphics[width=0.9\linewidth]{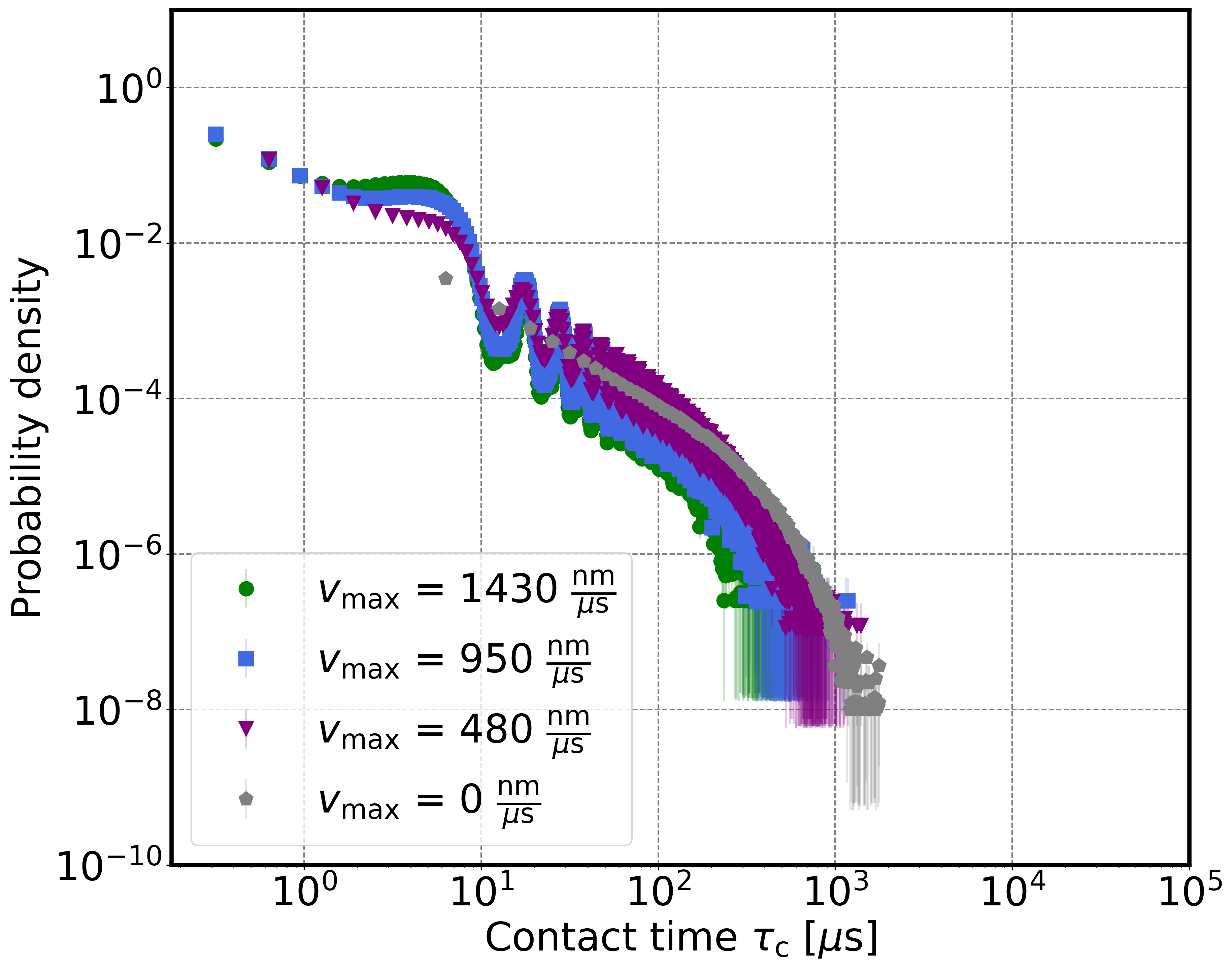}
    \caption{Probability density of hydrogel–NP contact times in a sticky hydrogel network with polymer volume fraction $\phi = 0.86\%$ and Lennard-Jones interaction strength $\epsilon = 0.5\kT$, measured at various US peak pressures $\Pmax$.
    Each contact time corresponds to a specific NP-hydrogel bead pair and quantifies the duration of individual adhesive interactions.
    }
    \label{fig:sticky_network_tc_epsilon_0.5}
\end{figure}

\begin{figure}[h!]
    \centering
    \includegraphics[width=0.9\linewidth]{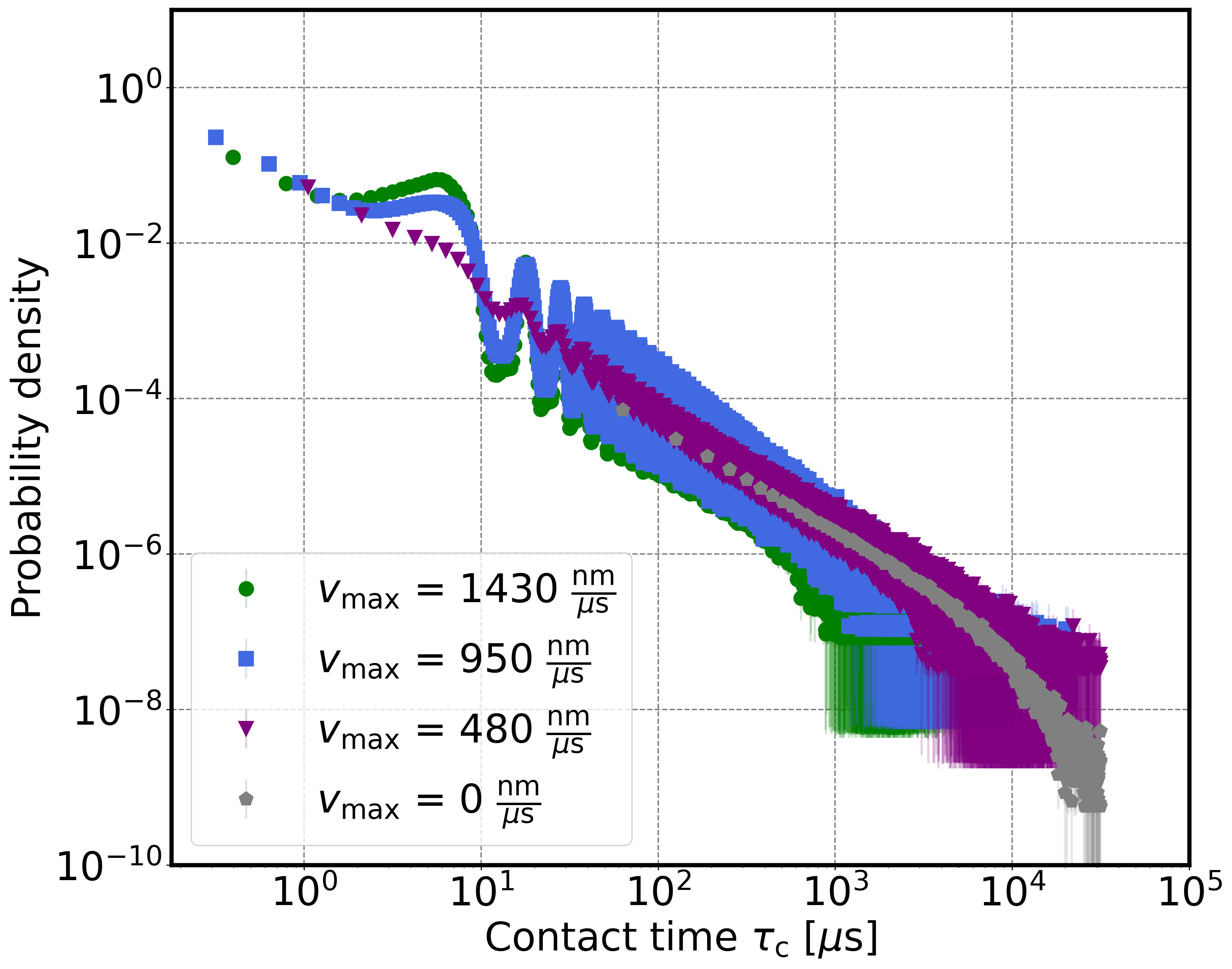}
    \caption{Probability density of hydrogel–NP contact times in a sticky hydrogel network with polymer volume fraction $\phi = 0.86\%$ and Lennard-Jones interaction strength $\epsilon = 1.5\kT$, measured at various US peak pressures $\Pmax$.
    Each contact time corresponds to a specific NP-hydrogel bead pair and quantifies the duration of individual adhesive interactions.}
    \label{fig:sticky_network_tc_epsilon_1.5}
\end{figure}

\clearpage

\section{Schematics of the "stick-and-release" mechanism}

\begin{figure}[h!]
    \centering
    \includegraphics[width=0.95\linewidth]{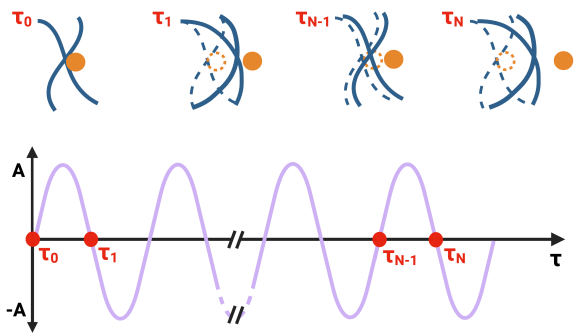}
\caption{The upper panels depict successive configurations of a nanoparticle (orange sphere) interacting with the hydrogel network (blue chains) at different time points, $\tau_0$–$\tau_N$ under stress caused by an acoustic US wave (bottom).
At the initial time $\tau_0$, the NP adheres to the hydrogel through attractive interactions.
Both the NP and the network oscillate in response to the US field, reaching maximum displacement at each half-period of the wave (e.g., $\tau_1$).
After multiple cycles, these oscillations progressively weaken and eventually break the NP–hydrogel contact ($\tau_{N-1}$), leading to NP release ($\tau_N$) and, consequently, enhanced diffusion, compared to a system without US.
    }
    \label{fig:schematics}
\end{figure}

\end{document}